\documentclass[authoryear,preprint,review,12pt]{elsarticle}

\usepackage{geometry}
\usepackage{amsthm, amsmath, amssymb, amsfonts, amsbsy}
\usepackage{mathrsfs} 
\usepackage{stmaryrd} 
\usepackage{mathtools} 
\allowdisplaybreaks
\usepackage{indentfirst}
\usepackage{graphicx}
\usepackage{tikz-cd}
\usepackage{makeidx}
  \makeindex

\usepackage{pdfpages} 
\usepackage{subcaption}
\usepackage{longtable}
\usepackage{multirow}
\usepackage{xcolor}
  \definecolor{b-}{rgb}{0, 0, 0.5}
  \definecolor{r|b-}{rgb}{0.75, 0, 0.25}
  \definecolor{r+}{rgb}{1, 0.8, 0.8}
  \definecolor{g|b-}{rgb}{0, 0.75, 0.25}
\usepackage{ulem}
\newcommand{\mydel}[1]{}
\newcommand{\myadd}[1]{{#1}}
\usepackage[
  colorlinks,
  urlcolor = black,
  linkcolor = b-,
  anchorcolor = r|b-,
  citecolor = g|b-,
  hypertexnames = false 
]{hyperref}
\usepackage{lastpage}
\usepackage{extramarks}
\usepackage[fit]{truncate}
\usepackage{fancyhdr}
\usepackage{float}
\usepackage{stmaryrd}
  \SetSymbolFont{stmry}{bold}{U}{stmry}{m}{n}

\renewcommand{\(}{\left(}
\renewcommand{\)}{\right)}
\renewcommand{\[}{\left\lbrack}
\renewcommand{\]}{\right\rbrack}
\newcommand{\lsbsb}{\left\llbracket}
\newcommand{\rsbsb}{\right\rrbracket}
\newcommand{\lbk}{\left\lbrace}
\newcommand{\rbk}{\right\rbrace}
\newcommand{\lag}{\left\langle}
\newcommand{\rag}{\right\rangle}
\newcommand{\lnm}{\left\Vert}
\newcommand{\rnm}{\right\Vert}
\newcommand{\lmdl}{\left\vert}
\newcommand{\rmdl}{\right\vert}

\newcommand{\larw}{\leftarrow}

\newcommand{\tp}{^{T}} 

\newcommand{\bigO}[1]{\mathcal{O} \( {#1} \)} 
\newcommand{\tr}[1]{\textup{Tr} \( {#1} \)} 

\DeclareMathOperator{\rank}{rank}

\newcommand{\mypower}[2]{%
  \ifthenelse{
    \( \equal{#2}{0} \)
  }{}{%
    \ifthenelse{
      \( \equal{#2}{1} \)
    }{#1}{{#1} ^{#2}}
  }
}
\newcommand{\mykgms}[3]{%
  \ifthenelse{
    \( \equal{#1}{0} \) \AND \( \equal{#2}{0} \) \AND \( \equal{#3}{0} \)
  }{%
    1}{}
  \ifthenelse{
    \( \equal{#1}{0} \) \AND \( \equal{#2}{0} \) \AND \( \NOT \equal{#3}{0} \)
  }{%
    \mypower{\textup{s}}{#3}}{}
  \ifthenelse{
    \( \equal{#1}{0} \) \AND \( \NOT \equal{#2}{0} \) \AND \( \equal{#3}{0} \)
  }{%
    \mypower{\textup{m}}{#2}}{}
  \ifthenelse{
    \( \equal{#1}{0} \) \AND \( \NOT \equal{#2}{0} \) \AND \( \NOT \equal{#3}{0} \)
  }{%
    \mypower{\textup{m}}{#2} \cdot \mypower{\textup{s}}{#3}}{}
  \ifthenelse{
    \( \NOT \equal{#1}{0} \) \AND \( \equal{#2}{0} \) \AND \( \equal{#3}{0} \)
  }{%
    \mypower{\textup{kg}}{#1}}{}
  \ifthenelse{
    \( \NOT \equal{#1}{0} \) \AND \( \equal{#2}{0} \) \AND \( \NOT \equal{#3}{0} \)
  }{%
    \mypower{\textup{kg}}{#1} \cdot \mypower{\textup{s}}{#3}}{}
  \ifthenelse{
    \( \NOT \equal{#1}{0} \) \AND \( \NOT \equal{#2}{0} \) \AND \( \equal{#3}{0} \)
  }{%
    \mypower{\textup{kg}}{#1} \cdot \mypower{\textup{m}}{#2}}{}
  \ifthenelse{
    \( \NOT \equal{#1}{0} \) \AND \( \NOT \equal{#2}{0} \) \AND \( \NOT \equal{#3}{0} \)
  }{%
    \mypower{\textup{kg}}{#1} \cdot \mypower{\textup{m}}{#2} \cdot \mypower{\textup{s}}{#3}}{}
}
\newcommand{\e}[1]{\times 10 ^{#1}}

\counterwithin{equation}{section}
\counterwithin{figure}{section}
\counterwithin{table}{section}

\newcommand{\gridA}{Grid-A}
\newcommand{\gridB}{Grid-B}
\newcommand{\gridC}{Grid-C}

\usepackage{lineno}

\journal{Journal of Computational Physics}

\begin{document}

\begin{frontmatter}



\title{Proper Orthogonal Decomposition-based Model-Order Reduction for Smoothed Particle Hydrodynamics Simulation}


\author[labelSUFE]{Lidong Fang}
\ead{fanglidong@sufe.edu.cn}
\author[labelZilong]{Zilong Song}
\ead{zilong.song@usu.edu}
\author[labelKirk]{Kirk Fraser}
\ead{kirk.fraser@cnrc-nrc.gc.ca}
\author[labelFields]{Faisal Habib}
\ead{fhabib@fields.utoronto.ca}
\author[labelChristopher]{Christopher Drummond}
\ead{christopher.drummond@nrc-cnrc.gc.ca}
\author[labelHuaxiong1,labelHuaxiong2,labelHuaxiong3]{Huaxiong Huang\texorpdfstring{\corref{cor1}}{}}
\ead{hhuang@yorku.ca}

\affiliation[labelSUFE]{
    organization={School of Mathematics, Shanghai University of Finance and Economics},
    city={Shanghai},
    postcode={200433}, 
    state={Shanghai},
    country={China}
}
\affiliation[labelFields]{
    organization={Mathematics, Analytics and Data Science Lab, Fields Institute for Research in Mathematical Sciences},
    city={Toronto},
    postcode={M5T 3J1}, 
    state={Ontario},
    country={Canada}
}
\affiliation[labelZilong]{
    organization={Department of Mathematics and Statistics, Utah State University},
    city={Logan},
    postcode={84322}, 
    state={Utah},
    country={USA}
}
\affiliation[labelKirk]{
    organization={National Research Council Canada},
    city={Saguenay},
    postcode={G7H 8C3}, 
    state={Quebec},
    country={Canada}
}
\affiliation[labelChristopher]{
    organization={NRC Institute for Information Technology},
    city={Ottawa},
    postcode={K1A 0R6}, 
    state={Ontario},
    country={Canada}
}
\affiliation[labelHuaxiong1]{
    organization={Zu Chongzhi Center, Duke Kunshan University},
    city={Suzhou},
    postcode={215316}, 
    state={Jiangsu},
    country={China}
}
\affiliation[labelHuaxiong2]{
    organization={BNU-HKBU United International College},
    city={Zhuhai},
    postcode={519087}, 
    state={Guangdong},
    country={China}
}
\affiliation[labelHuaxiong3]{
    organization={Department of Mathematics and Statistics, York University},
    city={Toronto},
    postcode={M3J 1P3}, 
    state={Ontario},
    country={Canada}
}

\cortext[cor1]{Corresponding author}


\begin{abstract}
  In this paper, we present a projection-based model-order reduction (MOR) technique for smoothed particle hydrodynamics (SPH) simulations, which is a mesh-free approach within the Lagrangian framework. Our approach utilizes the proper orthogonal decomposition (POD) technique to generate a subspace basis for the reduction process.
  The main objective of this study is to conduct an initial exploration of the feasibility of employing POD-based MOR (POD-MOR) in SPH simulations and to quantify the associated POD error. To illustrate the effectiveness of this approach, we consider the friction stir spot welding problem, which involves the coupling of flow equations and heat equation.
  Our findings reveal that, with the same degrees of freedom, POD-MOR significantly reduces computational error compared to the uniform reduction of particle numbers in SPH simulations.
  \myadd{
  Additionally, the acceleration technique of POD-MOR for SPH simulation via linearization and freezing coefficients has been shown to be effective while keeping the error small. We have also showed the effectiveness of POD-MOR in predictive settings in SPH simulations with different parameter values.  
  }
\end{abstract}


\begin{highlights}
  \item Feasibility and effectiveness of POD-based model-order reduction in SPH simulations: The study demonstrates the potential and applicability of using proper orthogonal decomposition (POD) for model-order reduction in smoothed particle hydrodynamics (SPH) simulations. By successfully reducing the complexity of SPH models through POD-based techniques, the research highlights the feasibility of this approach in achieving efficient and accurate representations.
  \item Advantage of POD-based model-order reduction over uniform coarsening: A key finding of the study is that POD-based model-order reduction (POD-MOR) outperforms the uniform reduction of particle numbers in SPH simulations. With the same degrees of freedom, POD-MOR yields significantly smaller errors, indicating its superiority over traditional coarsening techniques. This highlights the potential of POD-MOR as a more effective strategy for reducing computational costs while maintaining model accuracy in SPH simulations.
  \myadd{
  \item Acceleration of POD-based model-order reduction: The POD-MOR for SPH simulation can be accelerated by applying linearization and freezing certain terms. This approach significantly reduces CPU time with only small additional errors and is effective in predictive settings as well. 
  }
\end{highlights}


\end{frontmatter}



\section{Introduction}


Model order reduction (MOR) is highly valuable in engineering and scientific fields as they offer efficient simulations by significantly reducing the degrees of freedom (DoFs) while minimizing the error introduced in the reduction process \cite{lassila2014model, lucia2004reduced, peherstorfer2015dynamic}.
Physical simulations can be computationally costly due to the problem complexity, and in some decision-making problems such as parameter study or design optimization, multiple simulations are needed, and in such cases the MOR will be useful \cite{copeland2022reduced}.

The most popular MOR method is based on projections \cite{benner2015survey}, where the state variables are approximated in a low-dimensional subspace \cite{mojgani2017lagrangian}. 
A typical projection-based MOR has two stages: an offline dimensional compression stage that identifies the dominant modes in the dynamics, and an online evolution stage that computes the approximated solutions \cite{magargal2022lagrangian}.
One leading method for offline dimensional reduction is Proper Orthogonal Decomposition (POD) method \cite{sirovich1987turbulence1, sirovich1987turbulence2, sirovich1987turbulence3, holmes1996turbulence}, also known as the Karhunen–Lo\`eve expansion method \cite{pearson1901liii}, for subspace basis generation, and a non-exhaustive list of other methods can be found in \cite{taira2017modal}. On the other hand, either Galerkin projection or Petrov-Galerkin projection can be applied in the online evolution stage \cite{rowley2004model, rapun2010reduced, carlberg2017galerkin, choi2019space, carlberg2011efficient}.

POD is a powerful singular value decomposition-based method that can automatically identify the most dominant spatial modes within a given dataset and construct a reduced basis using these modes. By capturing the essential features of the system, POD enables the representation of complex phenomena with significantly fewer DoFs. This reduction not only reduces computational costs but also provides deeper insights into the underlying physics \cite{chatterjee2000introduction, berkooz1993proper, kerschen2005method}.

The applicability of POD becomes particularly evident when considering the heat equation on a fixed grid. The method is well-suited for this context due to the rapid decay of high-frequency modes. For example, the heat equation $f _{t} \( x, t \) = a ^{2} f _{xx}\( x, t \)$ with initial condition $f \( x, 0 \) = \cos \( kx \)$, $k \in \mathbb{N}$, and periodic boundary condition $f \( 0, t \) = f \( 2 \pi, t \), f _{x} \( 0, t \) = f _{x} \( 2 \pi, t \)$ on $x \in \[ 0, 2 \pi \]$, has the solution $f \( x, t \) = \exp \( -k a ^{2} t \) \cos \( kx \)$. As high-frequency modes (large $k$) decay quickly, the number of DoFs can be effectively reduced without notably compromising the accuracy of the simulations. Consequently, POD offers an attractive technique for reducing the computational complexity associated with heat transfer problems in various applications.


Another example is the two-dimensional (2D) Friction Stir Welding (FSW) problem solved by the finite difference method in Eulerian framework, where POD can be employed to achieve computational efficiency \cite{cao2022machine}. 
FSW has gained widespread application in industries such as automotive, aeronautical, and structural engineering, and it involves inserting a rotating hardened steel tool, comprising a pin and a shoulder, at the junction of two metal workpieces \cite{fraser2016mesh, fraser2017robust, fraser2018optimization, stubblefield2021meshfree}. By applying axial force, the welding pressure from the tool shoulder ensures continuous contact with the workpiece's top surface. This contact, combined with the friction force between the rotating tool and the workpieces, generates heat and induces local plastic deformation \cite{djurdjanovic2009heat}. Importantly, the temperature of the workpieces is carefully controlled to remain below the melting temperature, but the metal can still flow and blend near the tool so that the workpieces can be welded together. In cases where the tool advancement velocity is set to zero, the process is known as Friction Stir Spot Welding (FSSW), which is primarily used for point-to-point welding \cite{patil2017modeling, yang2012numerical, yang2015simulation, tartakovsky2006modeling}.


The mesh-free Lagrangian framework has recently been employed to address diverse problem domains, such as the cases involving complex boundary conditions and large deformations.
This makes mesh-free methods in the Lagrangian framework, such as Smoothed Particle Hydrodynamics (SPH) \cite{monaghan1992smoothed, monaghan2005smoothed, liu2010smoothed}, well-suited for FSW simulations. 
This raises the question of how MOR techniques, such as POD-based MOR (POD-MOR), behave in the Lagrangian framework, specifically in SPH simulations.

In \cite{mojgani2017lagrangian}, the POD-MOR for mesh-based Lagrangian simulation is established and proved to outperform the one for Eulerian framework in some one-dimensional convection-dominated solutions, since it is not likely to efficiently represent those solutions using summation of products of global Eulerian spatial and temporal basis functions.
In one spatial dimension, the algorithm requires interpolations back to Eulerian framework when the grid entanglement occurs due to the mesh-based nature.
With mesh-based Lagrangian framework, another recent work that applies POD to finite element method is \cite{copeland2022reduced}, where some further techniques like time-windowing and hyper-reduction are used in order to accelerate the computation of reduced model in two and three dimensions.

The mesh-free methods, such as SPH, are quite different from the mesh-based methods, due to the computational difficulties raised from mesh distortion and remeshing. 
Few studies have focused on the \myadd{POD-related analysis} for SPH simulations as far as we know, and a recent exploration is in \cite{magargal2022lagrangian}, where the authors run SPH simulations for a fluid dynamic system and generate POD modes by constructing a post-processing Lagrangian-to-Eulerian mapping. After the mapping, the interpolated Eulerian modes can effectively capture coherent model structures. 
The authors point out that the irregular SPH solution prohibits a traditional snapshot matrix decomposition approach to identify coherent and low-dimensional structures in the dynamical system.
\myadd{However, the MOR is not applied to the SPH simulation in that work.}

It is worth mentioning that the application of POD-MOR has been extended to other Lagrangian models such as molecular dynamics simulation \cite{lee2013proper}. However, it is important to clarify that investigating these applications falls beyond the scope of our current research.

In the context of our current research, the implementation of POD-MOR for SPH simulations has emerged as a critical and pressing project to investigate.
Furthermore, the behavior of errors when reducing the DoFs or  modifying the number of particles in SPH is not well-established. Hence, we are motivated to study whether POD can offer a more intelligent approach to select a smaller number of DoFs compared with uniformly reducing the number of particles.
As far as we know, this is the first systematic implementation of POD-MOR for SPH simulations.


In this paper, our focus is on a 2D FSSW problem, which is modeled using the SPH simulation.  By conducting various tests, we gain insights into the behavior of erros in SPH when the number of particles is varied. Our findings demonstrate that POD can be used to achieve a significant reduction in DoFs automatically while keeping the accuracy. The reduction achieved through POD is more efficient compared  with uniformly reducing particles in the system, leading to smaller errors for the field variables. Additionally, the POD modes provide valuable insights into how POD groups particles and efficiently reduces the DoFs.
\myadd{
Furthermore, we accelerate the POD-MOR for SPH simulation using a linearization approach with frozen terms, where the additional error is of similar order to the original POD error. 
}

The main contributions of this work are summarized as following:
\begin{itemize}
  \item We find out POD-MOR outperforms the uniform reduction of particle numbers in SPH simulations. With the same degrees of freedom, POD-MOR yields significantly smaller errors, emphasizing its superiority over traditional coarsening techniques.
  \item We have illustrated the POD modes in Lagrangian framework for SPH simulations, provide insights into grouping of particles and  effective representation of solutions in SPH simulations. 
  \myadd{
  \item The acceleration technique of POD-MOR for SPH simulation via linearization and freezing terms has been proven effective in reducing the computational cost, even in predictive settings.
  }
  \item Our study serves as a systematic implementation of POD-MOR for SPH simulations in specific scenarios.
\end{itemize}



The manuscript is arranged as follows. In \autoref{sec:model}, we introduce the mathematical modeling of FSW problem, the formulation of SPH simulation, and the POD-MOR method.
Numerical simulations are carried out in \autoref{sec:numerical}. The discussions and conclusions are provided in section \autoref{sec:conclusion}.

\section{Model}

\label{sec:model}

\subsection{Model of FSW Problem}

Assume in our computation, the spatial domain is $\Omega \subset \mathbb{R} ^{3}$, and the temporal interval is $\Theta \subset \mathbb{R}$. The boundary of $\Omega$ contains the inner boundary $\Gamma _{\textup{in}}$ on the lateral surface of tool pin and the outer boundary $\Gamma _{\textup{out}}$ far away from the tool. The unknown variables, i.e., velocity field $\mathbf{u}$, temperature field $T$, and density field $\rho$, are functions of the spatial position $\mathbf{x} \in \Omega$ and the time $t \in \Theta$. The modeling of FSW couples the flow equations with the heat transfer equation, given below. 

For the fluid flow equations, the conservation of mass gives
\begin{align}
  \mathrm{D} _{t} \rho
  = 
  -\rho \nabla \cdot \mathbf{u},
\end{align}
where $\nabla := \partial / \partial \mathbf{x}$ is the spatial derivative operator and $\mathrm{D} _{t} := \partial / \partial t + \mathbf{u} \cdot \nabla$ is the material time derivative operator. The conservation of linear momentum is
\begin{align}
  \rho \mathrm{D} _{t} \mathbf{u}
  & =
  \nabla \cdot \boldsymbol{\sigma},
\end{align}
where $\boldsymbol{\sigma} = -p \mathbf{I} + \mathbf{S}$ is the stress tensor, $p$ is the thermodynamic pressure, $\mathbf{S} = 2 \mu \[ \dot{\boldsymbol{\varepsilon}} - \frac{1}{3} \[ \nabla \cdot \mathbf{u} \] \mathbf{I} \]$ is the deviatoric stress tensor, $\dot{\boldsymbol{\varepsilon}} = \frac{1}{2} \[ \nabla \mathbf{u} + \[ \nabla \mathbf{u} \] \tp \]$ is the strain rate tensor, and $\mu$ is the dynamic viscosity.  The dynamic viscosity is assumed to be temperature dependent and based on Arrhenius chemical kinetics \cite{battezzati1989viscosity}
\begin{align}
  \mu = \mu_0 \exp \( \frac{E}{R T} \),
  \label{eq:model_viscosity}
\end{align}
where $E$ is the activation energy for viscous flow, $\mu _{0}$ is the viscosity at high-temperature-limit, $R$ is the gas constant.
We assume the equation of state is $p = c _{0} ^{2} \[ \rho - \rho _{0} \]$ \cite{fraser2017robust}, where $c _{0}$ is the speed of sound and $\rho _{0}$ is the reference density.

The heat equation is
\begin{equation}
\begin{aligned}
  \rho \mathrm{D} _{t} \[ C T \] 
  & =
  \nabla \cdot \[ k \nabla T \] 
  + 
  \beta \tr{\mathbf{S} \cdot \dot{\boldsymbol{\varepsilon}}}
  \nonumber \\
  & \quad 
  + 
  \eta \frac{F}{V} \lnm \mathbf{u} - \mathbf{u} _{\textup{shoulder}} \rnm \delta _{\mathbf{x} \in \Omega _{\textup{shoulder}}}
  + 
  h _{\textup{c}} \[ T _{0} - T \],
\end{aligned}
\end{equation}
\cite{fraser2017robust},
where $C$ is the specific heat capacity, 
$k$ is the heat conduction coefficient, 
$\beta$ is the fraction of plastic work converted to heat, 
$\eta$ is the fraction of frictional work converted to heat, 
$F$ is the force applied by the tool shoulder onto the workpiece, 
$V$ is the effective volume of the workpiece under the tool shoulder, 
$\mathbf{u} _{\textup{shoulder}} = \mathbf{u} _{0} + \omega \mathbf{z} \times \[ \mathbf{x} - \mathbf{x} _{0} \]$ is the velocity of the tool shoulder,
$\omega \mathbf{z}$ is the angular velocity of the tool which is parallel to the tool axis, 
$\mathbf{z}$ is the unit vector parallel to the tool axis, 
$\mathbf{x} _{0}$ is the position of the tool center, 
and $\mathbf{u} _{0} \equiv 0$ means the tool center is fixed as in FSSW,  
$\Omega _{\textup{shoulder}}$ is the tool shoulder part, 
$h _{\textup{c}}$ is the heat transfer coefficient, 
$T _{0}$ is the ambient temperature.

On the boundary, we set the nonslip boundary condition $\mathbf{u} = \mathbf{u} _{\textup{shoulder}}$ on $\Gamma _{\textup{in}}$ and the Dirichlet boundary condition $\mathbf{u} \equiv 0$ on $\Gamma _{\textup{out}}$.
Since we have already considered the main sources (plastic and frictional work) that generating heat and there is a uniform heat transfer to cool the workpiece down, we assume the adiabatic on the boundary for simplicity by ignoring other minor effects like heat exchange with the tool and the outer boundary.
We do not need any boundary conditions for density since it is solved by an initial-value problem.

In this study, considering the real-world metal workpieces are thin compared with the size of tool, it is reasonable to assume the physical quantities are uniform in the $\mathbf{z}$ direction that is perpendicular to workpieces.
Thus we can simplify the complex 3D physical problem associated with FSSW into a more manageable 2D framework so that the two workpieces to be welded are treated as two thin plates arranged next to each other \cite{seidel2003two,cao2022machine}. 
Actually we regard $\Omega \subset \mathbb{R} ^{2}$, and all variables are functions of the first two spatial coordinates and the time $t$. The velocity $\mathbf{u}$ is regarded as a function from $\Omega \times \Theta$ to $\mathbb{R} ^{2}$, so the strain and stress tensors are modified correspondingly.
In the 2D FSSW scenario, the rotating pin is represented by a circular object undergoing rotational motion with the boundary $\Gamma _{\textup{in}}$. The shoulder zone $\Omega _{\textup{shoulder}} \subset \mathbb{R} ^{2}$, in our 2D model, refers to the circular region where the workpiece comes into contact with the shoulder of the tool.

\subsection{SPH in Lagrangian Framework}

SPH is a mesh-free particle method based on Lagrangian formulation, and has been widely applied to different areas in engineering and science. In the FSSW problem, the position $\mathbf{x}$, velocity $\mathbf{u}$, temperature $T$, and density $\rho$ need to be solved for each particle through the governing equations in SPH formulation.

We follow the standard SPH formulation, which contains two main steps. The first step, kernel approximation, is to represent a function and its derivatives in continuous form as integrals over $\Omega$. The second step, particle approximation, is to rewrite the integrals over the computational domain as summations over discrete particles. In this way, the particle evolution equations can be derived as a dynamical systems of $\mathbf{x} _{i}$, $\mathbf{u} _{i}$, $T _{i}$, $\rho _{i}$, for all particle $i$, in our FSSW problem.

In the kernel approximation step, we define a 2D smoothing kernel function as a widely-used cubic spline function \cite{monaghan2005smoothed} $W _{h}: \mathbb{R} ^{2} \to \mathbb{R}$ as follows
\begin{align}
  W _{h} \( \mathbf{x} \)
  & :=
  \tilde{W} _{h} \( \lmdl \mathbf{x} \rmdl \)
  =
  \frac{10}{7 \pi h ^{2}}
  \left\{
  \begin{aligned}
    & 
    1 - \frac{3}{2} \[ \frac{\lmdl \mathbf{x} \rmdl}{h} \] ^{2} + \frac{3}{4} \[ \frac{\lmdl \mathbf{x} \rmdl}{h} \] ^{3}, 
    && 
    0 \leq \frac{\lmdl \mathbf{x} \rmdl}{h} \leq 1,
    \\
    & 
    \frac{1}{4} \[ 2 - \frac{\lmdl \mathbf{x} \rmdl}{h} \] ^{3}, 
    && 
    1 \leq \frac{\lmdl \mathbf{x} \rmdl}{h} \leq 2,
    \\
    & 
    0, 
    && 
    2 \leq \frac{\lmdl \mathbf{x} \rmdl}{h},
  \end{aligned}
  \right.
\end{align}
where $h > 0$ is the smoothing length. The kernel function $W _{h}$ has a finite support of 2D ball with radius $2h$, and approaches a Delta function as $h \to 0$.

For a scalar field function $f: \Omega \times \Theta \to \mathbb{R}$, we define the approximation function $\lag f \rag: \Omega \times \Theta \to \mathbb{R}$ as
\begin{align}
  \lag f \rag \( \mathbf{x}, t \)
  :=
  \int _{\Omega} f \( \mathbf{y}, t \) W _{h} \( \mathbf{x} - \mathbf{y} \) \, \mathrm{d} \mathbf{y}.
\end{align}
Moreover, using integration by parts and assuming $\int _{\partial \Omega} f \( \mathbf{y}, t \) W _{h} \( \mathbf{x} - \mathbf{y} \) \mathbf{n} \( \mathbf{y} \) \, \mathrm{d} \mathbf{y} = 0$, the spatial derivative $\nabla f$ can be approximated by $\lag \nabla f \rag$,
\begin{align}
  \lag \nabla f \rag \( \mathbf{x}, t \)
  & :=
  \int _{\Omega} f \( \mathbf{y}, t \) \nabla W _{h} \( \mathbf{x} - \mathbf{y} \) \, \mathrm{d} \mathbf{y}.
\end{align}
The second spatial derivative $\nabla \nabla f$ can be approximated by $\lag \nabla \nabla f \rag$ by assuming $\nabla \nabla \nabla \nabla f$ are bounded in the neighborhood of $\mathbf{x}$,
\begin{align}
  \lag \nabla \nabla f \rag \( \mathbf{x}, t \)
  & :=
  \int _{\Omega} \[ f \( \mathbf{x}, t \) - f \( \mathbf{y}, t \) \] \frac{\[ \mathbf{x} - \mathbf{y} \] \cdot \nabla W _{h} \( \mathbf{x} - \mathbf{y} \)}{\lmdl \mathbf{x} - \mathbf{y} \rmdl ^ {2}} \[ 4 \frac{\[ \mathbf{x} - \mathbf{y} \] \[ \mathbf{x} - \mathbf{y} \]}{\lmdl \mathbf{x} - \mathbf{y} \rmdl ^{2}} - \mathbf{I} \] \, \mathrm{d} \mathbf{y},
\end{align}
which is an extension of \cite{brookshaw1985method, espanol2003smoothed, monaghan2005smoothed}.

In the particle approximation step, we denote $\Lambda$ as the index set of particles that are filled in the computations domain $\Omega$. Given the particle position $\lbk \mathbf{x} _{i} \( t \) \rbk _{i \in \Lambda}$ and particle mass $\lbk m _{i} \rbk _{i \in \Lambda}$, we can approximate the integrals by summations over the position of particles,
\begin{align}
  \lag \lag f \rag \rag \( \mathbf{x}, t \)
  :=
  \sum _{j \in \Lambda}
  \frac{m _{j}}{\rho \( \mathbf{x} _{j} \( t \), t \)} 
  f \( \mathbf{x} _{j} \( t \), t \) 
  W _{h} \( \mathbf{x} - \mathbf{x} _{j} \( t \) \).
\end{align}
Based on the different chain rules of derivatives, we can approximate the spatial derivative as the following two options, for the sake of symmetry,
\begin{align}
  \lag \lag \nabla f \rag \rag \( \mathbf{x}, t \)
  :=
  \frac{1}{\rho \( \mathbf{x}, t \)}
  \sum _{j \in \Lambda}
  m _{j}
  \[ f \( \mathbf{x} _{j} \( t \), t \) - f \( \mathbf{x}, t \) \]
  \nabla W _{h} \( \mathbf{x} - \mathbf{x} _{j} \( t \) \),
\end{align}
or
\begin{align}
  \lag \lag \nabla f \rag \rag \( \mathbf{x}, t \)
  :=
  \rho \( \mathbf{x}, t \)
  \sum _{j \in \Lambda}
  m _{j}
  \[ \frac{f \( \mathbf{x} _{j} \( t \), t \)}{\[ \rho \( \mathbf{x} _{j} \( t \), t \) \] ^{2}} + \frac{f \( \mathbf{x}, t \)}{\[ \rho \( \mathbf{x}, t \) \] ^{2}} \]
  \nabla W _{h} \( \mathbf{x} - \mathbf{x} _{j} \( t \) \).
\end{align}
The second spatial derivative can be approximated as
\begin{align}
  \lag \lag \nabla \nabla f \rag \rag \( \mathbf{x}, t \)
  & :=
  \sum _{j \in \Lambda}
  \frac{m _{j}}{\rho \( \mathbf{x} _{j} \( t \), t \)}
  \[ f \( \mathbf{x}, t \) - f \( \mathbf{x} _{j} \( t \), t \) \] 
  \nonumber \\
  & \quad \quad
  \frac{\[ \mathbf{x} - \mathbf{x} _{j} \( t \) \] \cdot \nabla W _{h} \( \mathbf{x} - \mathbf{x} _{j} \( t \) \)}{\lmdl \mathbf{x} - \mathbf{x} _{j} \( t \) \rmdl ^ {2}} 
  \[ 4 \frac{\[ \mathbf{x} - \mathbf{x} _{j} \( t \) \] \[ \mathbf{x} - \mathbf{x} _{j} \( t \) \]}{\lmdl \mathbf{x} - \mathbf{x} _{j} \( t \) \rmdl ^{2}} - \mathbf{I} \].
\end{align}

So far, all approximations are represented in the Eulerian framework.
In order to write down formulations for SPH simulation, which are in the Lagrangian framework. We additionally derive a similar approximation $\lsbsb f \rsbsb _{i} \( t \) \sim \lag \lag f \rag \rag \( \mathbf{x} _{i} \( t \), t \)$ for an arbitrary Lagrangian particle function $f _{i} \( t \)$, ($i \in \Lambda$, we will omit this in the following), by defining,
\begin{align}
  \lsbsb f \rsbsb _{i} \( t \)
  :=
  \sum _{j \in \Lambda}
  \frac{m _{j}}{\rho _{j} \( t \)} 
  f _{j} \( t \) 
  W _{h} \( \mathbf{x} _{ij} \( t \) \).
\end{align}
The first spatial derivative is either
\begin{align}
  \lsbsb \nabla f \rsbsb _{i} \( t \)
  :=
  \frac{1}{\rho _{i} \( t \)}
  \sum _{j \in \Lambda}
  m _{j}
  f _{ji} \( t \)
  \nabla W _{h} \( \mathbf{x} _{ij} \( t \) \),
\end{align}
or
\begin{align}
  \lsbsb \nabla f \rsbsb _{i} \( t \)
  :=
  \rho _{i} \( t \)
  \sum _{j \in \Lambda}
  m _{j}
  \[ \frac{f _{j} \( t \)}{\[ \rho _{j} \( t \) \] ^{2}} + \frac{f _{i} \( t \)}{\[ \rho _{i} \( t \) \] ^{2}} \]
  \nabla W _{h} \( \mathbf{x} _{ij} \( t \) \).
\end{align}
The second spatial derivative is
\begin{align}
  \lsbsb \nabla \nabla f \rsbsb _{i} \( t \)
  & :=
  \sum _{j \in \Lambda}
  \frac{m _{j}}{\rho _{j} \( t \)}
  f _{ij} \( t \)
  \frac{\mathbf{x} _{ij} \( t \) \cdot \nabla W _{h} \( \mathbf{x} _{ij} \( t \) \)}{\lmdl \mathbf{x} _{ij} \( t \) \rmdl ^ {2}} 
  \[ 4 \frac{\mathbf{x} _{ij} \( t \) \mathbf{x} _{ij} \( t \)}{\lmdl \mathbf{x} _{ij} \( t \) \rmdl ^{2}} - \mathbf{I} \].
\end{align}
Here we denote $g _{ij} \( t \) := g _{i} \( t \) - g _{j} \( t \)$, for any particle function $g$. 

Now, the FSSW governing equations can be written in SPH formulation, where neighbor search is used to determine the neighbors within kernel support for every single particle.
In fact, any numerical ODE scheme can be used for temporal integration, and the time step $\Delta t$ should be chosen based on the Courant–Friedrichs–Lewy (CFL) condition with a CFL parameter $\alpha _{\textup{CFL}}$, see \autoref{sec:numerical_sphSetup} for details based on the SPH setup.
Here we drop the dependency on the temporal variable $t$, and denote $W _{ij} := W \( \mathbf{x} _{ij} \)$, and $\nabla W _{ij} := \nabla W \( \mathbf{x} _{ij} \)$. We use the notation $\sum _{j \sim i}$ for the summation over $j \in \Lambda$ and $\lmdl \mathbf{x} _{ij} \rmdl \le 2h$ so that the kernel function is nonzero. 

The conservation of mass derives the update of particle density,
\begin{align}
  \dot{\rho} _{i}
  =
  \rho _{i} 
  \sum _{j \sim i}
  \frac{m _{j}}{\rho _{j}}
  \mathbf{u} _{ij} \cdot \nabla W _{ij}.
  \label{eq:model_SPH_massConservation}
\end{align}

The conservation of linear momentum derives the update of particle velocity,
\begin{align}
  \dot{\mathbf{u}} _{i}
  & =
  - 
  \sum _{j \sim i}
  m _{j} 
  \[ \frac{p _{i} + p _{j}}{\rho _{i} \rho _{j}} \] 
  \nabla W _{ij} 
  \nonumber \\
  & \quad
  + 
  \sum _{j \sim i} 
  \frac{m _{j} \mu _{i}}{\rho _{i} \rho _{j}}
  \frac{\mathbf{x} _{ij} \cdot \nabla W _{ij}}{\lnm \mathbf{x} _{ij} \rnm ^{2}}
  \left[
  \frac{5}{3} \mathbf{u} _{ij}
  +
  \frac{4}{3} \mathbf{x} _{ij} \frac{\mathbf{u} _{ij} \cdot \mathbf{x} _{ij}}{\lnm \mathbf{x} _{ij} \rnm ^{2}}
  \right] 
  \nonumber \\
  & \quad
  + 
  \[ \sum _{j \sim i} \frac{m _{j}}{\rho _{j}} \mathbf{u} _{ji} \nabla W _{ij} \] 
  \cdot
  \[ \sum _{j \sim i} \frac{m _{j} \mu _{j}}{\rho _{i} \rho _{j}} \nabla W _{ij} \]
  \nonumber \\
  & \quad
  +  
  \[ \sum _{j \sim i} \frac{m _{j} \mu _{j}}{\rho _{i} \rho _{j}} \nabla W _{ij} \]
  \cdot
  \[ \sum _{j \sim i} \frac{m _{j}}{\rho _{j}} \mathbf{u} _{ji} \nabla W _{ij} \]
  \nonumber \\
  & \quad
  - 
  \frac{2}{3}
  \[ \sum _{j \sim i} \frac{m _{j} \mu _{j}}{\rho _{i} \rho _{j}} \nabla W _{ij} \]
  \[ \sum _{j \sim i} \frac{m _{j}}{\rho _{j}} \mathbf{u} _{ji} \cdot \nabla W _{ij} \]
  .
  \label{eq:model_SPH_linearMomentumConservation}
\end{align}
where we use the second spatial derivative instead of the strain and stress to avoid a large support domain.

The heat equation derives the update of particle temperature,
\begin{align}
  \[ \dot{C _{i} T _{i}} \]
  & 
  =
  \frac{1}{\rho _{i}}
  \sum _{j \sim i}
  \frac{m _{j}}{\rho _{j}}
  \frac{4 k _{i} k _{j}}{k _{i} + k _{j}}
  \frac{T _{ij} \mathbf{x} _{ij} \cdot \nabla W _{ij}}{\lnm \mathbf{x} _{ij} \rnm ^{2}}
  \nonumber \\
  & \quad
  +
  \frac{1}{\rho _{i}}
  \[ \sum _{j \sim i} \frac{m _{j}}{\rho _{j}} k _{j} \nabla W _{ij} \]
  \cdot
  \[ \sum _{j \sim i} \frac{m _{j}}{\rho _{j}} T _{j} \nabla W _{ij} \]
  \nonumber \\
  & \quad
  +
  \frac{\beta}{\rho _{i}}
  \tr{\mathbf{S} _{i} \cdot \dot{\boldsymbol{\varepsilon}} _{i}}
  \nonumber \\
  & \quad
  + 
  \frac{\eta F}{V \rho _{i}}
  \lnm \mathbf{u} _{i} - \mathbf{u} _{0} - \omega \mathbf{z} \times \[ \mathbf{x} - \mathbf{x} _{0} \] \rnm \delta _{\mathbf{x} _{i} \in \Omega _{\textup{shoulder}}}
  \nonumber \\
  & \quad
  + 
  \frac{h _{\textup{c}}}{\rho _{i}}
  \[ T _{0} - T _{i} \],
  \\
  \dot{\boldsymbol{\varepsilon}} _{i}
  & =
  \frac{1}{2}
  \sum _{j = 1} ^{N _{i}}
  \frac{m _{j}}{\rho _{j}}
  \[
  \mathbf{u} _{ji} \nabla W _{ij}
  +
  \nabla W _{ij} \mathbf{u} _{ji}
  \],
  \\
  \mathbf{S} _{i}
  & =
  2 \mu _{i}
  \[ \dot{\boldsymbol{\varepsilon}} _{i} - \frac{1}{3} \tr{\dot{\boldsymbol{\varepsilon}} _{i}} \mathbf{I} \].
  \label{eq:model_SPH_heatEquation}
\end{align}

Apart from the governing equations shown above, in Lagrangian framework, there is an additional equation that determines the particle motion, i.e., particle positions evolve along the particle velocity. By definition,
\begin{align}
  \dot{\mathbf{x}} _{i} = \mathbf{u} _{i}.
\end{align}

Moreover, some extra corrections are often adopted in the literature for regularization.
The particle shift from the high particle concentration to the low one \cite{lind2012incompressible} with a dimensionless parameter $\lambda$ to achieve a better particle distribution,
\begin{align}
  \mathbf{x} _{i}
  \leftarrow
  \mathbf{x} _{i}
  -
  \lambda 
  h _{i} ^{2} 
  \sum _{j \sim i}
  \frac{m _{j}}{\rho _{j}} 
  \nabla W _{ij}.
  \label{eq:SPH_particleShift}
\end{align}
The velocity is corrected by XSPH so that the velocity of each particle will be smoothed by its neighbors \cite{monaghan2002sph, monaghan1989problem}, with a dimensionless parameter $\zeta$,
\begin{align}
  \mathbf{u} _{i}
  \leftarrow
  \mathbf{u} _{i}
  +
  \zeta
  \sum _{j \sim i}
  \frac{2 m _{j}}{\rho _{i} + \rho _{j}}
  \mathbf{u} _{ji}
  W _{ij}.
  \label{eq:SPH_XSPH}
\end{align}
The density is re-normalized by Shepard filter to resolve the issue when there are not sufficient number of neighbor particles near the boundary \cite{johnson1996normalized, randles1996smoothed, shepard1968two}.
\begin{align}
  \rho _{i}
  \leftarrow
  \[ \sum _{j \sim i} m _{j} W _{ij} \]
  \[ \sum _{j \sim i} \frac{m _{j}}{\rho _{j}} W _{ij} \] ^{-1}.
  \label{eq:SPH_ShepardFilter}
\end{align}

In SPH simulation, we prescribe the particle velocity on $\Gamma _{\textup{in}}$ by circular motion with the pin, and fix the particle positions (with zero velocity) on $\Gamma _{\textup{out}}$.
To be consistent with particle positions, we choose non-slip boundary condition for velocity on $\Gamma _{\textup{in}}$ and zero velocity (Dirichlet boundary condition) on $\Gamma _{\textup{out}}$. The temperature of boundary particles are updated based on the evolution equation, where the boundary conditions are implicitly included. The density of boundary particles is also updated based on evolution equation.

Our primary focus in this study lies in POD-MOR and its application within the Lagrangian (SPH) system, by considering the FSSW problem as a benchmark problem for our simulations.

\subsection{Projection-based MOR and POD-MOR}

The projection-based MOR usually contains two stages, the offline dimensional compression stage that identifies the feature in the dynamics from data, and the online evolution stage where state variables are approximated in a low-dimensional subspace and the approximated solution can be solved \cite{magargal2022lagrangian}.

Assume we are solving a discrete time-dependent problem, where the numerical solution $\mathbf{w} ^{m}$ (can be any variable in SPH, e.g. position) at time step $m$ is an approximation of true solution $w \( \mathbf{x}, t ^{m} \)$,
\begin{align}
  \mathbf{w} ^{m} = \( w _{1} ^{m}, \cdots, w _{N} ^{m} \) \tp \approx \( w \( x _{1}, t ^{m} \), \cdots, w \( x _{N}, t ^{m} \) \) \tp \in \mathbb{R} ^{N},
\end{align}
where $N$ is the DoF of the discretized problem, $\mathbf{x} = \( x _{1}, \cdots, x _{N} \)$ is the discretized spatial coordinate, and $M$ is the number of temporal grid points, $\( t ^{0}, \cdots, t ^{M - 1} \)$ is the set of discretized temporal steps.

To explain the projection-based MOR, the numerical scheme is assumed to be one-step for simplicity, 
\begin{align}
  \mathbf{R} \( \mathbf{w} ^{m}, \mathbf{w} ^{m - 1} \) =\mathbf{0}, \quad m = 1, \cdots, M - 1,
\end{align}
but the generalization to multi-step schemes is straightforward.

During the offline stage, we need to generate a matrix $\mathbf{U} \in \mathbb{R} ^{N \times k}$, $k \ll N$, where its columns form a basis of solution subspace, so that we can approximate numerical solution $\mathbf{w} ^{m}$ by reduced-order solution $\tilde{\mathbf{w}} _{\mathbf{U}} ^{m}$ via
\begin{align}
  \mathbf{w} ^{m} \approx \tilde{\mathbf{w}} _{\mathbf{U}} ^{m} = \mathbf{w} ^{0} + \mathbf{U} \cdot \mathbf{U} \tp \cdot \[ \mathbf{w} ^{m} - \mathbf{w} ^{0} \] = \mathbf{w} ^{0} + \mathbf{U} \cdot \mathbf{a} ^{m}, \quad m = 1, \cdots, M - 1,
\end{align}
where $\mathbf{a} ^{m} \in \mathbb{R} ^{k}$, $m = 1, \cdots, M - 1$, are reduced coordinates in subspace basis. We set $\mathbf{w} ^{0}= \mathbf{0}$ in later derivations. 

In our study, we use POD to generate subspace basis $\mathbf{U} \in \mathbb{R} ^{N \times k}$ from $M$ snapshots $\mathbf{X} = \( \mathbf{w} ^{0}, \cdots, \mathbf{w} ^{M - 1} \) \in \mathbb{R} ^{N \times M}$ by solving the low-rank approximation \cite{chatterjee2000introduction, berkooz1993proper, kerschen2005method}
\begin{align}
  \min _{\tilde{\mathbf{X}}: \rank \( \tilde{\mathbf{X}} \) = k} \lnm \mathbf{X} - \tilde{\mathbf{X}} \rnm _{\textup{Frobenius}}.
\end{align}
This is equivalent to applying singular value decomposition (SVD) to $\mathbf{X}$,
\begin{align}
  \mathbf{X} = \hat{\mathbf{U}} \cdot \hat{\boldsymbol{\Sigma}} \cdot \hat{\mathbf{V}} \tp,
  \quad
  \tilde{\mathbf{X}} = \hat{\mathbf{U}} _{:, 1 : k} \cdot \hat{\boldsymbol{\Sigma}} _{1 : k, 1 : k} \cdot \[ \hat{\mathbf{V}} _{:, 1 : k} \] \tp,
\end{align}
and we choose $\mathbf{U} := \hat{\mathbf{U}} _{:, 1 : k}$ as the first $k$ columns of $\hat{\mathbf{U}}$,
i.e., POD automatically identifies the most dominant spatial modes so that it will provide deeper insights into the underlying physics.

Since we have the over-determined system in solving the reduced coordinates, in the online stage, we multiply a test matrix $\boldsymbol{\Phi} \in \mathbb{R} ^{N \times k}$ and get a system  as \begin{align}
  \boldsymbol{\Phi} \tp \cdot \mathbf{R} \( \tilde{\mathbf{w}} _{\mathbf{U}} ^{m}, \tilde{\mathbf{w}} _{\mathbf{U}} ^{m - 1} \) = \mathbf{0}.
\end{align}
Here, we choose $\boldsymbol{\Phi} = \mathbf{U}$, i.e., we are using the Galerkin projection during the online stage. The projection-based MOR just solves $\mathbf{a} ^{m} \in \mathbb{R} ^{k}$ through the above equation.
If the operator $\mathbf{R}$ commutes with matrix multiplication, we can solve $\mathbf{a} ^{m}$ efficiently since the basis matrices will be eliminated by $\mathbf{U} \tp \cdot \mathbf{U} = \mathbf{I} \in \mathbb{R} ^{k \times k}$, and we just need to deal with the reduced system $\mathbf{R} \( \mathbf{a} ^{m}, \mathbf{a} ^{m - 1} \) = \mathbf{0}$.

However, in general cases, a direct computation of
\begin{align}
  \mathbf{U} \tp \cdot \mathbf{R} \( \mathbf{U} \cdot \mathbf{a} ^{m},  \mathbf{U} \cdot \mathbf{a} ^{m - 1} \) = \mathbf{0}
\end{align}
is not that simple.
For the mass conservation \autoref{eq:model_SPH_massConservation} in SPH as an example, the reduced formula reads, for all $i \in \Lambda$,
\begin{align}
  & \quad
  \sum _{p = 1} ^{k ^{\rho}} U _{i p} ^{\rho} \dot{a} _{p} ^{\rho}
  =
  \sum _{p = 1} ^{k ^{\rho}} U _{i p} ^{\rho} a _{p} ^{\rho}
  \sum _{j \sim i}
  \frac{m _{j}}{\sum _{q = 1} ^{k ^{\rho}} U _{j q} ^{\rho} a _{q} ^{\rho}}
  \sum _{\alpha = 1} ^{2} 
  \sum _{r = 1} ^{k ^{u _{\alpha}}} \[ U _{i r} ^{u _{\alpha}} - U _{j r} ^{u _{\alpha}} \] a _{r} ^{u _{\alpha}}
  \frac{\[ z _{ij} \] _{\alpha}}{\lmdl \mathbf{z} _{ij} \rmdl} 
  \tilde{W}' \( \lmdl \mathbf{z} _{ij} \rmdl \),
\end{align}
where
\begin{align}
  \[ z _{ij} \] _{\alpha} 
  := 
  \sum _{s = 1} ^{k ^{x _{\alpha}}} \[ U _{i s} ^{x _{\alpha}} - U _{j s} ^{x _{\alpha}} \] a _{s} ^{x _{\alpha}},
  \quad
  \lmdl \mathbf{z} _{ij} \rmdl ^{2}
  :=
  \sum _{\alpha = 1} ^{2} \[ z _{ij} \] _{\alpha} ^{2}.
\end{align}
Here $U$, $a$, and $k$ with superscripts indicate the basis matrices, the reduced coordinates, and the DoF of the reduced coordinates of the corresponding variables.
Then the Galerkin projection multiplies $U _{i q} ^{\rho}$ to the left of each term with index $i$ and then sum over all $i \in \Lambda$. Thus, for $q = 1, \cdots, k ^{\rho}$, the formulation becomes
\begin{align}
  \dot{a} _{q} ^{\rho}
  =
  \sum _{i \in \Lambda} \sum _{p = 1} ^{k ^{\rho}} U _{i q} ^{\rho} U _{i p} ^{\rho} a _{p} ^{\rho}
  \sum _{j \sim i}
  \frac{m _{j}}{\sum _{q = 1} ^{k ^{\rho}} U _{j q} ^{\rho} a _{q} ^{\rho}}
  \sum _{\alpha = 1} ^{2} 
  \sum _{r = 1} ^{k ^{u _{\alpha}}} \[ U _{i r} ^{u _{\alpha}} - U _{j r} ^{u _{\alpha}} \] a _{r} ^{u _{\alpha}}
  \frac{\[ z _{ij} \] _{\alpha}}{\lmdl \mathbf{z} _{ij} \rmdl} 
  \tilde{W}' \( \lmdl \mathbf{z} _{ij} \rmdl \).
  \label{eq:model_POD_reducedMassConservation}
\end{align}
From the above expression, we can observe the difficulty in accelerating the computation of the RHS. In fact, we are not able to cancel out all appearance of basis $\mathbf{U}$ since $\mathbf{U} ^{\rho}$, $\mathbf{U} ^{x _{1}}$, and $\mathbf{U} ^{x _{2}}$ appear even in the denominators and the kernel function. Moreover, different basis $\mathbf{U} ^{\rho}$, $\mathbf{U} ^{u _{1}}$, $\mathbf{U} ^{u _{2}}$, $\mathbf{U} ^{x _{1}}$, and $\mathbf{U} ^{x _{2}}$ are coupled. So the computation still requires pulling-back to the system of $N = \lmdl \Lambda \rmdl$ particles, but if we use
\begin{align}
  \mathbf{w} ^{m} \larw \mathbf{U} \cdot \mathbf{U} \tp \cdot \mathbf{w} ^{m} = \mathbf{U} \cdot  \mathbf{a} ^{m},
  \label{eq:model_POD_directProjection}
\end{align}
the computational cost can not be reduced.

\myadd{
In order to achieve a speed-up of the SPH simulation with POD-MOR, we linearize the governing equations by freezing some of the terms.
For example on the RHS of \autoref{eq:model_POD_reducedMassConservation}, we only keep the term $a _{r} ^{u _{\alpha}}$ as the main variable, but treat all other terms as freezing terms, so that
\begin{align}
  \dot{a} _{q} ^{\rho}
  =
  \sum _{i \in \Lambda} U _{i q} ^{\rho}
  \sum _{j \sim i}
  \frac{\rho _{i} m _{j}}{\rho _{j}}
  \sum _{\alpha = 1} ^{2} 
  \sum _{r = 1} ^{k} \[ U _{i r} ^{u _{\alpha}} - U _{j r} ^{u _{\alpha}} \] a _{r} ^{u _{\alpha}}
  \nabla _{\alpha} W _{ij}
  =
  \sum _{\alpha = 1} ^{2} 
  \sum _{r = 1} ^{k}
  A _{q, \alpha, r} a _{r} ^{u _{\alpha}}
  .
  \label{eq:model_POD_linearizedMassConservation}
\end{align}
The freezing terms, $\rho _{i}, \rho _{j}, \nabla _{\alpha} W _{ij}$, as well as $A _{q, \alpha, r}$, will be updated every a few SPH iterations, so in most of the SPH iterations, the RHS of \autoref{eq:model_POD_linearizedMassConservation} is just a linear function of the reduced coordinate $a _{r} ^{u _{\alpha}}$. When the linear operator $A _{q, \alpha, r}$ is calculated once, the computational cost of the matrix-vector multiplication is cheaper than the counterpart in the full model. 
}

\myadd{
For the conservation of linear momentum \autoref{eq:model_SPH_linearMomentumConservation} and the heat equation \autoref{eq:model_SPH_heatEquation}, we can apply the similar linearization approaches as follows.
}

\myadd{
\begin{align}
  \[ \begin{array}{c} \dot{a} _{r} ^{u _{1}} \\ \dot{a} _{s} ^{u _{2}} \end{array} \]
  & =
  - \sum _{i, j \in \Lambda}
  \frac{m _{j}}{\rho _{i} \rho _{j}}
  \sum _{q = 1} ^{k ^{\rho}} c _{0} ^{2} \[ U _{i q} ^{\rho} + U _{j q} ^{\rho} \] a _{q} ^{\rho}
  \[ \begin{array}{c} U _{i r} ^{u _{1}} \nabla _{1} W _{ij} \\ U _{i s} ^{u _{2}} \nabla _{2} W _{ij} \end{array} \]
  \nonumber \\
  & +
  \frac{5}{3}
  \sum _{i, j \in \Lambda}
  \frac{m _{j} \mu _{j}}{\rho _{i} \rho _{j}}
  \frac{\[ x _{1} \] _{i j} \nabla _{1} W _{ij} + \[ x _{2} \] _{i j} \nabla _{2} W _{ij}}{\[ x _{1} \] _{i j} ^{2} + \[ x _{2} \] _{i j} ^{2}}
  \[ \begin{array}{c} U _{i r} ^{u _{1}} \sum _{q = 1} ^{k ^{u _{1}}} \[ U _{i q} ^{u _{1}} - U _{j q} ^{u _{1}} \] a _{q} ^{u _{1}} \\ U _{i s} ^{u _{2}} \sum _{q = 1} ^{k ^{u _{2}}} \[ U _{i q} ^{u _{2}} - U _{j q} ^{u _{2}} \] a _{q} ^{u _{2}} \end{array} \]
  \nonumber \\
  & +
  \frac{4}{3}
  \sum _{i, j \in \Lambda}
  \frac{m _{j} \mu _{j}}{\rho _{i} \rho _{j}}
  \frac{\[ x _{1} \] _{i j} \nabla _{1} W _{ij} + \[ x _{2} \] _{i j} \nabla _{2} W _{ij}}{\[ x _{1} \] _{i j} ^{2} + \[ x _{2} \] _{i j} ^{2}}
  \[ \begin{array}{c} U _{i r} ^{u _{1}} \[ x _{1} \] _{i j} \\ U _{i s} ^{u _{2}} \[ x _{2} \] _{i j} \end{array} \]
  \frac{\sum _{\alpha = 1} ^{2} \sum _{q = 1} ^{k ^{u _{\alpha}}} \[ U _{i q} ^{u _{\alpha}} - U _{j q} ^{u _{\alpha}} \] a _{q} ^{u _{\alpha}} \[ x _{\alpha} \] _{i j}}{\[ x _{1} \] _{i j} ^{2} + \[ x _{2} \] _{i j} ^{2}}
  \nonumber \\
  & -
  \sum _{i, j, k \in \Lambda}
  \frac{m _{j} m _{k} \mu _{k}}{\rho _{i} \rho _{j} \rho _{k}}
  \[ \begin{array}{c} U _{i r} ^{u _{1}} \sum _{q = 1} ^{k ^{u _{1}}} \[ U _{i q} ^{u _{1}} - U _{j q} ^{u _{1}} \] a _{q} ^{u _{1}} \\ U _{i s} ^{u _{2}} \sum _{q = 1} ^{k ^{u _{2}}} \[ U _{i q} ^{u _{2}} - U _{j q} ^{u _{2}} \] a _{q} ^{u _{2}} \end{array} \]
  \sum _{\alpha = 1} ^{2} \[ \nabla _{\alpha} W _{ij} \nabla _{\alpha} W _{ik} \]
  \nonumber \\
  & -
  \sum _{i, j, k \in \Lambda}
  \frac{m _{j} m _{k} \mu _{k}}{\rho _{i} \rho _{j} \rho _{k}}
  \sum _{\alpha = 1} ^{2} \[ \nabla _{\alpha} W _{ik} \sum _{q = 1} ^{k ^{u _{\alpha}}} \[ U _{i q} ^{u _{\alpha}} - U _{j q} ^{u _{\alpha}} \] a _{q} ^{u _{\alpha}} \]
  \[ \begin{array}{c} U _{i r} ^{u _{1}} \nabla _{1} W _{ij} \\ U _{i s} ^{u _{2}} \nabla _{2} W _{ij} \end{array} \]
  \nonumber \\
  & +
  \frac{2}{3}
  \sum _{i, j, k \in \Lambda}
  \frac{m _{j} m _{k} \mu _{k}}{\rho _{i} \rho _{j} \rho _{k}}
  \[ \begin{array}{c} U _{i r} ^{u _{1}} \nabla _{1} W _{ik} \\ U _{i s} ^{u _{2}} \nabla _{2} W _{ik} \end{array} \]
  \sum _{\alpha = 1} ^{2} \[ \sum _{q = 1} ^{k ^{u _{\alpha}}} \[ U _{i q} ^{u _{\alpha}} - U _{j q} ^{u _{\alpha}} \] a _{q} ^{u _{\alpha}} \nabla _{\alpha} W _{ij} \]
  .
  \label{eq:model_POD_linearizedLinearMomentumConservation}
  \\
  \dot{a} _{q} ^{T}
  & =
  \sum _{i, j \in \Lambda}
  \frac{m _{j}}{\rho _{i} \rho _{j} C _{i}}
  \frac{4 k _{i} k _{j}}{k _{i} + k _{j}}
  \frac{U _{i q} ^{T} \sum _{p = 1} ^{k ^{T}} \[ U _{i p} ^{T} - U _{j p} ^{T} \] a _{p} ^{T} \sum _{\alpha = 1} ^{2} \[ x _{\alpha} \] _{ij} \nabla _{\alpha} W _{ij}}{\[ x _{1} \] _{i j} ^{2} + \[ x _{2} \] _{i j} ^{2}}
  \nonumber \\
  & +
  \sum _{i, j, k \in \Lambda}
  \frac{m _{j} m _{k} k _{j}}{\rho _{i} \rho _{j} \rho _{k} C _{i}}
  U _{i q} ^{T}
  \sum _{p = 1} ^{k ^{T}}
  U _{k p} ^{T} a _{p} ^{T}
  \sum _{\alpha = 1} ^{2} \[ \nabla _{\alpha} W _{ij} \nabla _{\alpha} W _{ik} \]
  \nonumber \\
  & +
  \beta
  \sum _{i \in \Lambda}
  \frac{1}{\rho _{i} C _{i}} 
  U _{i q} ^{T}
  \sum _{\alpha, \alpha' = 1} ^{2}
  \[ S _{\alpha \alpha'} \] _{i} \[ \dot{\varepsilon} _{\alpha' \alpha} \] _{i}
  \nonumber \\
  & +
  \eta
  \sum _{i \in \Lambda}
  \frac{F}{\rho _{i} C _{i} V} 
  U _{i q} ^{T}
  \lnm \mathbf{u} _{i} - \mathbf{u} _{0} - \omega \mathbf{z} \times \[ \mathbf{x} - \mathbf{x} _{0} \] \rnm \delta _{\mathbf{x} _{i} \in \Omega _{\textup{shoulder}}}
  \nonumber \\
  & -
  \sum _{i \in \Lambda}
  \frac{h _{\textup{c} i}}{\rho _{i} C _{i}} 
  U _{i q} ^{T}
  \sum _{p = 1} ^{k ^{T}}
  U _{i p} ^{T} a _{p} ^{T}
  .
  \label{eq:model_POD_linearizedLinearHeatEquation}
\end{align}
}

\myadd{
Such a linearization approach with freezing coefficients will introduce additional error to the SPH simulation, so we need to balance the acceleration with the additional error. As long as the linearization error is smaller than or similar to the POD error, and at the same time, the acceleration is significant, this acceleration will be valuable in practice.
}

Since we focus on the question whether POD can automatically find the reduced model for SPH simulation in this study, we will not further optimize the computational cost at this moment.
In fact, there are a few existing techniques we can explore to speed up the computation, for example, hyper-reduction methods \cite{chaturantabut2010nonlinear, carlberg2013gnat, farhat2015structure, choi2020sns}, local basis \cite{amsallem2012nonlinear}, domain decomposition \cite{lucia2001reduced}, basis splitting \cite{carlberg2015adaptive}, and machine learning-based approaches. But that will be left as future investigation.


In SPH simulation of the 2D FSSW problem, the variable $\mathbf{w} ^{m}$ will be chosen as four Lagrangian variables --- position $\mathbf{x} ^{\alpha} \( t ^{m} \)$, $\alpha = 1, 2$, velocity $\mathbf{u} ^{\alpha} \( t ^{m} \)$, $\alpha = 1, 2$, temperature $\mathbf{T} \( t ^{m} \)$, and density $\boldsymbol{\rho} \( t ^{m} \)$.


\section{Numerical Results}
\label{sec:numerical}

In this section, we will setup SPH simulations and briefly discuss the choice of SPH parameters. After that, POD-MOR is applied to SPH simulation, and we can study the POD error for heat equation, flow equations, and the coupled FSSW governing equations.

Throughout the numerical simulations, we will show the following interesting observations to the readers.
\begin{itemize}
  \item We find a parameter set where the SPH simulation shows regular motion and the SPH error numerically decays as time step size is decreasing.
  \item POD-MOR works well with the heat equation in Lagrangian framework, and it is not sensitive to different types of prescribed motion.
  \item POD-MOR provides an effective approach for model reduction of SPH simulations. With the same DoFs, POD-MOR produces less error than uniformly reducing particle number in SPH simulations.
  \item We illustrate the POD modes in Lagrangian framework, and the geometric organization shows why POD-MOR can reduce the DoFs, and how it can be more efficient than uniformly reducing the number of particles.
  \myadd{
  \item By applying the linearization approach, the POD-MOR can be accelerated without introducing large additional error.
  \item We show that the POD-MOR can be applied with general situations with different parameter values. The POD error is small, even when the POD data is gathered from another system with different tool rotational speed in FSSW.
  }
\end{itemize}

\subsection{SPH Setup}
\label{sec:numerical_sphSetup}

During SPH simulation, we apply material parameters based on the material of AA6061 \cite{cao2022machine} and Aluminum. \autoref{tab:numerical_sphSetup_parameters} shows the default setting of our simulations.

The effective volume of workpiece under shoulder can be expressed by $V = \pi \[ r _{\textup{s}} ^{2} - r _{\textup{p}} ^{2} \] r _{\textup{thickness}}$, where $r _{\textup{thickness}}$ is the effective thickness of the workpieces under the shoulder. From the parameters in \autoref{tab:numerical_sphSetup_parameters}, the corresponding $r _{\textup{thickness}} \approx 0.007 \[ \mykgms{0}{1}{0} \]$.

For the sake of numerical efficiency, we choose $c _{0} = 50 \[ \mykgms{0}{1}{-1} \]$, which is large enough (compared with the largest flow velocity $\omega r _{\textup{p}} \approx 0.628 \[ \mykgms{0}{1}{-1} \]$) to achieve weakly compressibility so that the density variation will not be large. 


\begin{table}[!htp]
  \centering
  \begin{tabular}{|l|c|l|}
    \hline
    Parameter & Notation & Value\\
    \hline
    Domain radius & $r _{\textup{d}}$ & $0.045 \[ \mykgms{0}{1}{0} \]$\\
    Pin radius & $r _{\textup{p}}$ & $0.0075 \[ \mykgms{0}{1}{0} \]$\\
    Shoulder radius & $r _{\textup{s}}$ & $0.025 \[ \mykgms{0}{1}{0} \]$\\
    Shoulder force & $F$ & $2.5 \e{4} \[ \mykgms{1}{1}{-2} \]$\\
    Volume of workpiece under shoulder & $V$ & $1.25 \e{-5} \[ \mykgms{0}{3}{0} \]$\\
    Tool angular velocity & $\omega$ & $8\pi \[ \mykgms{0}{0}{-1} \] = 240 \[ \textup{RPM} \]$\\
    Tool advancing speed & $U$ & $0 \[ \mykgms{0}{1}{-1} \]$\\
    \hline
    Reference density & $\rho _{0}$ & $2.7 \e{3} \[ \mykgms{1}{-3}{0} \]$ \\
    Speed of sound (artificial) & $c _{0}$ & $50 \[ \mykgms{0}{1}{-1} \]$ \\
    \hline
    Specific heat capacity & $C$ & $870 \[ \mykgms{0}{2}{-2} \cdot \textup{K} ^{-1} \]$\\
    Heat conduction coefficient & $k$ & $170 \[ \mykgms{1}{1}{-3} \cdot \textup{K} ^{-1} \]$\\
    Plastic work parameter & $\beta$ & $0.8$ \\
    Shoulder pressure parameter & $\eta$ & $0.5$ \\
    Ambient temperature & $T _{0}$ & $293.15 \[ \textup{K} \]$ \\
    Boundary heat transfer coefficient & $h _{\textup{c}}$ & $1 \e{4} \[ \mykgms{1}{-1}{-3} \cdot \textup{K} ^{-1} \]$ \\
    \hline
    Activation energy (Aluminum) & $E$ & $1.65 \e{4} \[ \mykgms{1}{2}{-2} \cdot \textup{mol} ^{-1} \]$\\
    Initial viscosity (Aluminum) & $\mu_0$ & $1.49 \e{-4} \[ \mykgms{1}{-1}{-1} \]$\\
    Gas constant & $R$ &$8.3144 \[ \mykgms{1}{2}{-2} \cdot \textup{K} ^{-1} \cdot \textup{mol} ^{-1} \]$ \\
    \hline
    Particle diffusion parameter & $\lambda$ & $0.002$ \\
    Xsph coefficient & $\zeta$ & $0.4$ \\
    \hline
  \end{tabular}
  \caption{Default value of parameters used in numerical simulations.}
  \label{tab:numerical_sphSetup_parameters}
\end{table}

\begin{figure}[!htp]
  \centering
  \begin{subfigure}{0.6\textwidth}
    \centering
    \includegraphics[width = \linewidth]{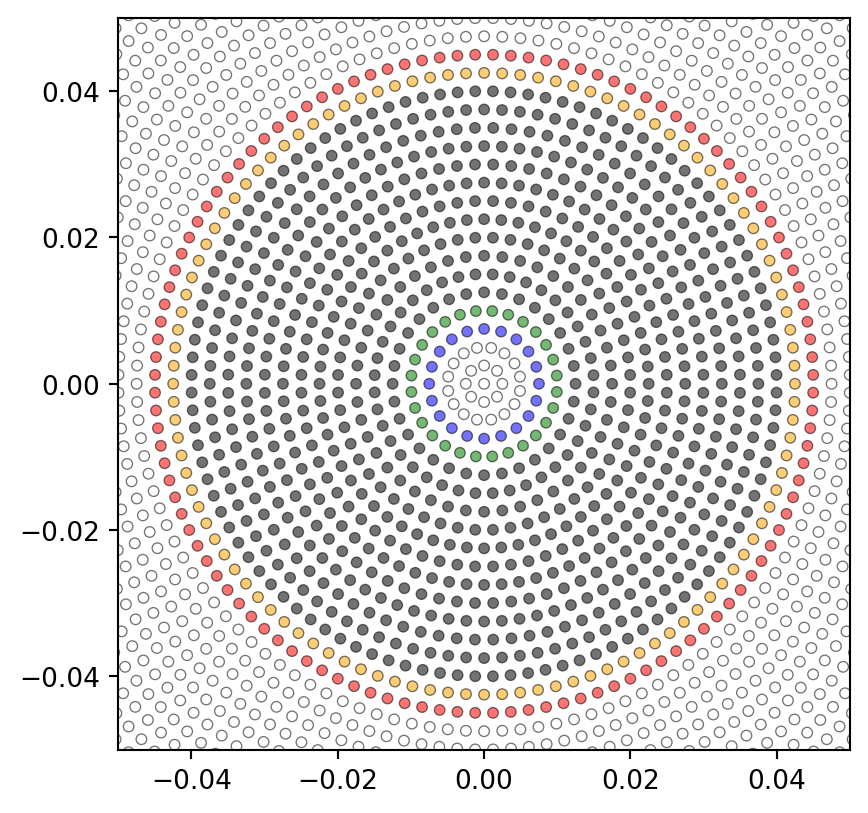}
  \end{subfigure}
  \caption{Typical initial distribution of smoothed particles. The blue and green particles represent inner boundary, the red and orange particles represent outer boundary, the gray particles represent the bulk particles, and the white particles are discarded. The index set $\Lambda$ contains all blue, green, red, orange, and gray particles.}
  \label{fig:numerical_sphSetup_mesh_extended_l18}
\end{figure}

The computational domain $\Omega$ represents the contacting workpieces to be welded without any gaps, which is simplified as an annulus between two concentric circles, since the central hole will be filled by the tool pin in FSSW. 
To generate initial particle distribution for a typical initial setup, for example as \autoref{fig:numerical_sphSetup_mesh_extended_l18}, with particle spacing about $r = 2.5 \e{-3} \[ \mykgms{0}{1}{0} \]$, we can first determine the radius of particle layers as $0, r, 2r, 3r, \cdots$, and then choose proper number of particles on each circular layer so that the distance between nearest particle pairs are close to $r$. After that, we discard particles outside of $\Omega$, i.e., the white particles in \autoref{fig:numerical_sphSetup_mesh_extended_l18}, and the remaining particles are indexed by $\Lambda$. In this example, there are a total of $N _{\textup{all}} = \lmdl \Lambda \rmdl = 1084$ particles located on $16$ concentric circles, that are almost uniformly distributed in $\Omega$.

We choose the smoothing length $h = 1.2 r$ throughout the simulations, and to ensure enough neighbors for particles near boundary, we use two layers of particles as boundary condition for velocity, for both inner and outer boundaries, since $2r < 2h < 3r$.
As in \autoref{fig:numerical_sphSetup_mesh_extended_l18}, we use different colors to distinguish boundary particles and bulk particles. In such a case, there are $N _{\textup{bulk}} = 812$ particles remaining in the bulk (gray particles in \autoref{fig:numerical_sphSetup_mesh_extended_l18}).
In practice, we prescribe $\lmdl \mathbf{u} \( \mathbf{x} \) \rmdl = \omega r _{\textup{p}} ^{2} \lmdl \mathbf{x} \rmdl ^{-1}$ for the velocity on the second layer of particles, based on the analytic steady-state solution of incompressible Navier-Stokes equation with uniform viscosity for a cylinder rotating in an infinite body of fluid, see section 7.1.3 of \cite{babu2022fundamentals}.

In \autoref{tab:numerical_sphSetup_mesh_particleNumSpacing}, we provide a list of initial particle distributions we are using in the numerical simulations. \gridC{} is just the example we mentioned above, and \gridA{}, \gridB{} contain less number of particles.
\begin{table}[!htp]
  \centering
  \begin{tabular}{|c|c|c|c|c|}
    \hline
    name & particle spacing $r$ & smoothing length $h$ & $N _{\textup{bulk}} / N _{\textup{all}}$ & mass $m$
    \\
    \hline
    \gridA & $7.5 \e{-3} \[ \mykgms{0}{1}{0} \]$ & $9 \e{-3} \[ \mykgms{0}{1}{0} \]$ & $48 / 144$ & $0.139 \[ \mykgms{1}{0}{0} \]$ 
    \\
    \hline
    \gridB & $3.75 \e{-3} \[ \mykgms{0}{1}{0} \]$ & $4.5 \e{-3} \[ \mykgms{0}{1}{0} \]$ & $320 / 504$ & $0.0365 \[ \mykgms{1}{0}{0} \]$ 
    \\
    \hline
    \gridC & $2.5 \e{-3} \[ \mykgms{0}{1}{0} \]$ & $3 \e{-3} \[ \mykgms{0}{1}{0} \]$ & $812 / 1084$ & $0.0164 \[ \mykgms{1}{0}{0} \]$ 
    \\
    \hline
  \end{tabular}
  \caption{Information of initial particle distributions.}
  \label{tab:numerical_sphSetup_mesh_particleNumSpacing}
\end{table}

To assign mass $\lbk m _{i} \rbk _{i \in \Lambda}$ to all SPH particles, we choose uniform mass $m _{i} \equiv m _{0}$ based on the material density $\rho _{0}$, the volume of the workpiece $V _{\textup{workpiece}} = \pi \[ r _{\textup{d}} ^{2} - r _{\textup{p}} ^{2} \] r _{\textup{thickness}}$, and the effective number of SPH particles $N _{\textup{eff}}$, so that $\rho _{0} V _{\textup{workpiece}} = N _{\textup{eff}} m _{0}$. Here the effective number of SPH particles $N _{\textup{eff}}$ includes the boundary effect that each inner most and outer most particles contribute as half while the rest are all counted by the whole.
The mass of particles are also listed in \autoref{tab:numerical_sphSetup_mesh_particleNumSpacing}.

A typical SPH simulation time for the FSSW process is $2 \[ \mykgms{0}{0}{1} \]$, and the typical CFL parameter with respect to the flow equations is $\alpha _{\textup{CFL}} = 0.1$ so that typically for \gridC{}, $\Delta t := \alpha _{\textup{CFL}} h / \[ c _{0} + \max _{i \in \Lambda} \lmdl \mathbf{u} _{i} \rmdl \] \approx 6 \e{-6} \[ \mykgms{0}{0}{1} \]$ ($\Delta t$ is larger for \gridA{} and \gridB), and we note that the heat diffusion is a much slower process so the upper bound of numerical time step size is only restricted by the flow equations. 
Particle functions will be saved at certain time instances $\Theta _{\delta} \subset \Theta$ with temporal data-saving spacing $\delta$, for the sake of result comparison and also be used as POD dataset.
The first-order explicit Euler scheme is used for temporal integration in this benchmark problem.


\subsection{Error Indicator}
\label{sec:numerical_errorIndicator}

In this section, we explain how to measure the difference between two different SPH simulations. 

\begin{figure}[!htp]
  \centering
  \begin{subfigure}{0.6\textwidth}
    \centering
    \includegraphics[width = \linewidth]{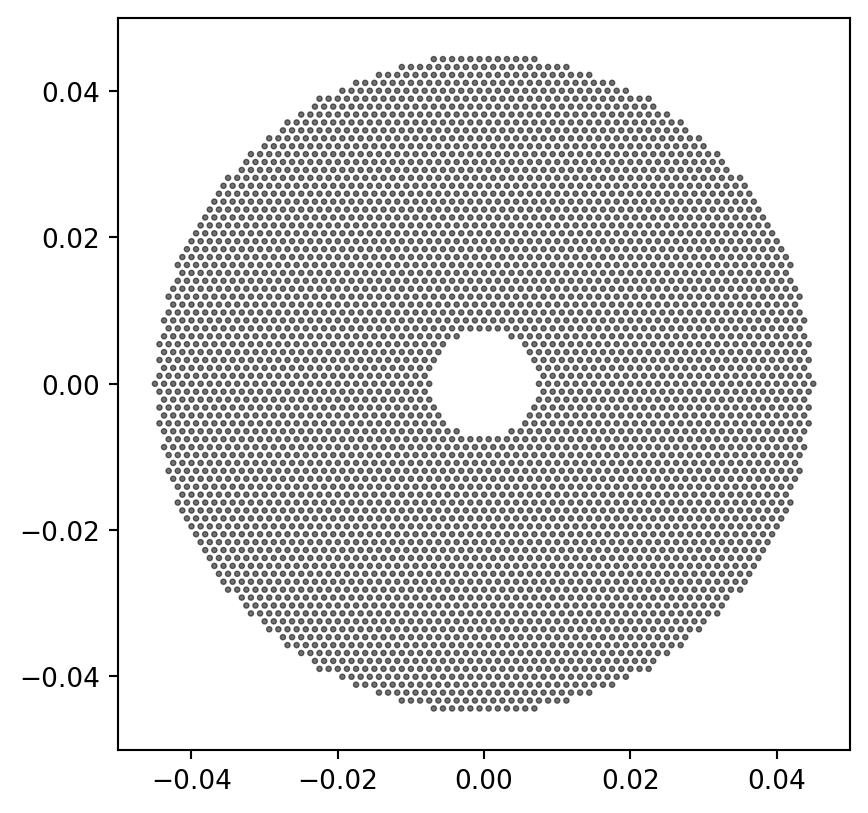}
  \end{subfigure}
  \caption{The fixed grid for interpolation.}
  \label{fig:numerical_errorIndicator_mesh_interpolation_l30}
\end{figure}

Since the particle position is varying in the Lagrangian framework, it is not straightforward to directly compare particle functions based on their indices, let alone the comparison of two systems with different number of particles.
However, we can interpolate particle functions at the fixed grid points $\mathbf{Y} := \lbk \mathbf{y} _{j} \rbk _{j \in \Lambda _{\epsilon}} \subset \Omega$ naturally based on SPH kernel functions and then compare the interpolated functions as in Eulerian framework.
In our study, we pick the uniformly refined grid $\mathbf{Y} = \mathbf{Y} _{\epsilon}$ within $\Omega$ that is generated by a perfect 2D triangular lattice with spacing $\epsilon$. Those fixed grid points are indexed by $\Lambda _{\epsilon}$.
\autoref{fig:numerical_errorIndicator_mesh_interpolation_l30} shows the fixed grid points with spacing $\epsilon = 1.25 \e{-3} \[ \mykgms{0}{1}{0} \]$, which will be used in our numerical experiments.

For a general particle function $f$, we try to compare solutions $\lbk \hat{f} _{i} \( t \) \rbk _{i \in \hat{\Lambda}, t \in \Theta _{\delta}}$ and $\lbk \tilde{f} _{i} \( t \) \rbk _{i \in \tilde{\Lambda}, t \in \Theta _{\delta}}$ from two different simulations. Possibly, the number of particles in two simulations are different, and so do kernel functions $W _{\hat{h}}$ and $W _{\tilde{h}}$. 
In Eulerian framework, we denote $\hat{f}$ and $\tilde{f}$ as the underlying smooth field functions to be approximated by SPH. From now on, the conventional SPH notations are used, but with an additional $\hat{\cdot}$ or $\tilde{\cdot}$ to distinguish two simulations correspondingly.

We derive the following error indicator of $\hat{f}$ with respect to $\tilde{f}$ 
by assuming sufficient smoothness of $\hat{f}$ and $\tilde{f}$ and the boundedness of $\lnm \tilde{f} \( \cdot, \cdot \) \rnm _{L ^{2} \( \Omega, \Theta \)} ^{-1}$,
\begin{align}
  & \quad
  \frac{\lnm \hat{f} \( \cdot, \cdot \) - \tilde{f} \( \cdot, \cdot \) \rnm _{L ^{2} \( \Omega, \Theta \)}}{\lnm \tilde{f} \( \cdot, \cdot \) \rnm _{L ^{2} \( \Omega, \Theta \)}}
  & \stackrel{\epsilon, \delta \to 0}{=}
  \frac{\[ \sum _{i \in \Lambda _{\epsilon}} \sum _{t \in \Theta _{\delta}} \[ \hat{f} \( \mathbf{y} _{i}, t \) - \tilde{f} \( \mathbf{y} _{i}, t \) \] ^{2} \] ^{1 / 2}}{\[ \sum _{i \in \Lambda _{\epsilon}} \sum _{t \in \Theta _{\delta}} \[ \tilde{f} \( \mathbf{y} _{i}, t \) \] ^{2}\] ^{1 / 2}} + \bigO{\epsilon} + \bigO{\delta},
\end{align}
where
\begin{align}
  \hat{f} \( \mathbf{y} _{i}, t \)
  & \stackrel{\hat{h} \to 0}{=}
  \sum _{j \in \hat{\Lambda}} \frac{\hat{m} _{j}}{\hat{\rho} _{j}} \hat{f} _{j} \( t \) W _{\hat{h}} \( \mathbf{y} _{i} - \hat{\mathbf{x}} _{j} \) + \bigO{\hat{h} ^{2}},
  \\
  \tilde{f} \( \mathbf{y} _{i}, t \)
  & \stackrel{\tilde{h} \to 0}{=}
  \sum _{j \in \tilde{\Lambda}} \frac{\tilde{m} _{j}}{\tilde{\rho} _{j}} \tilde{f} _{j} \( t \) W _{\tilde{h}} \( \mathbf{y} _{i} - \tilde{\mathbf{x}} _{j} \) + \bigO{\tilde{h} ^{2}}.
\end{align}

The above error indicator takes the relative $L ^{2}$ norm in both the spatial domain $\Omega$ and the temporal domain $\Theta$, and the discrete relative $L ^{2}$ norm can be rewritten in terms of the particle functions $\hat{f} _{j}$ and $\tilde{f} _{j}$ in snapshots of SPH simulations. 

Since particle position comparison makes no sense when particle number is different in two simulations, in our following numerical experiments, we will focus on the velocity error, the temperature error, and the density error, that can be measured by the above error indicator.


\subsection{More About SPH Parameters}
\label{sec:numerical_parameter}

There are two artificial SPH parameters --- particle diffusion parameter $\lambda$ in \autoref{eq:SPH_particleShift} and XSPH coefficient $\zeta$ in \autoref{eq:SPH_XSPH}, and the SPH simulation results are sensitive to the choice of these two parameter values.
Determining the best parameter values of $\lambda, \zeta$ to achieve accurate SPH simulations will be tough (out of the scope of the current work), however, we are able to find out an acceptable choice of the values based on simulation results.

From our simulations, we find out the POD-MOR SPH simulations is likely to be unstable, particularly for the velocity field, when the original SPH simulation shows irregular motion rather than a regular pattern.
So, when choosing the parameters $\lambda$ and $\zeta$, we mainly focus on the following two factors.

For viscous flows with similar settings, the fluctuation in radial displacement is not common. So one factor we introduce is the maximum radial motion among all particle trajectories, which is defined as
\begin{align}
  \max _{t \in \Theta _{\Delta t}, i \in \Lambda} \lmdl \lmdl \mathbf{x} _{i} \( t \) \rmdl - \lmdl \mathbf{x} _{i} \( 0 \) \rmdl \rmdl.
\end{align}
In some regular flow motion, this value is small.
With different choices of $\lambda$ and $\zeta$, we run some SPH simulations on a fine spatial resolution (\gridC) and a coarse temporal resolution ($\Delta t = 1 \e{-5} \[ \mykgms{0}{0}{1} \]$) for $2 \[ \mykgms{0}{0}{1} \]$, and list some of the maximum radial motions in \autoref{tab:numerical_parameter_lambdaZeta_l18}.
Both too small or too large parameters result in large radial motion, which means the particle trajectories are irregular.
We may choose $\lambda = 2 \e{-3}$ and $\zeta = 4 \e{-1}$ based on the result, considering there is a theoretic upper bound $\zeta \le 0.5$ \cite{fraser2017robust}.

\begin{table}[!htp]
  \begin{center}
    \begin{tabular}{|c|c|c|c|}
      \hline
      $\lambda \quad \backslash \quad \zeta$ & $0$ & $2 \e{-1}$ & $4 \e{-1}$
      \\
      \hline
      $0$        & $1.7 \e{-2} \[ \textup{m} \]$ & $3.1 \e{-3} \[ \textup{m} \]$ & $3.0 \e{-3} \[ \textup{m} \]$ 
      \\
      \hline
      $2 \e{-3}$ & $8.7 \e{-3} \[ \textup{m} \]$ & $4.0 \e{-4} \[ \textup{m} \]$ & $3.7 \e{-4} \[ \textup{m} \]$ 
      \\
      \hline
      $5 \e{-3}$ & $1.0 \e{-2} \[ \textup{m} \]$ & $2.3 \e{-3} \[ \textup{m} \]$ & $2.5 \e{-3} \[ \textup{m} \]$ 
      \\
      \hline
    \end{tabular}
  \end{center}
  \caption{Measure the maximum radial motion among all particles, for different combinations of $\lambda$ and $\zeta$. Initial particle distribution is \gridC{}, $\Theta = \[ 0, 2 \] \[ \textup{s} \]$, $\Delta t = 1 \e{-5} \[ \textup{s} \]$.}
  \label{tab:numerical_parameter_lambdaZeta_l18}
\end{table}

The same test is done with \gridA{} and \gridB{}, where the overall radial motion is weaker for coarser particle distribution, but we still find our choice $\lambda = 2 \e{-3}$ and $\zeta = 4 \e{-1}$ is good among different trials.


The other factor is the self convergence with respect to small time step size.
We run SPH simulations with a range of different time step sizes 
\begin{align}
  \Delta t \in \lbk 1 \e{-7}, 2 \e{-7}, 5 \e{-7}, 1 \e{-6}, 2 \e{-6}, 5 \e{-6}, 1 \e{-5} \rbk \[ \mykgms{0}{0}{1} \],
\end{align}
and compare the solution from the finest temporal resolution $\Delta t = 1 \e{-7} \[ \mykgms{0}{0}{1} \]$ with other ones.
The comparison is calculated by the error indicator explained in \autoref{sec:numerical_errorIndicator}.

\begin{figure}[!htp]
  \centering
  \begin{subfigure}{0.6\textwidth}
    \centering
    \includegraphics[width = \linewidth]{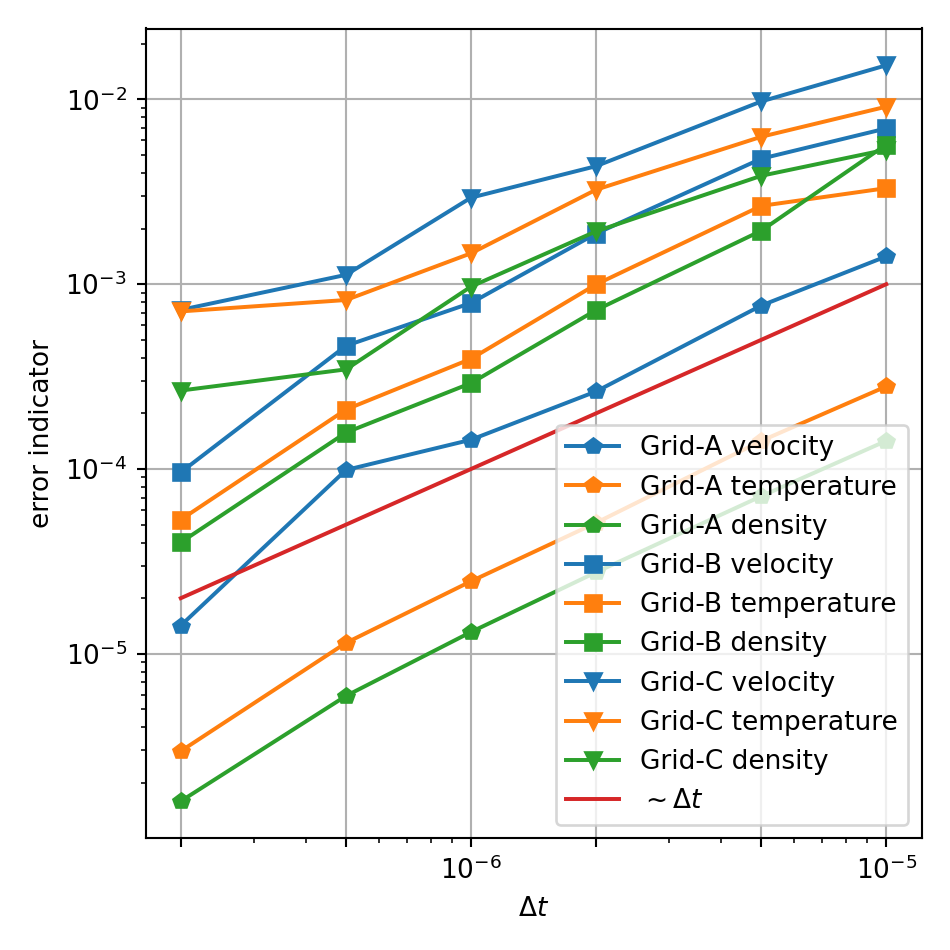}
  \end{subfigure}
  \caption{The error of velocity, temperature, and density with respect to $\Delta t = 1 \e{-7} \[ \textup{s} \]$. Initial particle distribution is \gridA{}, \gridB{}, and \gridC{}, $\Theta = \[ 0, 2 \] \[ \textup{s} \]$, $\lambda = 2 \e{-3}$, $\zeta = 4 \e{-1}$.}
  \label{fig:numerical_temporalCFL1m_Pshift2mVshift400m}
\end{figure}

Based on the chosen $\lambda$ and $\zeta$, we are able to observe the self-convergence with respect to time step size approaching to a small value in \autoref{fig:numerical_temporalCFL1m_Pshift2mVshift400m} with all \gridA{}, \gridB{}, and \gridC{}. 
The almost linear convergence is observed, which is theoretically optimal since our numerical temporal integration is first-order. This confirms our choice of $\lambda$ and $\zeta$ is acceptable for current simulation setup.

\begin{figure}[!htp]
  \centering
  \begin{subfigure}{0.6\textwidth}
    \centering
    \includegraphics[width = \linewidth]{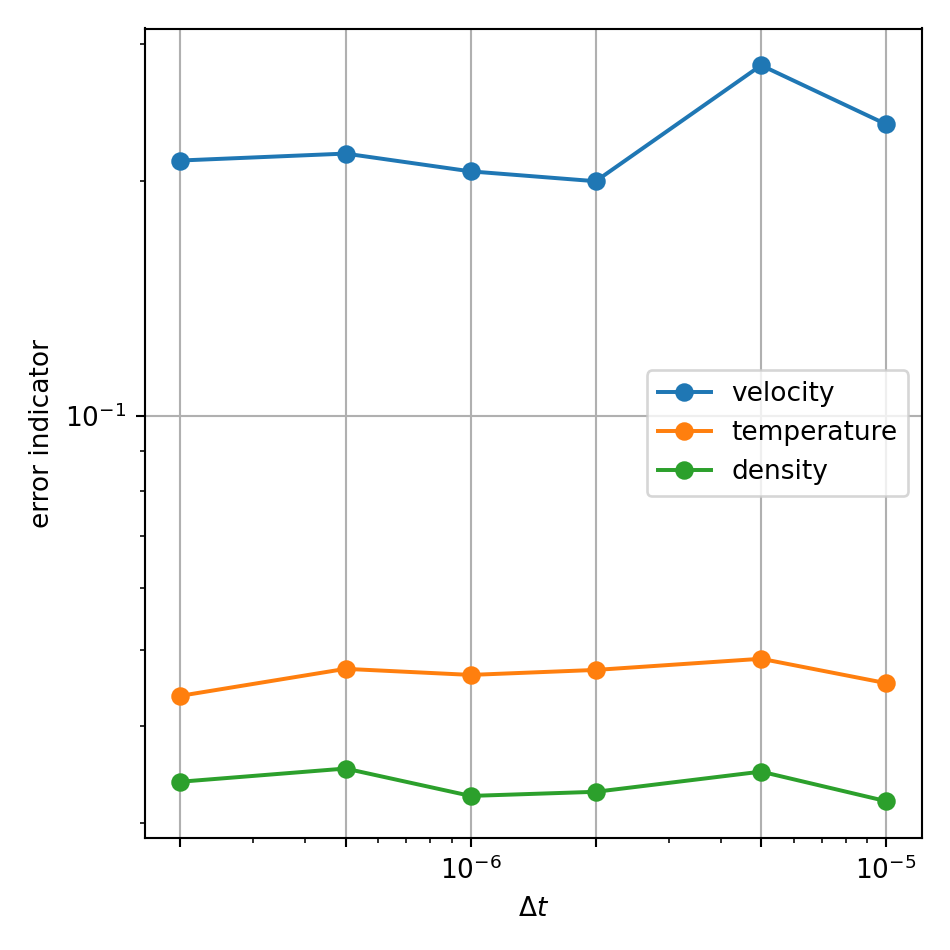}
  \end{subfigure}
  \caption{The error of velocity, temperature, and density with respect to $\Delta t = 1 \e{-7} \[ \textup{s} \]$. Initial particle distribution is \gridB{}, $\Theta = \[ 0, 2 \] \[ \textup{s} \]$, $\lambda = 0$, $\zeta = 0$.}
  \label{fig:numerical_temporalCFL1m_Pshift0Vshift0}
\end{figure}

\begin{figure}[!htp]
  \centering
  \begin{subfigure}{0.5\textwidth}
    \centering
    \includegraphics[width = \linewidth]{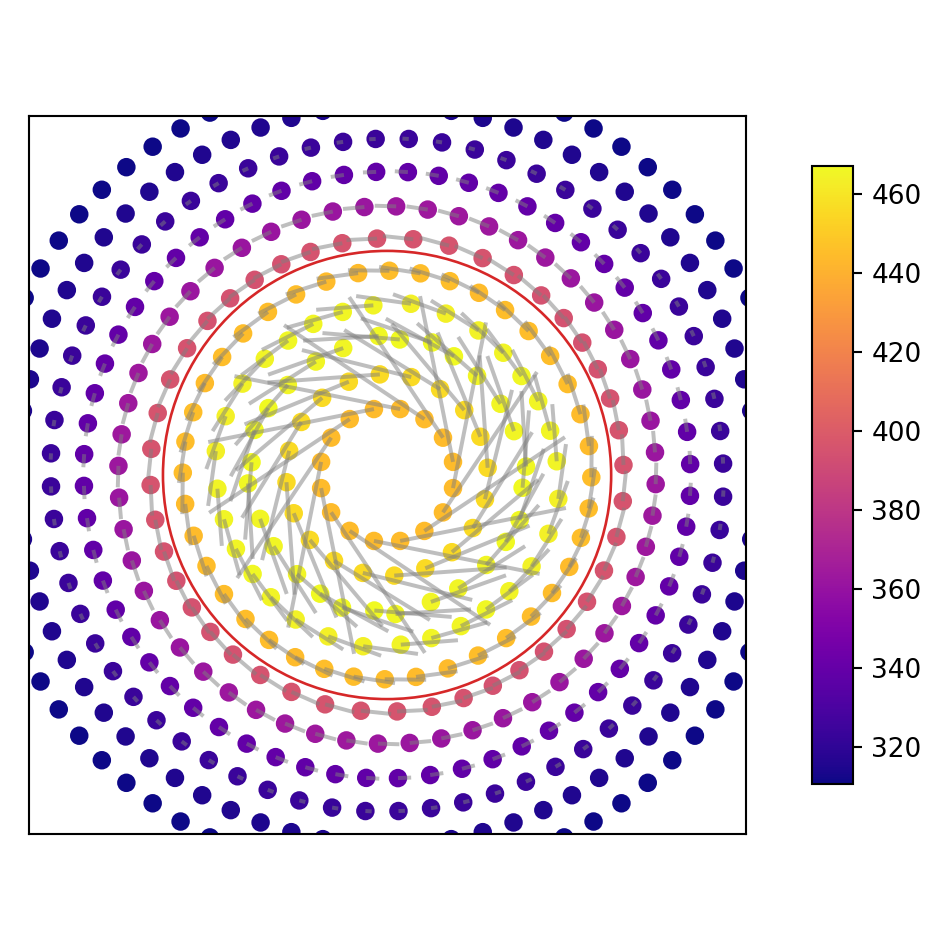}
  \end{subfigure}
  \begin{subfigure}{0.5\textwidth}
    \centering
    \includegraphics[width = \linewidth]{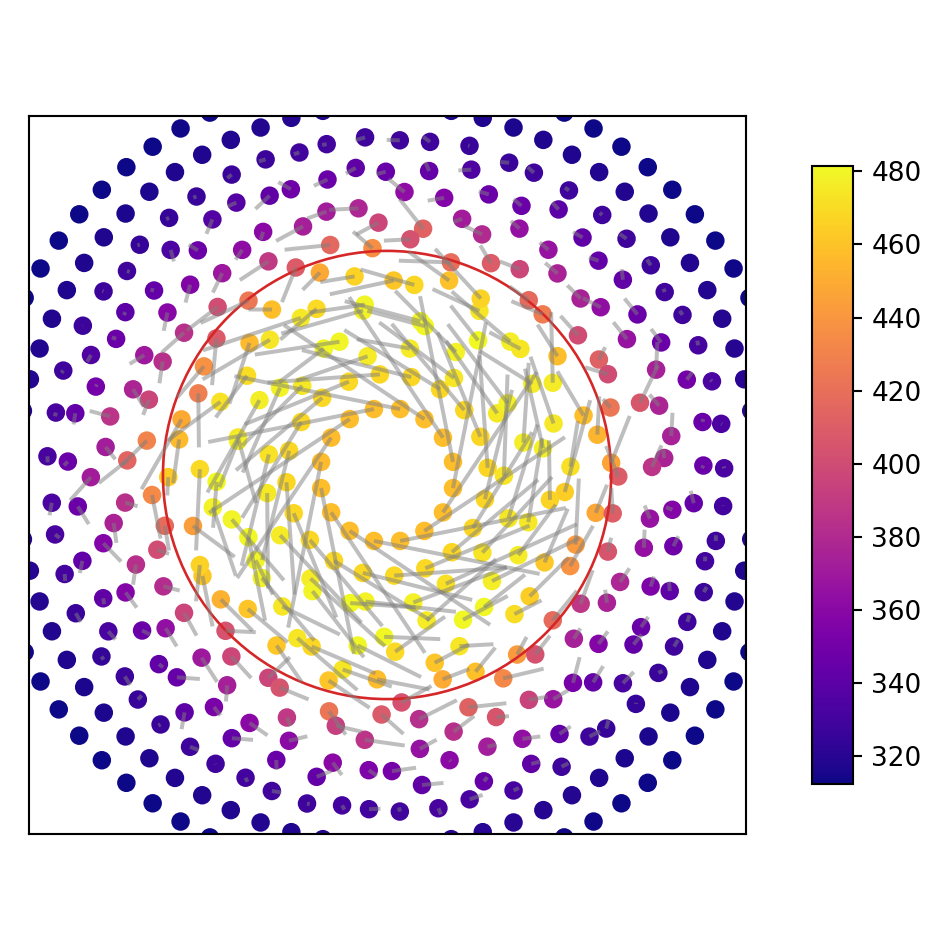}
  \end{subfigure}
  \caption{The snapshot of smoothed particles at the end of simulation, where particle velocity $\mathbf{u}$ is marked as gray segments on particles, and the color shows particle temperature $T$ in unit $\[ \textup{K} \]$. Simulation time $t = 2 \[ \textup{s} \]$. Initial particle distribution is \gridB{}, $\Delta t = 1 \e{-5} \[ \textup{s} \]$, $\lambda = 2 \e{-3}$, $\zeta = 4 \e{-1}$ (top), $\lambda = 0$, $\zeta = 0$ (bottom). The red circle is the boundary of tool shoulder.}
  \label{fig:numerical_snapshott2}
\end{figure}

For comparison, no convergence can be found when we choose another set of $\lambda$ and $\zeta$, for example $\lambda = \zeta = 0$ in \autoref{fig:numerical_temporalCFL1m_Pshift0Vshift0}.
The reason behind this is still the irregular motion, which is clearly revealed in \autoref{fig:numerical_snapshott2}, where a comparison between two snapshots under different choices of $\lambda$ and $\zeta$ is shown. Starting from the same initial particle distribution \gridB{}, and with the same numerical method, the particle motion at simulation time $t = 2 \[ \mykgms{0}{0}{1} \]$ is almost concentric with $\lambda = 2 \e{-3}$ and $\zeta = 4 \e{-1}$, while it is irregular in the bulk with $\lambda = \zeta = 0$.

Moreover, the velocity error is always dominant compared with the temperature error and the density error for all particle distributions \gridA{}, \gridB{}, and \gridC{}, which can be confirmed from \autoref{fig:numerical_snapshott2} where velocity profiles differ a lot but temperature profiles are close.

In the following numerical simulations, we will fix the two parameters $\lambda = 2 \e{-3}$ and $\zeta = 4 \e{-1}$ as our acceptable choice.
We note that the choice of $\lambda$ and $\zeta$ could depend on other parameters in SPH simulation.

\subsection{POD Error with Prescribed Motion}

In the following subsections, we explore the error induced from using the POD method in SPH simulation. To do this, we apply the POD-MOR with respect to particle functions $\mathbf{x} ^{\alpha} \( t \)$, $\mathbf{u} ^{\alpha} \( t \)$, $\mathbf{T} \( t \)$, and $\boldsymbol{\rho} \( t \)$, $\alpha = 1, 2$, using certain number of POD modes in SPH simulations, and compare the evolution of particle functions with the ones from the original SPH simulation.

From the previous study, we find the velocity error is always dominant compared with the temperature and density error, moreover, particle position is not easy to track due to possible irregular motion.
In order to study the effect of POD-MOR, we start with the simplest case, where we prescribe the motion of particles (position are velocity are prescribed, and then density can be computed) and only solve the heat equation. The POD will be applied only to the temperature. 
\myadd{
The directic projection \autoref{eq:model_POD_directProjection} is implemented.
}

First we prescribe the particle motion by the following equations in polar coordinate system $\( r, \theta \)$,
\begin{align}
  r \( t \) 
  & = 
  r \( 0 \) + r _{\textup{amp}} \( \frac{r _{\textup{pin}}}{r \( 0 \)} \) ^{\alpha _{\textup{dis}}} \sin \( 2 \pi t \) \cos \( N _{\textup{ang}} \theta \( t \) \),
  \\
  \theta \( t \)
  & =
  \theta \( 0 \) + \omega _{\textup{pin}} \( \frac{r _{\textup{pin}}}{r \( 0 \)} \) ^{\alpha _{\textup{ang}}} t.
\end{align}
The above equations describes the circular motion with oscillations, where $r _{\textup{amp}}$, $\alpha _{\textup{dis}}$, $N _{\textup{ang}}$, and $\alpha _{\textup{ang}}$ are parameters to be chosen.

We fix the initial particle distribution \gridC{}, and run the simulation for $2 \[ \mykgms{0}{0}{1} \]$ with $\Delta t = 1 \e{-5} \[ \mykgms{0}{0}{1} \]$. We calculate the temperature error when using different number of temperature POD modes.

\autoref{tab:numerical_motion_equation} shows the case of $r _{\textup{amp}} = 0.005 \[ \mykgms{0}{1}{0} \]$, $\alpha _{\textup{dis}} = 1$, $N _{\textup{ang}} = 4$, $\alpha _{\textup{ang}} = 2$, and the case $r _{\textup{amp}} = 0.005 \[ \mykgms{0}{1}{0} \]$, $\alpha _{\textup{dis}} = 2$, $N _{\textup{ang}} = 2$, $\alpha _{\textup{ang}} = 2$.
The original particle system \gridC{} has a total of $1084$ particles (temperature DoFs), that means, with POD-MOR, we can reduce the model to $5\%$ DoFs with $2.5\e{-5}$ error, which is a huge reduction.
We conclude that the POD error for temperature is always small.
\begin{table}[!htp]
  \begin{center}
    \begin{tabular}{|c|c|}
      \hline
      \# T modes & T error
      \\
      \hline
      $200$ & $6.740 \e{-9}$
      \\
      \hline
      $100$ & $1.027 \e{-6}$
      \\
      \hline
      $50$ & $2.529 \e{-5}$
      \\
      \hline
    \end{tabular}
    \begin{tabular}{|c|c|}
      \hline
      \# T modes & T error
      \\
      \hline
      $200$ & $5.253 \e{-8}$
      \\
      \hline
      $100$ & $3.288 \e{-6}$
      \\
      \hline
      $50$ & $2.524 \e{-5}$
      \\
      \hline
    \end{tabular}
  \end{center}
  \caption{The temperature error (T error) with different number of temperature POD modes (\# T modes). Initial particle distribution \gridC{}, $\Theta = \[ 0, 2 \] \[ \textup{s} \]$, $\Delta t = 1 \e{-5} \[ \textup{s} \]$. Prescribed motion by equations, with $r _{\textup{amp}} = 0.005 \[ \textup{m} \]$, $\alpha _{\textup{dis}} = 1$, $N _{\textup{ang}} = 4$, $\alpha _{\textup{ang}} = 2$ (left), $r _{\textup{amp}} = 0.005 \[ \textup{m} \]$, $\alpha _{\textup{dis}} = 2$, $N _{\textup{ang}} = 2$, $\alpha _{\textup{ang}} = 2$ (right).}
  \label{tab:numerical_motion_equation}
\end{table}

In order to show results with more realistic prescribed motions, we first run a full SPH simulation and save particle motion to a file, and then solve the heat equation by SPH simulation based on the saved particle motion with POD-MOR. The POD error is calculated from the comparison of these two solutions of temperature.

Here for the coupled simulation, again we fix the initial particle distribution \gridC{}, SPH parameters $\lambda = 2 \e{-3}$, $\zeta = 4 \e{-1}$, and run the simulation for $2 \[ \mykgms{0}{0}{1} \]$ with $\Delta t = 1 \e{-5} \[ \mykgms{0}{0}{1} \]$. In addition, we also choose the SPH parameters $\lambda = \zeta = 0$ in order to generate the irregular particle motion for comparison.
In \autoref{tab:numerical_motion_full}, we list the temperature error with respect to the number of temperature POD modes we used. Even with a non-optimal set of SPH parameters $\lambda$ and $\zeta$, the temperature error is still bounded by $2.1 \e{-4}$ when using less than $5\%$ POD modes ($N _{\textup{all}} = 1084$).
\begin{table}[!htp]
  \begin{center}
    \begin{tabular}{|c|c|}
      \hline
      \# T modes & T error
      \\
      \hline
      $200$ & $5.107 \e{-9}$
      \\
      \hline
      $100$ & $2.312 \e{-6}$
      \\
      \hline
      $50$ & $3.467 \e{-5}$
      \\
      \hline
    \end{tabular}
    \begin{tabular}{|c|c|}
      \hline
      \# T modes & T error
      \\
      \hline
      $200$ & $7.941 \e{-6}$
      \\
      \hline
      $100$ & $3.736 \e{-5}$
      \\
      \hline
      $50$ & $2.099 \e{-4}$
      \\
      \hline
    \end{tabular}
  \end{center}
  \caption{The temperature error (T error) with different number of temperature POD modes (\# T modes). Initial particle distribution \gridC, $\Theta = \[ 0, 2 \] \[ \textup{s} \]$, $\Delta t = 1 \e{-5} \[ \textup{s} \]$. Prescribed motion from full simulation. SPH parameters $\lambda = 2 \e{-3}$, $\zeta = 4 \e{-1}$ (left), $\lambda = \zeta = 0$ (right).}
  \label{tab:numerical_motion_full}
\end{table}

Our numerical results indicate that POD is able to reduce the DoFs for heat equations in Lagrangian framework, and the error is not sensitive to the particle motion.
This conclusion is consistent with the results in Eulerian framework where high-frequency modes decay fast in heat equations.


\subsection{POD Error with Prescribed Temperature}
\label{sec:numerical_temperature}

In this subsection, we study the POD error for flow equations only. In the FSSW setting, we can prescribe the temperature, and the unknowns are position, velocity, and density. 
\myadd{
The directic projection \autoref{eq:model_POD_directProjection} is implemented.
}

The particle temperature can be prescribed as functions of particle position. We generate the following three cases. The uniform temperature case $T _{\textup{u}} \( \mathbf{x} \) \equiv T _{0}$, the high temperature case $T _{\textup{h}} \( \mathbf{x} \) = T _{0} + b \[ \lmdl \mathbf{x} \rmdl - r _{\textup{d}} \] ^{2}$, and the low temperature case $T _{\textup{l}} \( \mathbf{x} \) = T _{\textup{h}} \( \mathbf{x} \) - T _{\textup{diff}}$, where we choose $b = 1 \e{5} \[ \mykgms{0}{-1}{0} \cdot \textup{K} \]$ and $T _{\textup{diff}} = 140 \[ \textup{K} \] \approx b \[ r _{\textup{p}} - r _{\textup{d}} \] ^{2}$.
Both the high and low temperature profiles are radial symmetric with respect to the origin, and we use a simple quadratic relation to mimic the temperature decay profile as it appears in coupled FSSW simulation.

If we consider using less particles (uniformly coarser initial grid) as another approach of MOR, we can ask the question whether the POD-MOR that using same number of DoFs (as the coarser initial grid) can achieve less error.

We run the SPH simulation that contains only flow equations with initial particle distribution \gridB{} and \gridC{} for $2 \[ \mykgms{0}{0}{1} \]$ with time step size $\Delta t = 1 \e{-5} \[ \mykgms{0}{0}{1} \]$. We fix the SPH parameters $\lambda = 2 \e{-3}$, $\zeta = 4 \e{-1}$, and use different prescribed temperature profiles.
In \autoref{tab:numerical_temperature}, we compare the the POD error (between the POD-MOR SPH solution and the original SPH solution of \gridC{}) with the coarsening error (between the original SPH solutions of \gridB and \gridC{}). In POD-MOR we apply $320$ ($N _{\textup{bulk}}$ in \gridB{} $\sim 39\% N _{\textup{bulk}}$ in \gridC{}) position POD modes, $320$ velocity POD modes, and $504$ ($N _{\textup{all}}$ in \gridB{} $\sim 46\% N _{\textup{all}}$ in \gridC{}) density POD modes.

From the magnitude of the error indicators and the number of POD modes we used, it seems that the POD-MOR is more challenging in the prescribed temperature case ($1.2 \e{-6}$ error with $46\% N _{\textup{all}}$ modes from \autoref{tab:numerical_temperature}) than the prescribed motion case ($5.1 \e{-9}$ error with $18\% N _{\textup{all}}$ modes from \autoref{tab:numerical_motion_full}).
However, POD-MOR shows the huge advantage over uniformly coarsening the system.
With the same DoFs, POD achieves less error than uniformly decreasing the particle number, because POD uses a better strategy in picking combination of particles than just uniformly decreasing number of particles in the computational domain $\Omega$.

An interesting observation on POD error is, the nonuniform temperature profiles result in lower POD error since the particle motion is more localized, i.e., restricted in the region near the tool, furthermore, the low temperature case has the lowest POD error since the motion will be more localized than the high temperature case due to the nonlinearity of dynamic viscosity \autoref{eq:model_viscosity} with respect to temperature.

One of the reasons that uniformly decreasing particle number causes a larger error may come from the mismatch of initial particle distribution. When we apply velocity boundary condition on two layers of particles, such mismatch will introduce a large interpolation error at time $t = 0 \[ \mykgms{0}{0}{1} \]$, where all particles except the ones on two inner layers are assigned to the zero velocity.
The other reason may be from the complexity of SPH algorithm itself. When choosing a different number of particles, different SPH artificial parameters $\lambda$ and $\zeta$ lead to potentially different physical systems. Moreover, with our choice of SPH artificial parameters, the irregular nature of the system can not be entirely eliminated, so the solution may be sensitive to any modifications in SPH setup. To this extent, directly reducing particle numbers in SPH simulation may not be a good choice, but POD-MOR provides us an effective method to reduce the model.

\begin{table}[!htp]
  \begin{center}
    \begin{tabular}{|c|c|c|c|}
      \hline
      & \multicolumn{3}{c|}{uniform coarsening error}
      \\
      \hline
      T profile & V error & T error & D error
      \\
      \hline
      uniform & $4.257 \e{-2}$ & $1.942 \e{-2}$ & $1.945 \e{-2}$
      \\
      \hline
      high    & $4.214 \e{-2}$ & $1.987 \e{-2}$ & $1.948 \e{-2}$
      \\
      \hline
      low     & $1.022 \e{-1}$ & $2.068 \e{-2}$ & $1.968 \e{-2}$
      \\
      \hline
      \hline
      & \multicolumn{3}{c|}{POD error}
      \\
      \hline
      T profile & V error & T error & D error
      \\
      \hline
      uniform & $5.616 \e{-5}$ & $2.144 \e{-5}$ & $2.144 \e{-5}$
      \\
      \hline
      high    & $2.587 \e{-5}$ & $1.815 \e{-5}$ & $1.896 \e{-5}$
      \\
      \hline
      low     & $1.240 \e{-6}$ & $9.113 \e{-7}$ & $1.048 \e{-6}$
      \\
      \hline
    \end{tabular}
  \end{center}
  \caption{Velocity error (V error), temperature error (T error), and density error (D error) with prescribed temperature profile (T profile). Initial particle distribution \gridC{}, $\Theta = \[ 0, 2 \] \[ \textup{s} \]$, $\Delta t = 1 \e{-5} \[ \textup{s} \]$, $\lambda = 2 \e{-3}$, $\zeta = 4 \e{-1}$. Under different prescribed temperature profiles, we can compare the solution of \gridC{} with the solutions of \gridB{} (top) or with the POD error with the same DoFs (bottom).}
  \label{tab:numerical_temperature}
\end{table}

Although we have not applied POD-MOR on particle temperature, the temperature error still exists. This is because the mismatch of particles position will introduce the interpolation error. For the case we fix temperature as constant on particles, the temperature error will be very close to the density error, since the density is almost uniform during evolution, and the interpolation error is dominant in the total error of density.

In \autoref{tab:numerical_temperature_}, the same test is done between initial particle distribution \gridB{} and \gridA{}, we note that $N _{\textup{bulk}}$ in \gridA{} $\sim 15\% N _{\textup{bulk}}$ in \gridB{}, $N _{\textup{all}}$ in \gridA{} $\sim 29\% N _{\textup{all}}$ in \gridB{}. Similar conclusions can be made, but the POD error is larger since we are using less percentage of DoFs compared with the above example.

\begin{table}[!htp]
  \begin{center}
    \begin{tabular}{|c|c|c|c|}
      \hline
      & \multicolumn{3}{c|}{uniform coarsening error}
      \\
      \hline
      T profile & V error & T error & D error
      \\
      \hline
      uniform & $1.250 \e{-1}$ & $4.847 \e{-2}$ & $4.858 \e{-2}$
      \\
      \hline
      high    & $1.250 \e{-1}$ & $4.549 \e{-2}$ & $4.858 \e{-2}$
      \\
      \hline
      low     & $1.706 \e{-1}$ & $4.387 \e{-2}$ & $4.865 \e{-2}$
      \\
      \hline
      \hline
      & \multicolumn{3}{c|}{POD error}
      \\
      \hline
      T profile & V error & T error & D error
      \\
      \hline
      uniform & $4.662 \e{-2}$ & $2.366 \e{-2}$ & $2.366 \e{-2}$
      \\
      \hline
      high    & $4.999 \e{-2}$ & $3.088 \e{-2}$ & $2.967 \e{-2}$
      \\
      \hline
      low     & $9.095 \e{-3}$ & $4.683 \e{-3}$ & $4.063 \e{-3}$
      \\
      \hline
    \end{tabular}
  \end{center}
  \caption{Velocity error (V error), temperature error (T error), and density error (D error) with prescribed temperature profile (T profile). Initial particle distribution \gridB{}, $\Theta = \[ 0, 2 \] \[ \textup{s} \]$, $\Delta t = 1 \e{-5} \[ \textup{s} \]$, $\lambda = 2 \e{-3}$, $\zeta = 4 \e{-1}$. Under different prescribed temperature profiles, we can compare the solution of \gridB{} with the solutions of \gridA{} (top) or with the POD error with the same DoFs (bottom).}
  \label{tab:numerical_temperature_}
\end{table}


\subsection{POD Error with Full FSSW Model}

In the previous subsections, we show that the POD-MOR can significantly reduce the DoFs of the model within acceptable error with heat equation and with flow equations for SPH simulations. In this subsection, with the coupled FSSW governing equations, the POD-MOR still works well on SPH simulation
\myadd{
when the directic projection \autoref{eq:model_POD_directProjection} is implemented.
}

\begin{figure}[!htp]
  \centering
  \begin{subfigure}{0.5\textwidth}
    \centering
    \includegraphics[width = \linewidth]{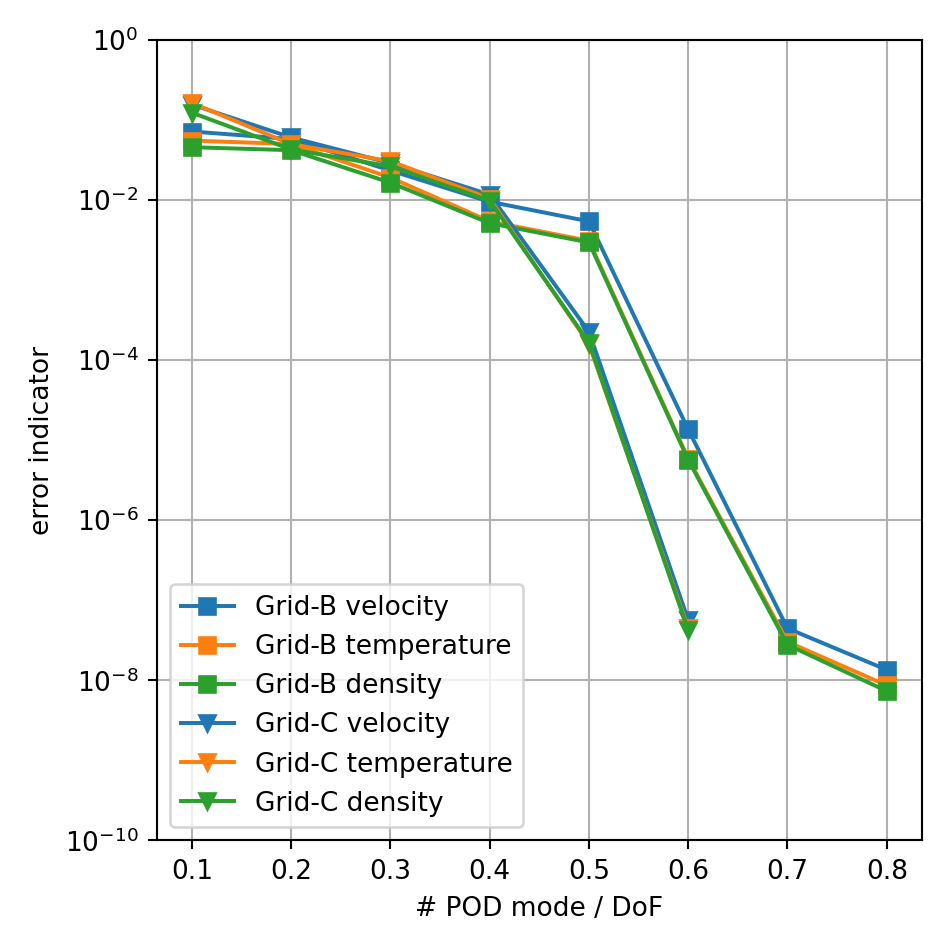}
  \end{subfigure}
  \begin{subfigure}{0.5\textwidth}
    \centering
    \includegraphics[width = \linewidth]{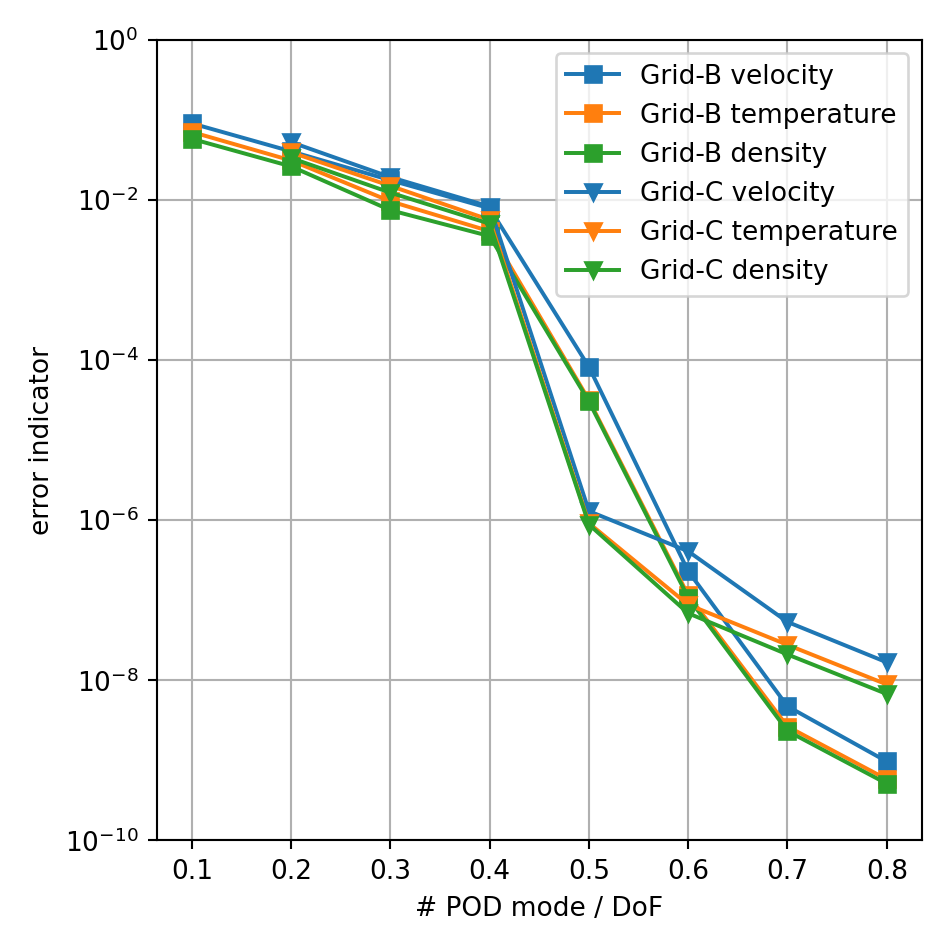}
  \end{subfigure}
  \caption{POD error with coupled model. Initial particle distribution \gridB{} and \gridC{}, $\Theta = \[ 0, 2 \] \[ \textup{s} \]$, $\lambda = 2 \e{-3}$, $\zeta = 4 \e{-1}$. $\Delta t = 1 \e{-6} \[ \textup{s} \]$ (top), $\Delta t = 1 \e{-5} \[ \textup{s} \]$ (bottom).}
  \label{fig:numerical_POD}
\end{figure}

We focus on the particle distributed as \gridB{} and \gridC{}, and run SPH simulation for $2 \[ \mykgms{0}{0}{1} \]$ with time step size $\Delta t = 1 \e{-6} \[ \mykgms{0}{0}{1} \]$ and $\Delta t = 1 \e{-5} \[ \mykgms{0}{0}{1} \]$.
In \autoref{fig:numerical_POD}, we plot the POD error with respect to the percentage of POD modes used over the total DoF, where uniform percentage ($10\%, 20\%, \cdots$) of position, velocity, temperature, and density modes are picked for each data point.
We observe a good alignment of all velocity, temperature, and density POD error for each initial particle distribution, and a fast decay of POD error when we use about $50\%$ DoFs as POD modes for position, velocity, temperature, and density. Overall, the more POD modes we use, the less POD error we can achieve.

Comparing with the results of temporal discretization error, the POD error is dominant when we use less than $40\%$ POD modes, since we can achieve about $1\%$ total error by using $40\%$ position, velocity, temperature, and density POD modes. If we accept $10\%$ total error, we just need $10\%$ DoFs as the number of POD modes.

Furthermore, we notice that the two plots in \autoref{fig:numerical_POD} are similar in magnitude, that means in practice, we can just apply POD-MOR on a coarse temporal mesh.

We note that the reduction in DoFs is not that remarkable compared with the Eulerian case in \cite{cao2022machine} and it indicates that the application of POD-MOR for SPH simulation is more challenging than in the Eulerian framework.
However, from \autoref{tab:numerical_full}, we can still conclude that POD provides an effective method of reduction for SPH simulations, whereas uniformly reducing number of particles introduces larger error, just as in the case with prescribed temperature.

\begin{table}[!htp]
  \begin{center}
    \begin{tabular}{|c|c|c|c|}
      \hline
      & \multicolumn{3}{c|}{uniform coarsening error}
      \\
      \hline
      $\Delta t$ & V error & T error & D error
      \\
      \hline
      $1 \e{-5} \[ \mykgms{0}{0}{1} \]$ & $7.514 \e{-2}$ & $2.664 \e{-2}$ & $2.042 \e{-2}$
      \\
      \hline
      $1 \e{-6} \[ \mykgms{0}{0}{1} \]$ & $7.330 \e{-2}$ & $2.769 \e{-2}$ & $2.055 \e{-2}$
      \\
      \hline
      \hline
      & \multicolumn{3}{c|}{POD error}
      \\
      \hline
      $\Delta t$ & V error & T error & D error
      \\
      \hline
      $1 \e{-5} \[ \mykgms{0}{0}{1} \]$ & $2.498 \e{-3}$ & $1.265 \e{-3}$ & $8.357 \e{-4}$
      \\
      \hline
      $1 \e{-6} \[ \mykgms{0}{0}{1} \]$ & $9.856 \e{-4}$ & $8.003 \e{-4}$ & $7.820 \e{-4}$
      \\
      \hline
    \end{tabular}
  \end{center}
  \caption{Velocity error (V error), temperature error (T error), and density error (D error) with full model with different time step size $\Delta t$. Initial particle distribution \gridB{}, $\Theta = \[ 0, 2 \] \[ \textup{s} \]$, $\lambda = 2 \e{-3}$, $\zeta = 4 \e{-1}$. Under different prescribed temperature profiles, we can compare the solution of \gridC{} with the solutions of \gridB{} (top) or with the POD error with the same DoFs (bottom).}
  \label{tab:numerical_full}
\end{table}


\subsection{POD Modes}

Besides the previous detailed analysis of POD error, we  now study the geometric organization of particles and physical interpretation of POD modes, i.e., how the combination of particles are chosen by POD algorithm. Since we do not have rigorous theory for error analysis, the illustrations will be useful to show why POD can actually reduce the DoFs, and how it is more efficient than uniformly reducing the number of particles.

\begin{figure}[!htp]
  \centering
  \begin{subfigure}{0.4\textwidth}
    \centering
    \includegraphics[width = \linewidth]{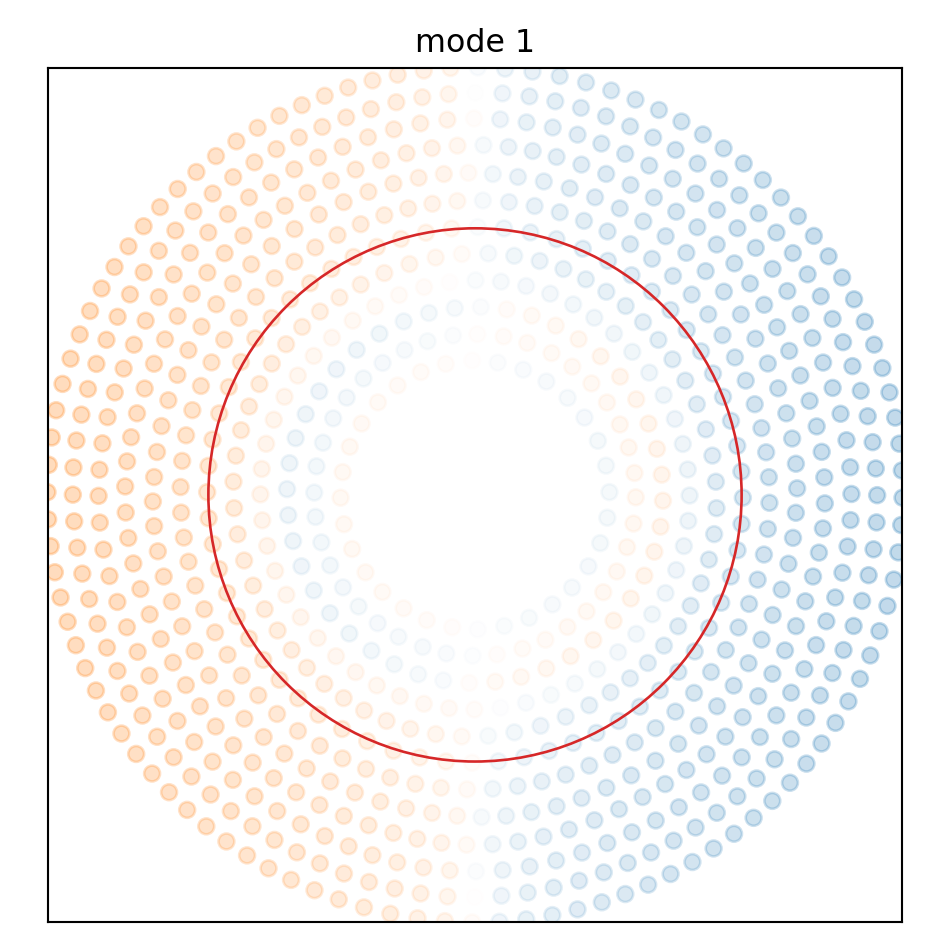}
  \end{subfigure}
  \begin{subfigure}{0.4\textwidth}
    \centering
    \includegraphics[width = \linewidth]{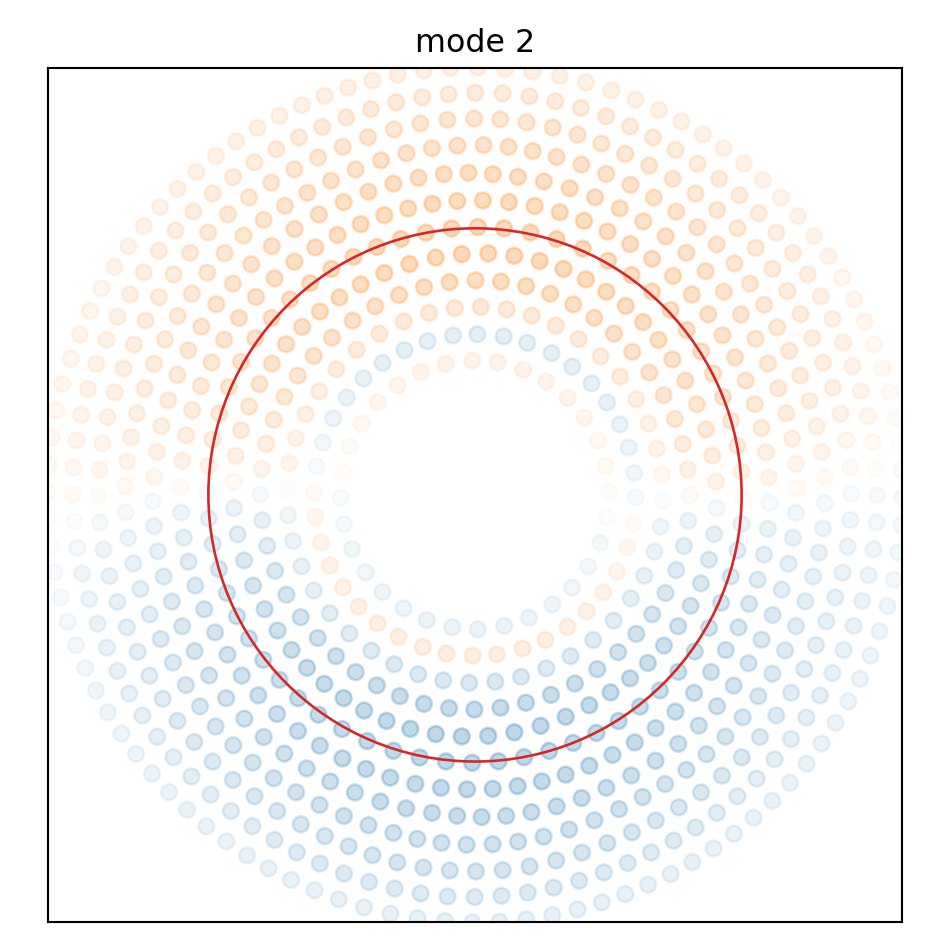}
  \end{subfigure}
  \\
  \begin{subfigure}{0.4\textwidth}
    \centering
    \includegraphics[width = \linewidth]{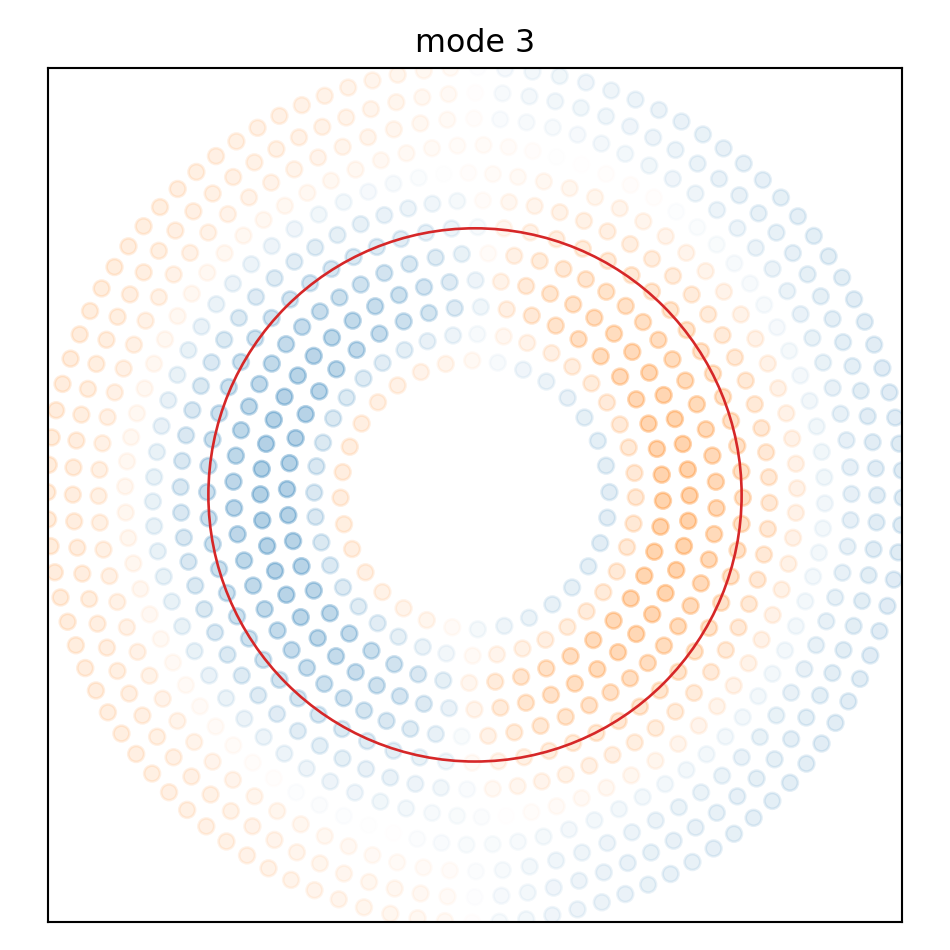}
  \end{subfigure}
  \begin{subfigure}{0.4\textwidth}
    \centering
    \includegraphics[width = \linewidth]{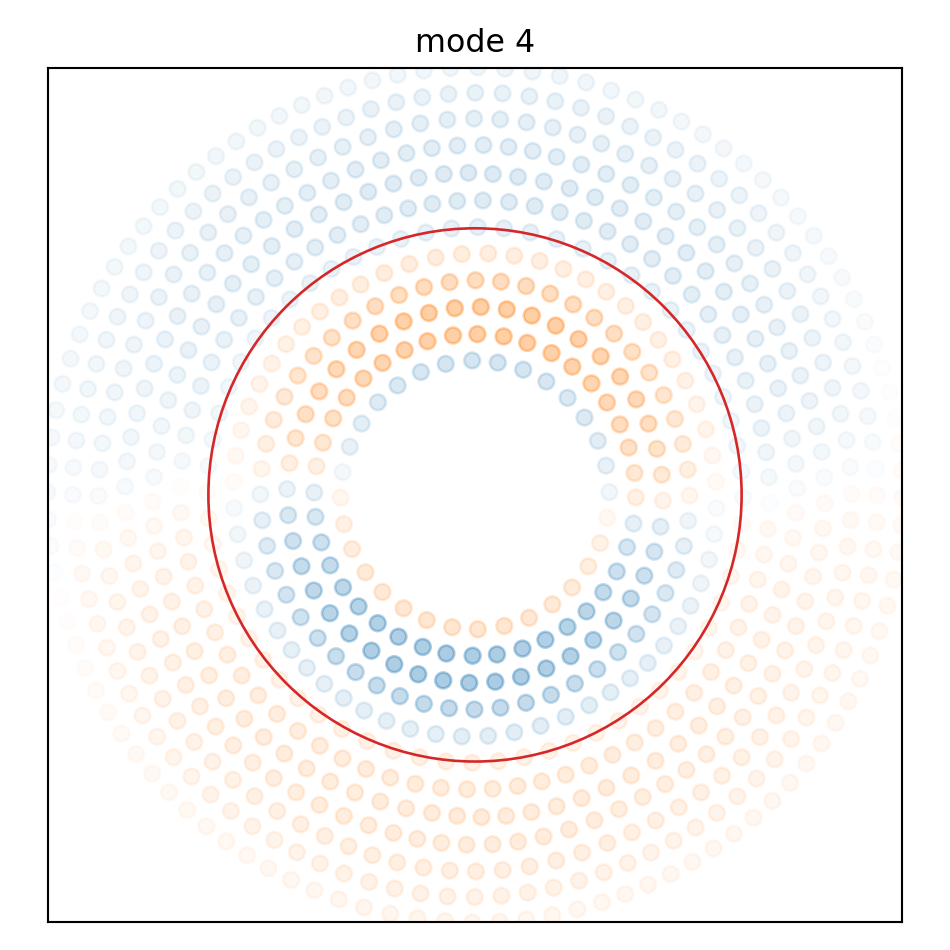}
  \end{subfigure}
  \\
  \begin{subfigure}{0.4\textwidth}
    \centering
    \includegraphics[width = \linewidth]{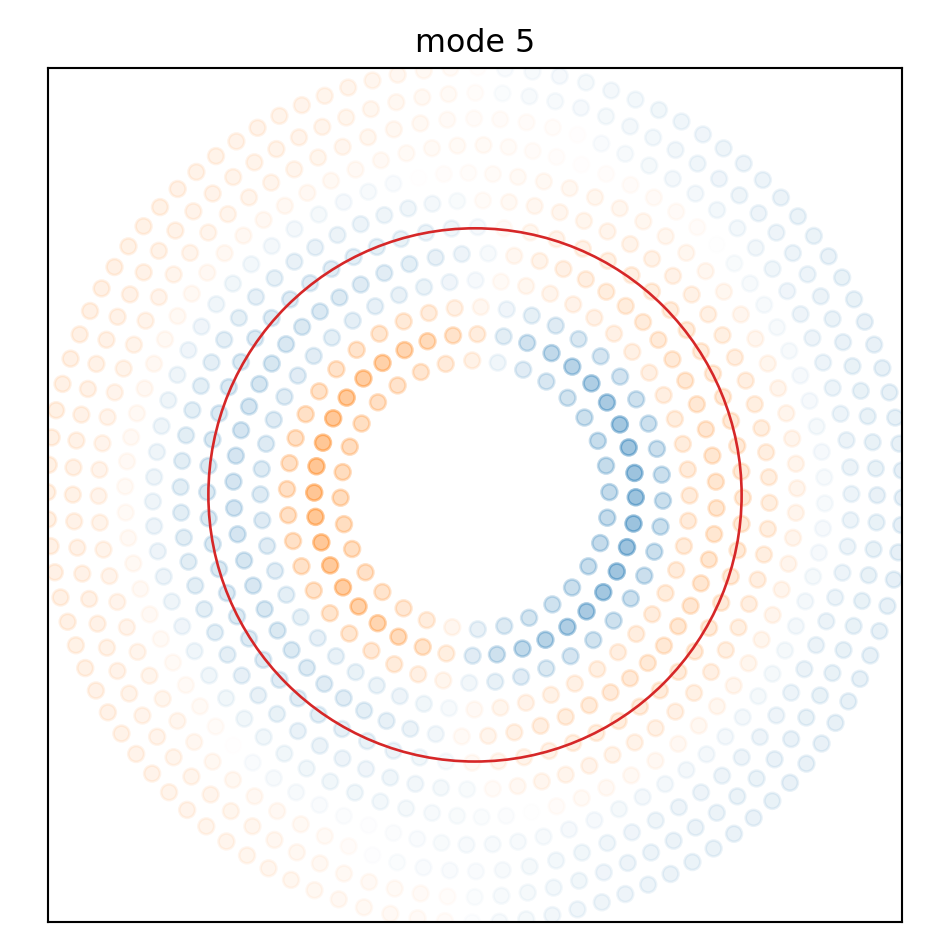}
  \end{subfigure}
  \begin{subfigure}{0.4\textwidth}
    \centering
    \includegraphics[width = \linewidth]{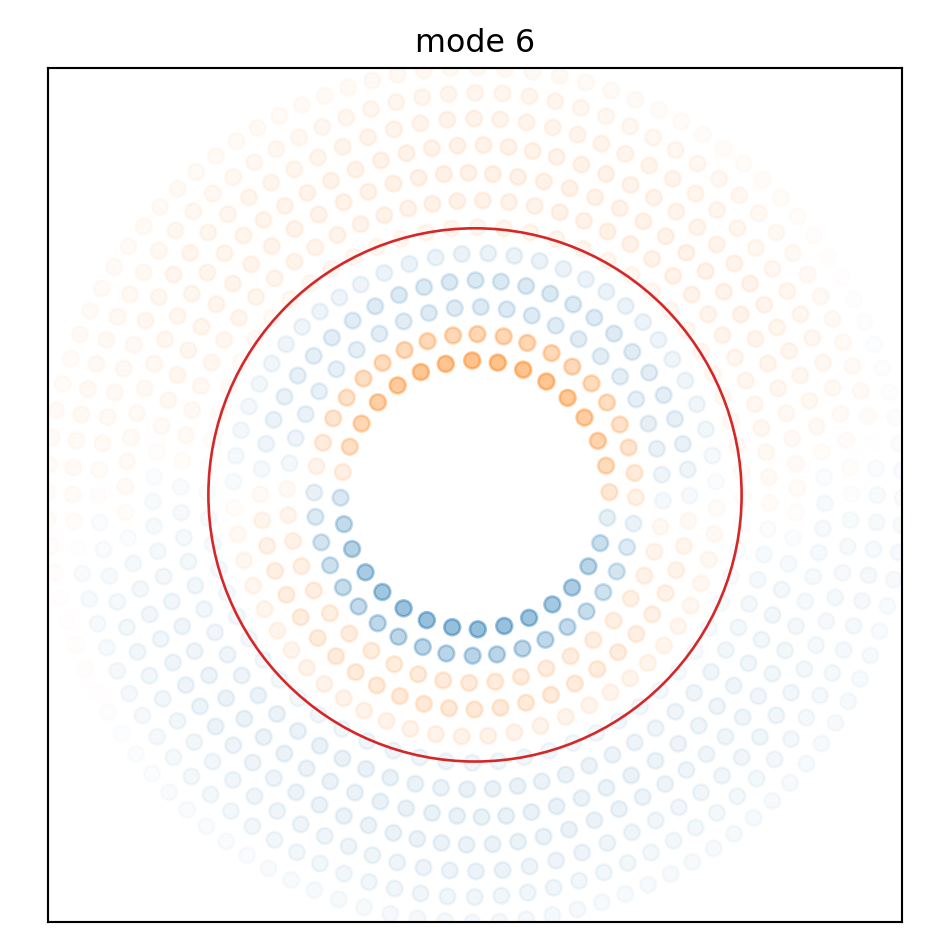}
  \end{subfigure}
  \caption{First six POD modes of position $x _{1}$.}
  \label{fig:PODmode_illustation_position_x}
\end{figure}

\begin{figure}[!htp]
  \centering
  \begin{subfigure}{0.4\textwidth}
    \centering
    \includegraphics[width = \linewidth]{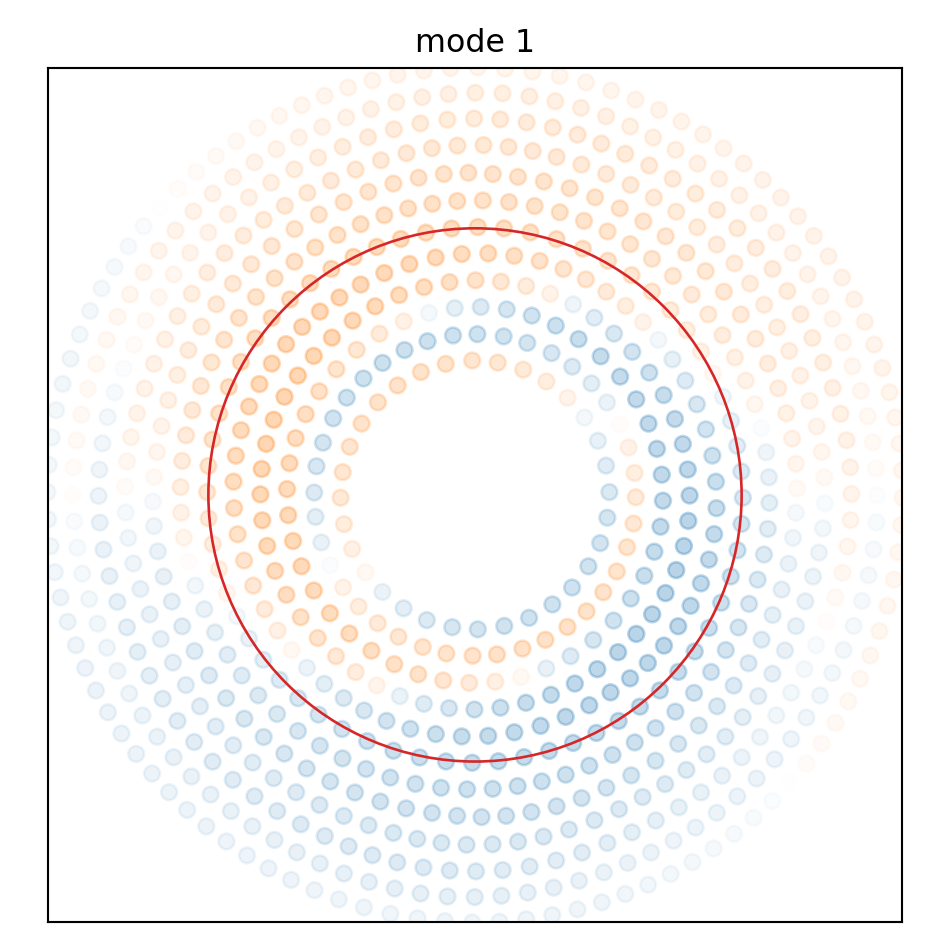}
  \end{subfigure}
  \begin{subfigure}{0.4\textwidth}
    \centering
    \includegraphics[width = \linewidth]{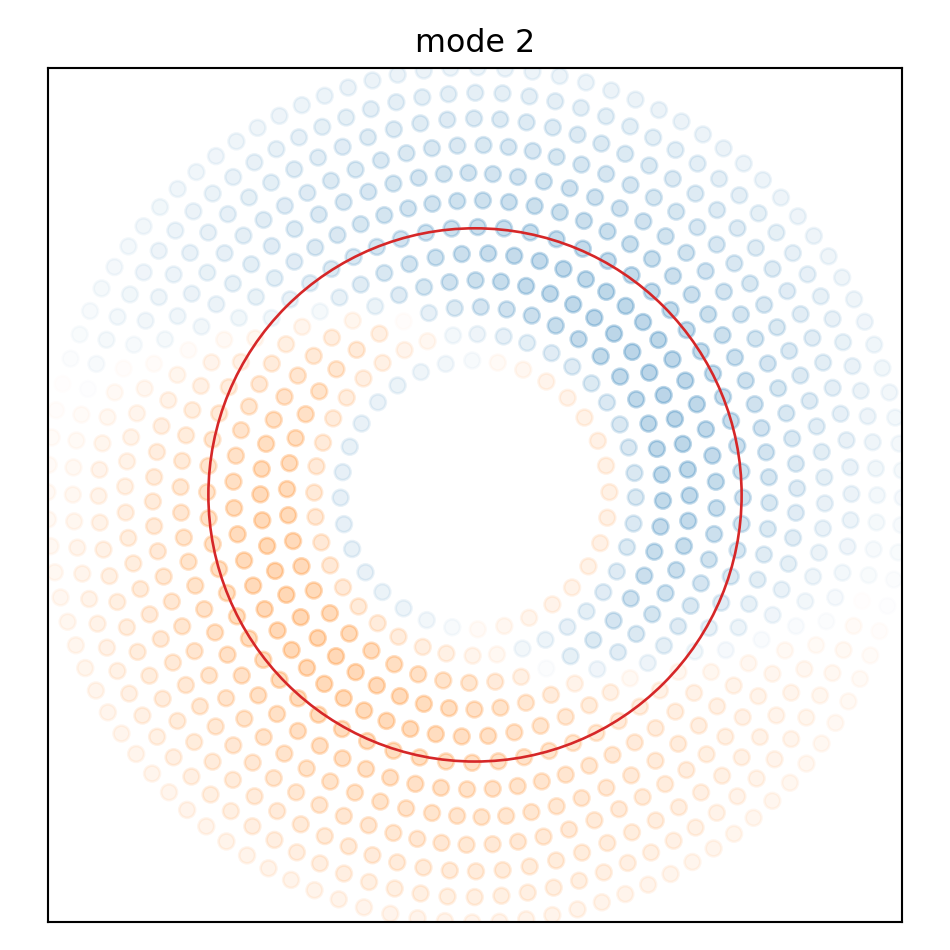}
  \end{subfigure}
  \\
  \begin{subfigure}{0.4\textwidth}
    \centering
    \includegraphics[width = \linewidth]{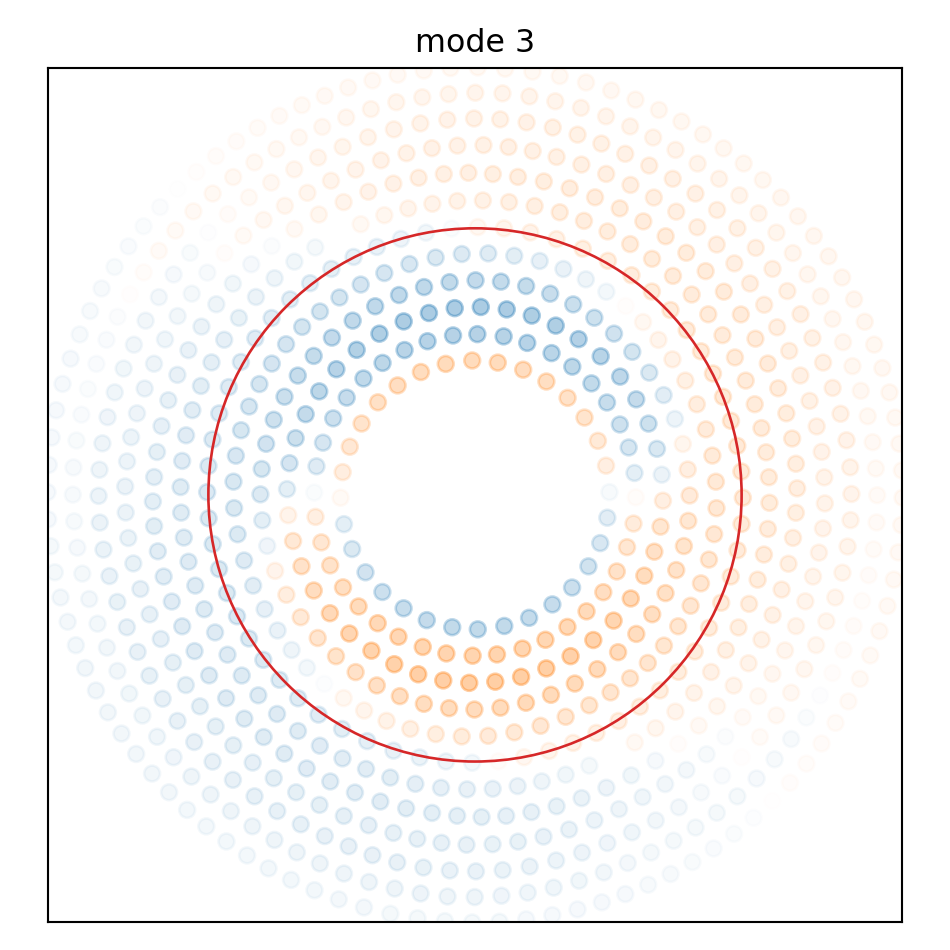}
  \end{subfigure}
  \begin{subfigure}{0.4\textwidth}
    \centering
    \includegraphics[width = \linewidth]{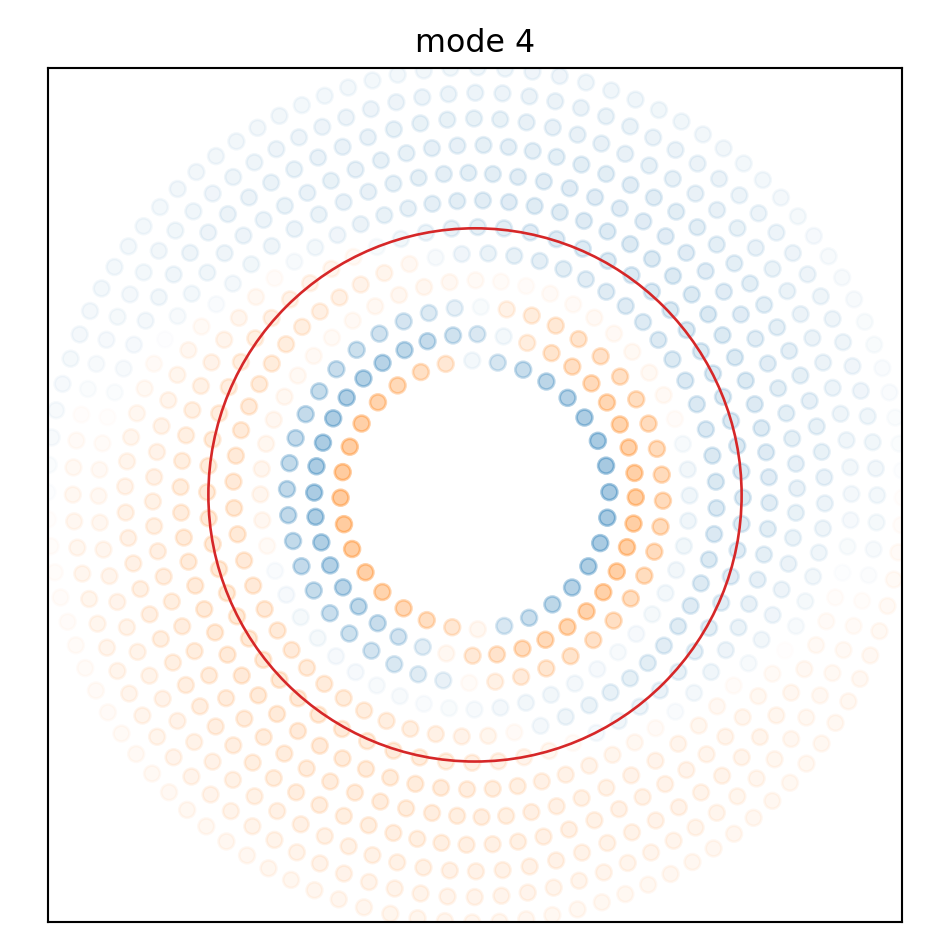}
  \end{subfigure}
  \\
  \begin{subfigure}{0.4\textwidth}
    \centering
    \includegraphics[width = \linewidth]{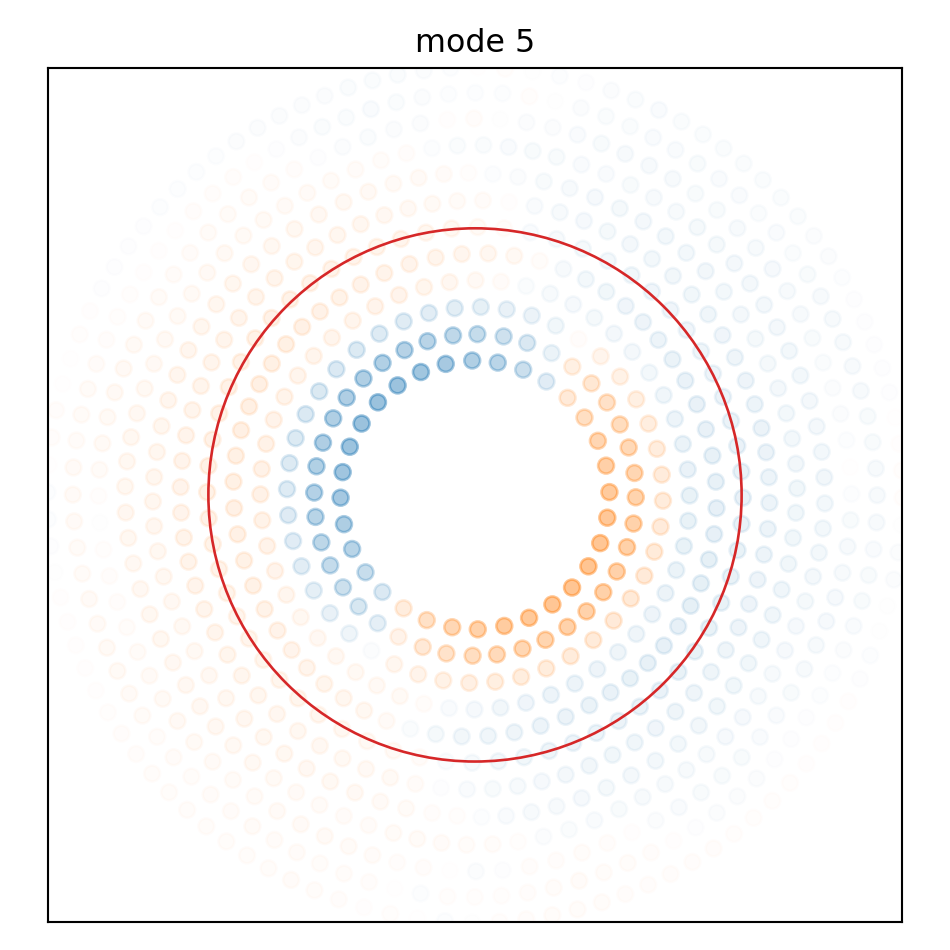}
  \end{subfigure}
  \begin{subfigure}{0.4\textwidth}
    \centering
    \includegraphics[width = \linewidth]{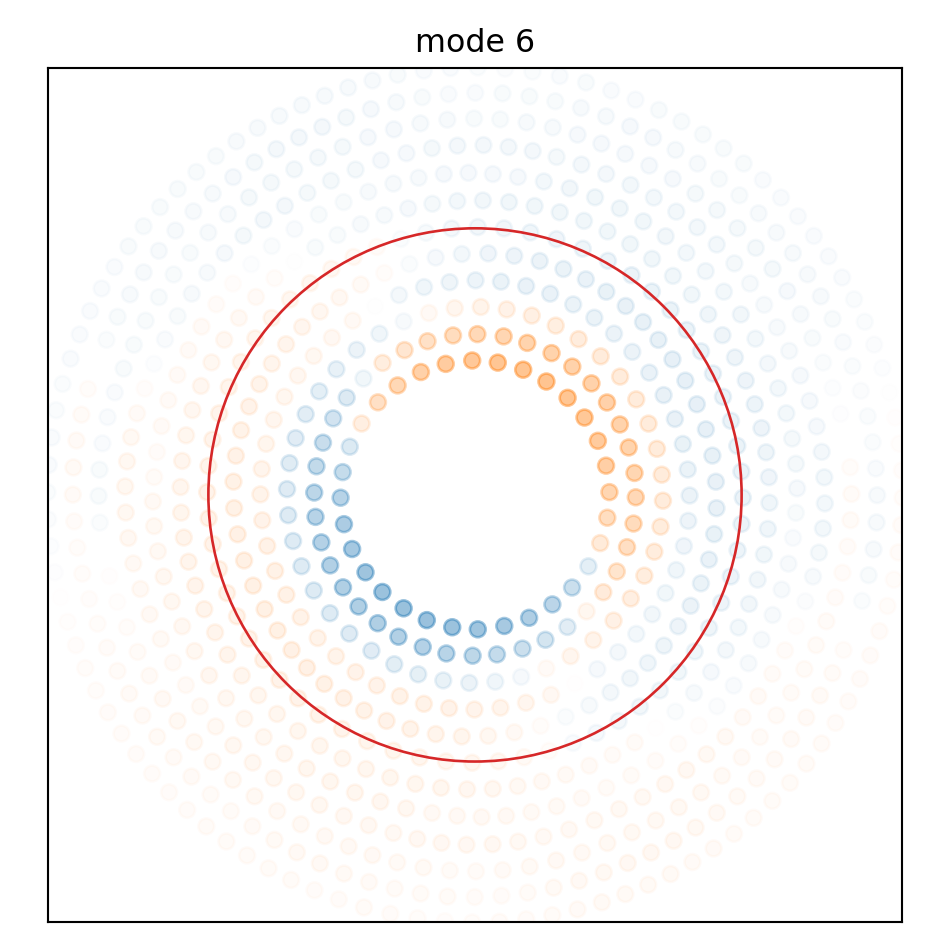}
  \end{subfigure}
  \caption{First six POD modes of velocity $u _{1}$.}
  \label{fig:PODmode_illustation_velocity_x}
\end{figure}

\begin{figure}[!htp]
  \centering
  \begin{subfigure}{0.4\textwidth}
    \centering
    \includegraphics[width = \linewidth]{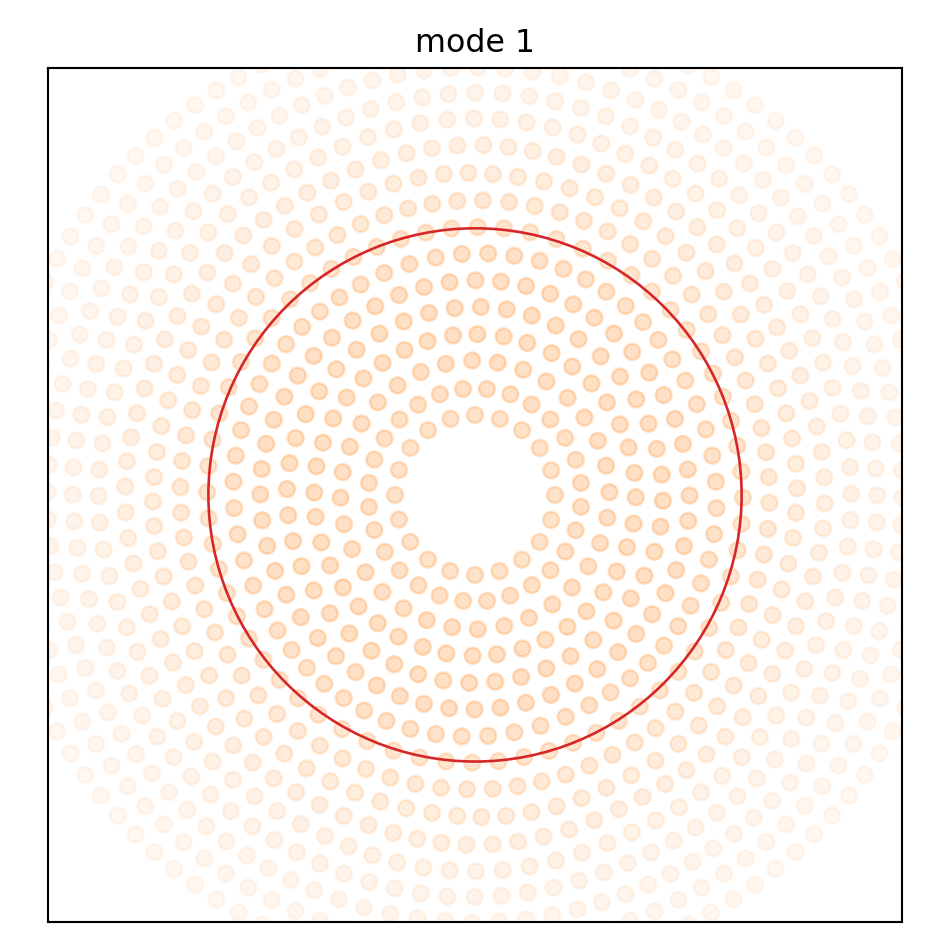}
  \end{subfigure}
  \begin{subfigure}{0.4\textwidth}
    \centering
    \includegraphics[width = \linewidth]{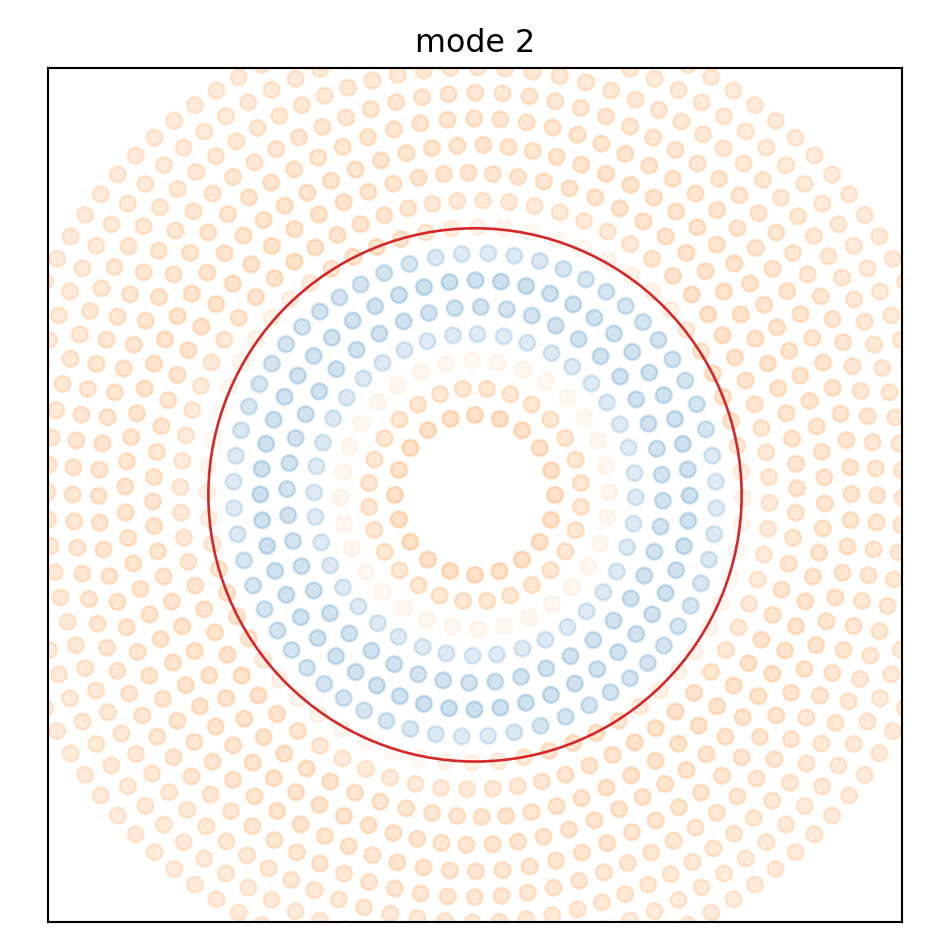}
  \end{subfigure}
  \\
  \begin{subfigure}{0.4\textwidth}
    \centering
    \includegraphics[width = \linewidth]{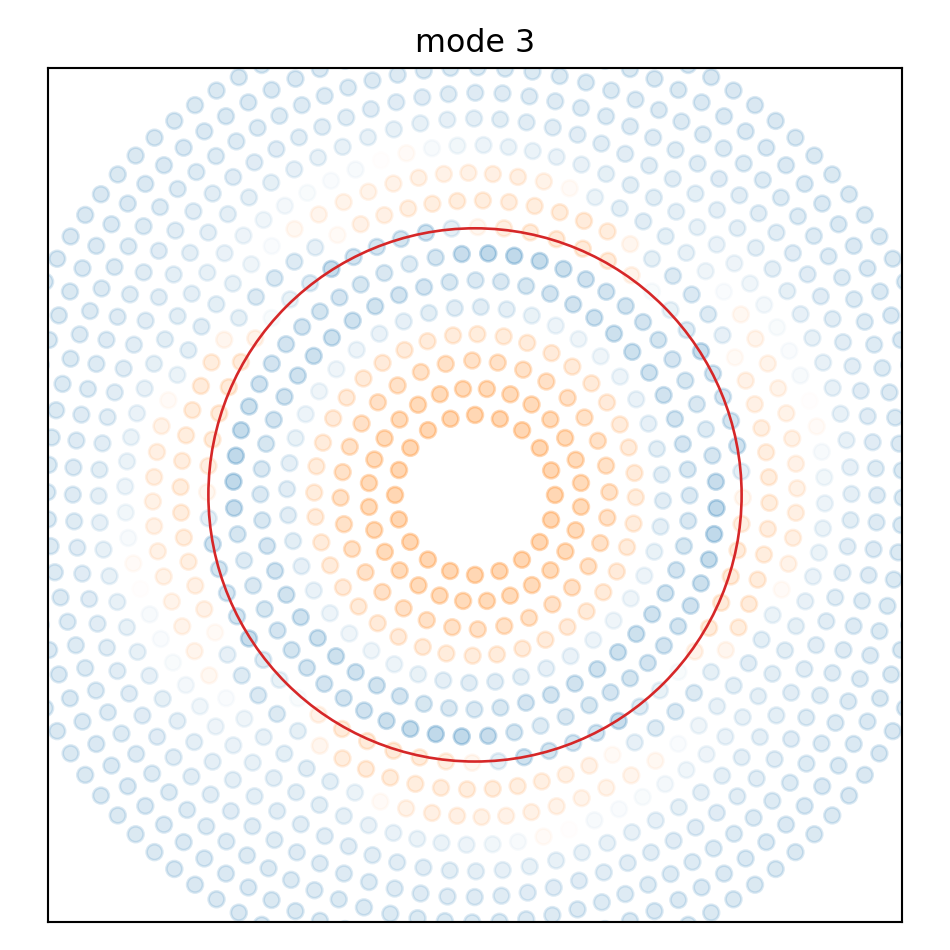}
  \end{subfigure}
  \begin{subfigure}{0.4\textwidth}
    \centering
    \includegraphics[width = \linewidth]{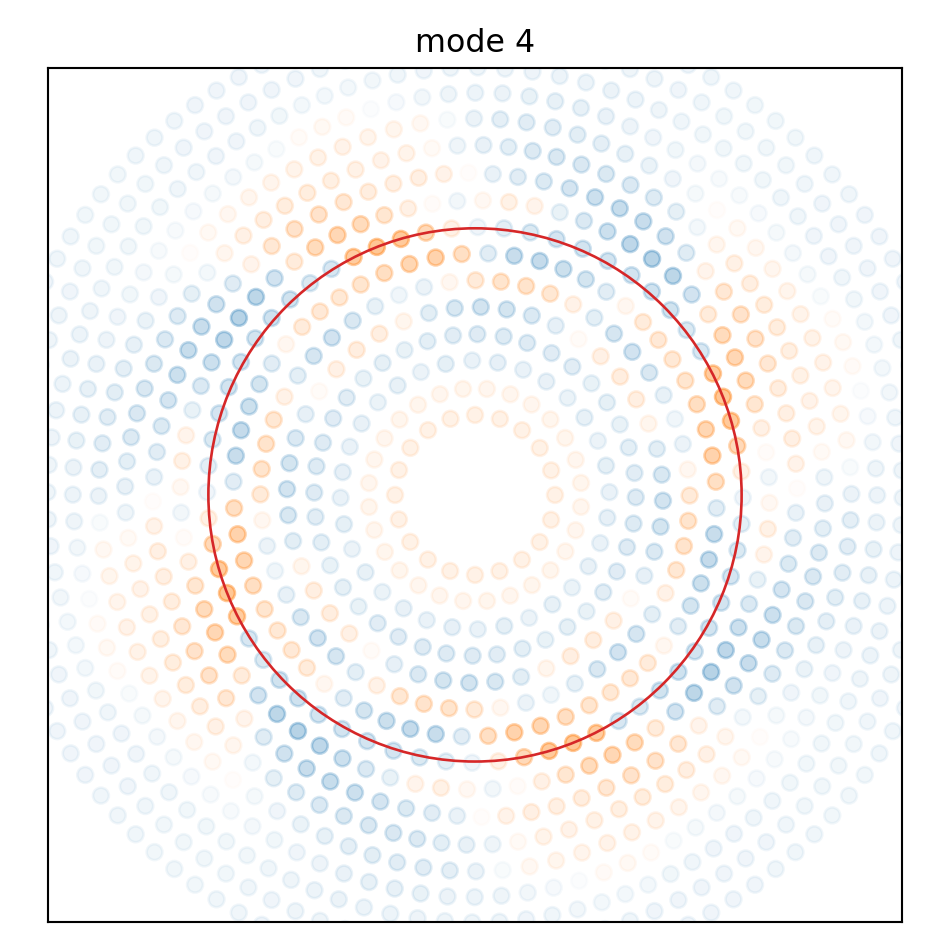}
  \end{subfigure}
  \\
  \begin{subfigure}{0.4\textwidth}
    \centering
    \includegraphics[width = \linewidth]{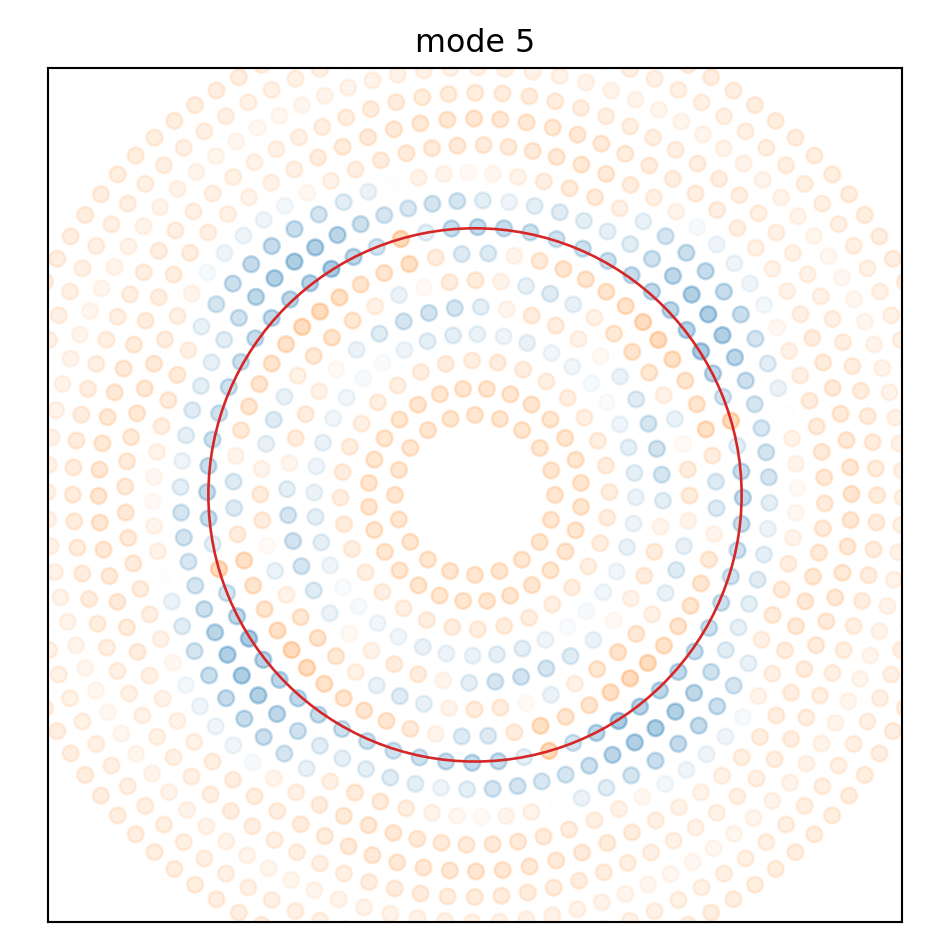}
  \end{subfigure}
  \begin{subfigure}{0.4\textwidth}
    \centering
    \includegraphics[width = \linewidth]{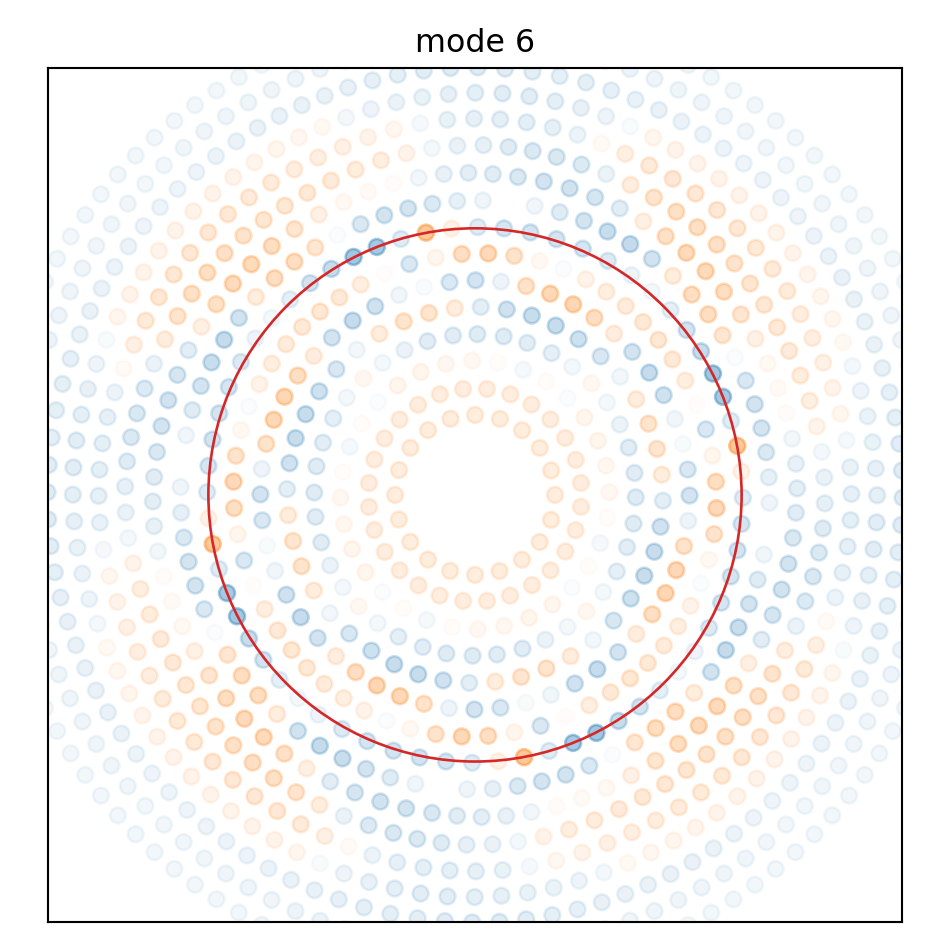}
  \end{subfigure}
  \caption{First six POD modes of temperature.}
  \label{fig:PODmode_illustation_temperature}
\end{figure}

\begin{figure}[!htp]
  \centering
  \begin{subfigure}{0.4\textwidth}
    \centering
    \includegraphics[width = \linewidth]{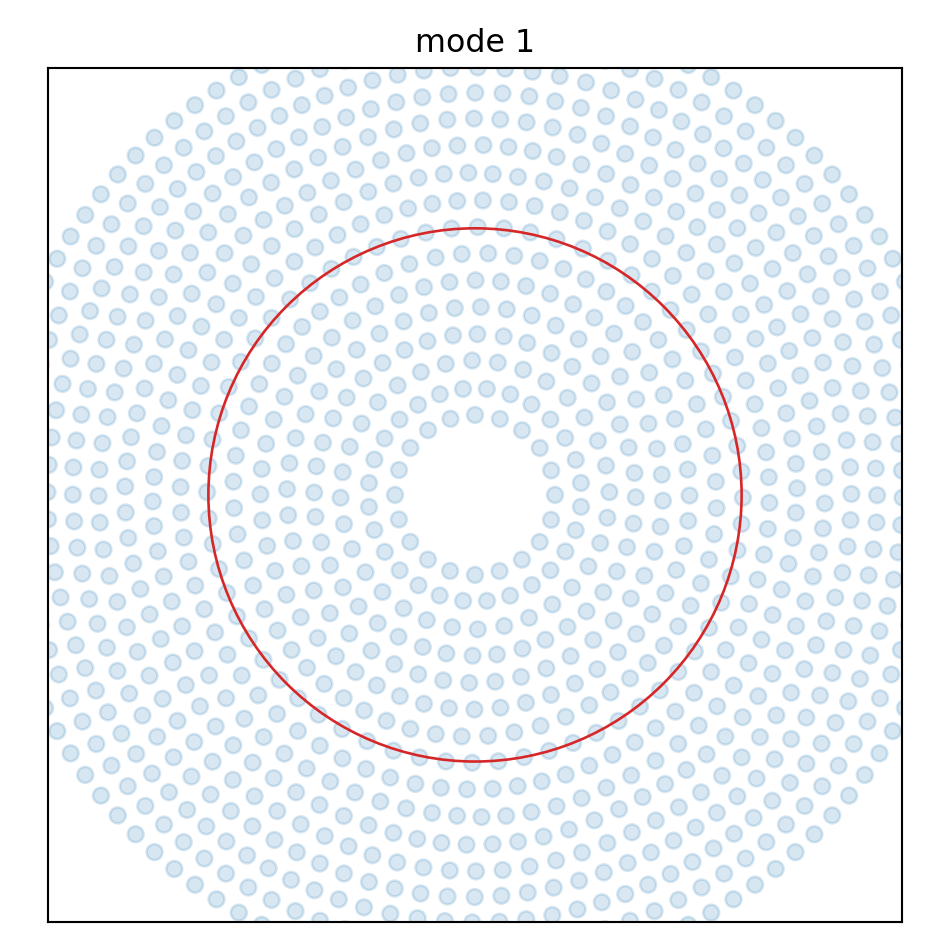}
  \end{subfigure}
  \begin{subfigure}{0.4\textwidth}
    \centering
    \includegraphics[width = \linewidth]{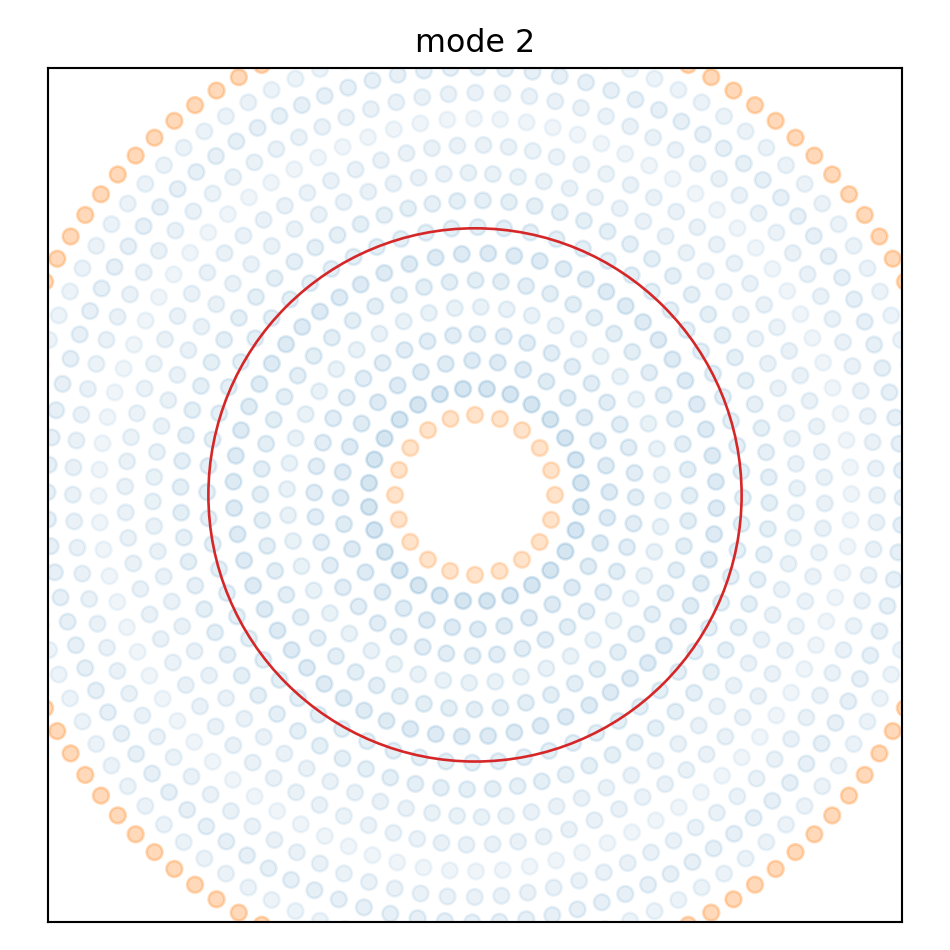}
  \end{subfigure}
  \\
  \begin{subfigure}{0.4\textwidth}
    \centering
    \includegraphics[width = \linewidth]{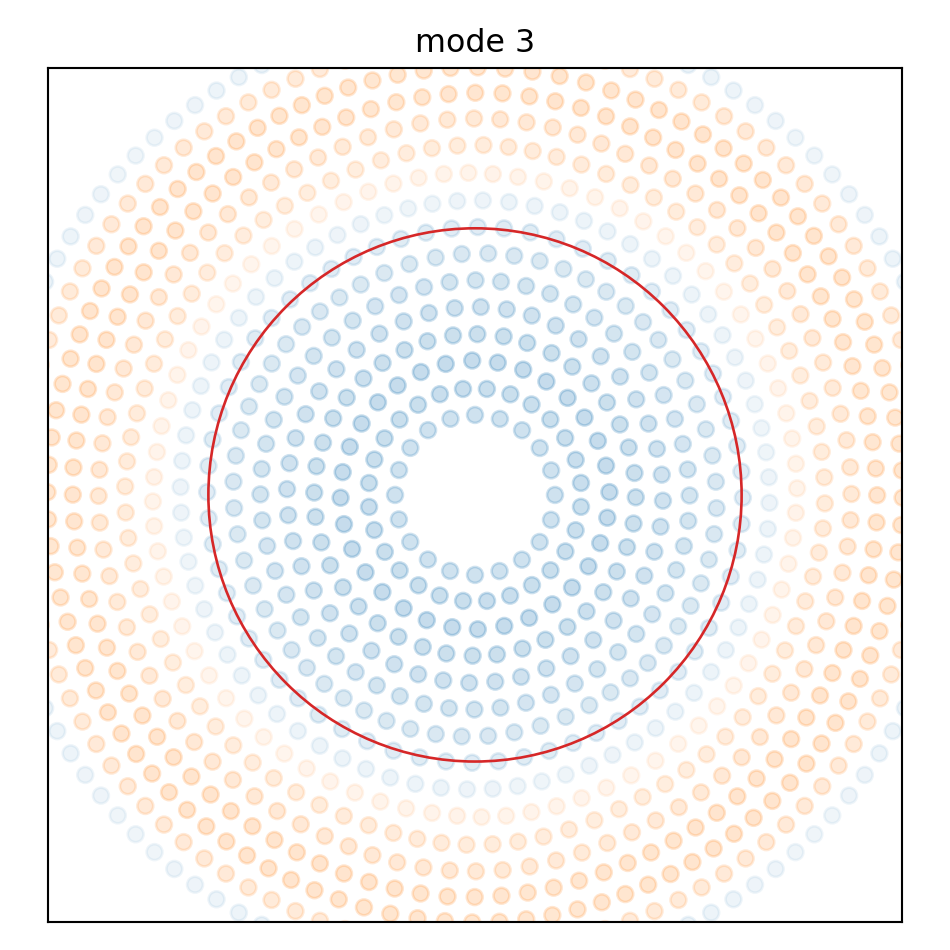}
  \end{subfigure}
  \begin{subfigure}{0.4\textwidth}
    \centering
    \includegraphics[width = \linewidth]{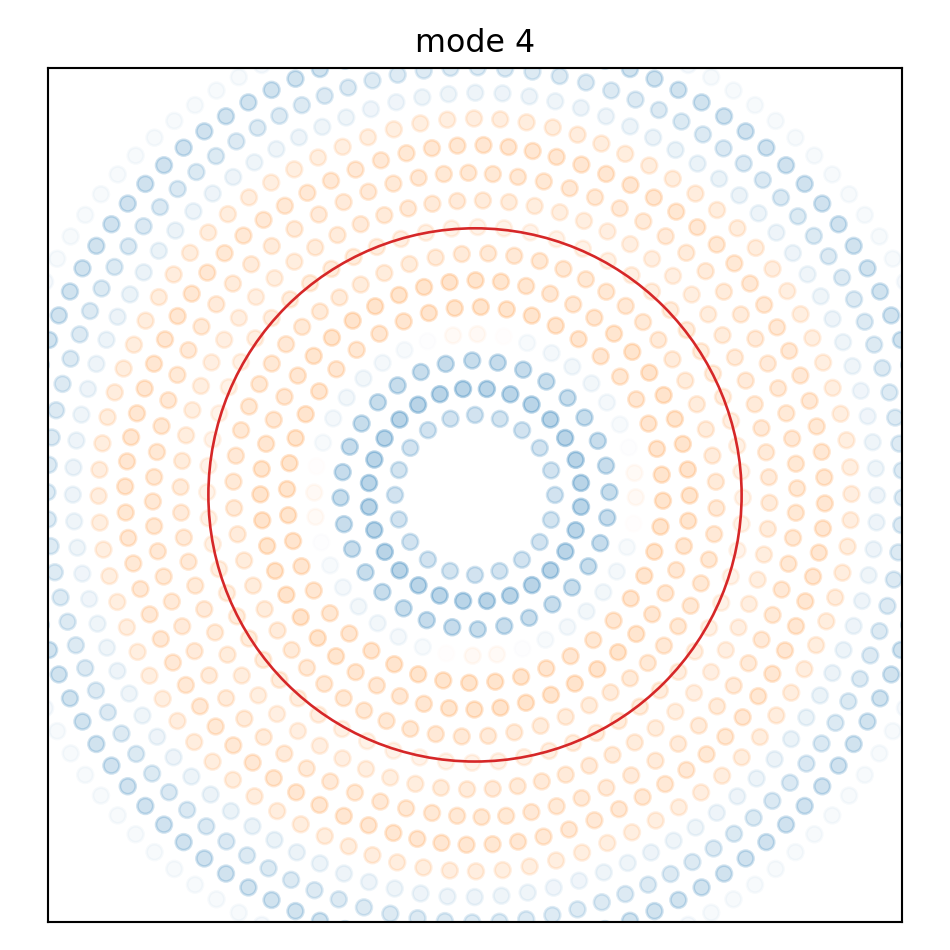}
  \end{subfigure}
  \\
  \begin{subfigure}{0.4\textwidth}
    \centering
    \includegraphics[width = \linewidth]{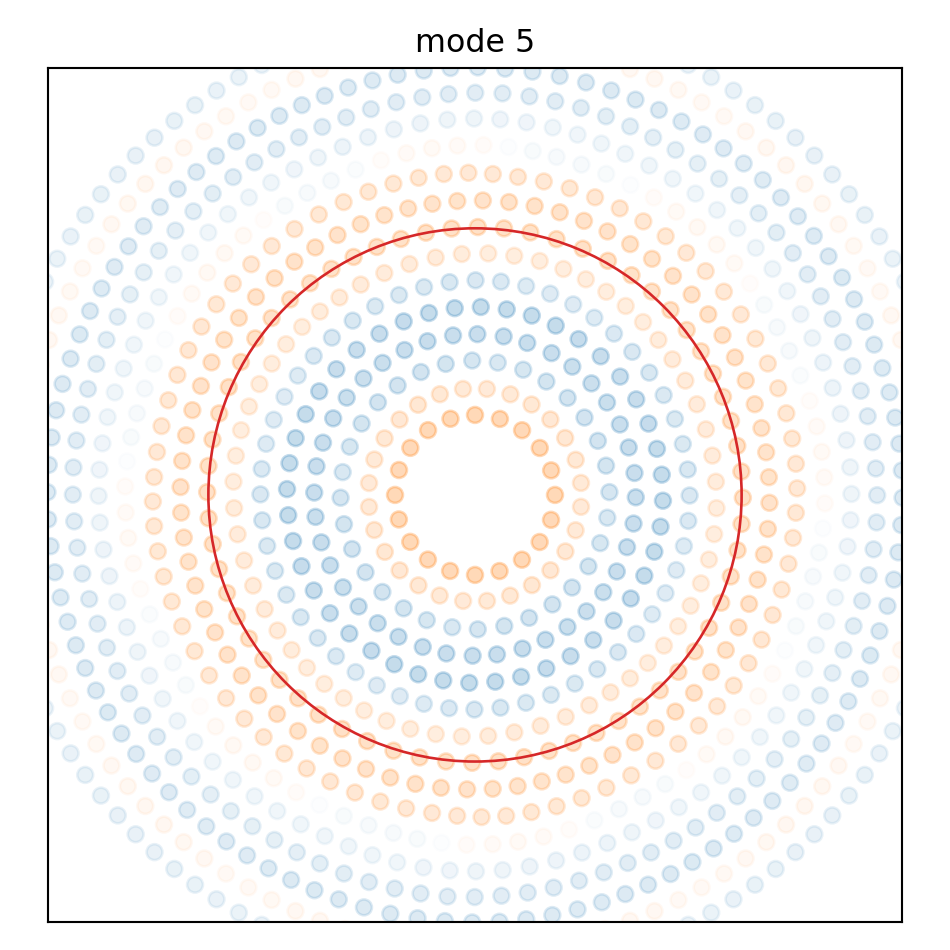}
  \end{subfigure}
  \begin{subfigure}{0.4\textwidth}
    \centering
    \includegraphics[width = \linewidth]{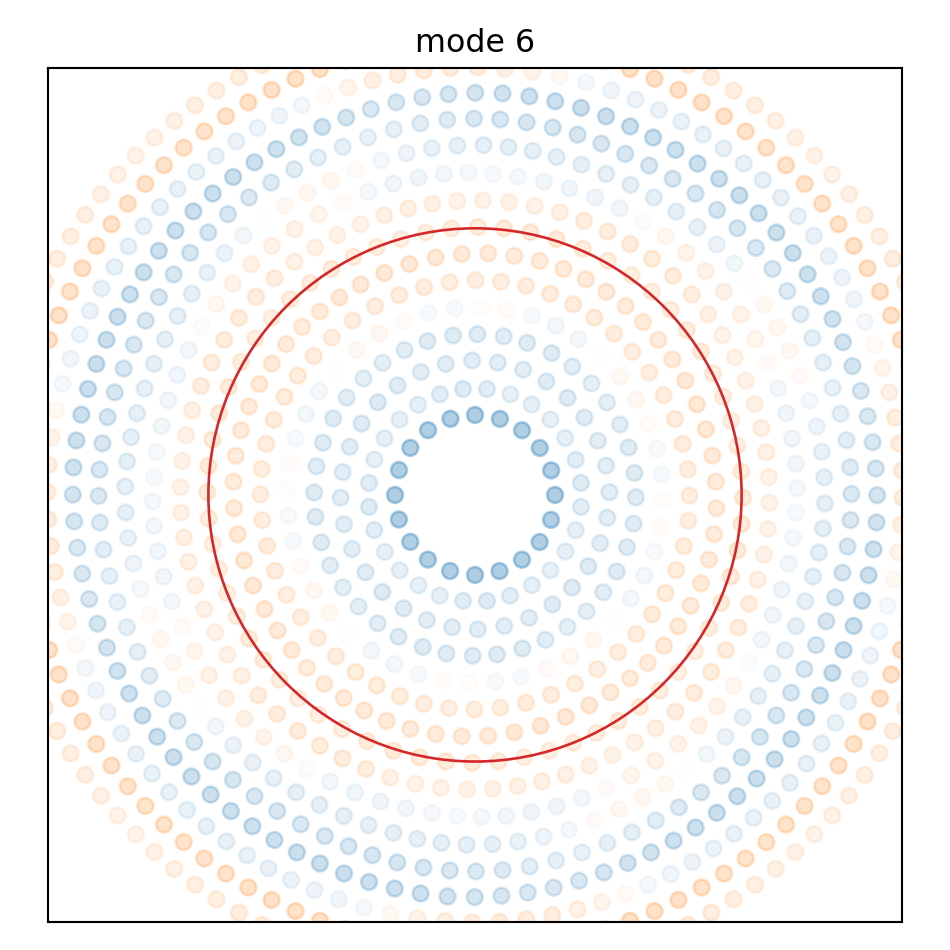}
  \end{subfigure}
  \caption{First six POD modes of density.}
  \label{fig:PODmode_illustation_density}
\end{figure}

Here we take initial particle distribution \gridC{} as an example, and choose SPH parameters $\lambda = 2 \e{-3}$, $\zeta = 4 \e{-1}$, and run the simulation for $2 \[ \mykgms{0}{0}{1} \]$ with $\Delta t = 1 \e{-5} \[ \mykgms{0}{0}{1} \]$.
We show the first six modes for particle position, velocity, temperature, and density in \autoref{fig:PODmode_illustation_position_x}, \autoref{fig:PODmode_illustation_velocity_x}, \autoref{fig:PODmode_illustation_temperature}, and \autoref{fig:PODmode_illustation_density}, respectively.
Since each POD mode is a weight over all particles from $\Lambda$, we can use the color opacity to represent such weight distribution. The weight can be either positive or negative on each particle, so we use two different colors to illustrate the sign. 
For better illustration, we plot the modes where particles are in the location $\mathbf{x} \( t ^{\star} \)$, $t ^{\star} = 1 \[ \mykgms{0}{0}{1} \]$. Since during POD-MOR, we uniformly pick snapshots on $\Theta = \[ 0, 2 \] \[ \mykgms{0}{0}{1} \]$, such illustration at the temporal midpoint shows a clearer pattern.

In \autoref{fig:PODmode_illustation_position_x}, the first few position modes (in the first spatial direction, $x _{1}$) focus on the outer particles and then on the inner particles since the position value is large for outer particles. For inner particles, the color on different particle layers has the phase shift, that is due to the different angular velocity during the simulation, the inner particles have larger angular velocity.

In \autoref{fig:PODmode_illustation_velocity_x}, the first few velocity modes (in the first spatial direction, $u _{1}$) focus more on the center particles compared with the position modes since the velocity field decays from the center to the boundary.

In \autoref{fig:PODmode_illustation_temperature}, the first temperature mode shows the global trend of the temperature profile that is higher near the center and lower near the boundary, while the second temperature mode indicates that the frictional work generates the heat within the region under the shoulder (inside the red circle in the plot). The following temperature modes show higher frequency in the radial direction and they are inhomogeneous in angular direction.

In \autoref{fig:PODmode_illustation_density}, the first density mode is almost homogeneous since the overall system is weakly compressible, while the second mode shows the density drop at the inner and outer boundaries due to insufficient neighbors for those particles. The following modes have almost radial symmetry but have higher frequency in the radial direction.

\myadd{
\subsection{Acceleration of Full FSSW Model}
}

\myadd{
The numerical results presented above rely on the direct projection method described in \autoref{eq:model_POD_directProjection} within the POD-MOR framework. While this approach effectively reduces the degrees of freedom (DoF) of the system, it does not inherently accelerate computation. 
In this subsection, we implement the linearization approach outlined in \autoref{eq:model_POD_linearizedMassConservation}, \autoref{eq:model_POD_linearizedLinearMomentumConservation}, and \autoref{eq:model_POD_linearizedLinearHeatEquation} by freezing certain terms during the SPH iterations.
}

\myadd{
We use the initial particle distribution \gridB{} as an example, selecting SPH parameters $\lambda = 2 \e{-3}$, $\zeta = 4 \e{-1}$, and running the simulation for $2 \[ \mykgms{0}{0}{1} \]$ with a time step $\Delta t = 1 \e{-5} \[ \mykgms{0}{0}{1} \]$. We fix a uniform $10\%$ of the modes for position, velocity, temperature, and density.
}

\myadd{
In \autoref{fig:PODacceleration_error}, we plot the POD error as a function of the number of freezing steps $f$. Here, a freezing step $f$ indicates that we update the frozen terms once every $f$ steps, with $f = 1$ representing updates at every step. The error at $f = 2$ is nearly identical to that at $f = 1$, while the error at $f = 5$ is almost double that at $f = 1$. This suggests that, with $10\%$ of POD modes in this configuration, the POD error remains acceptable for freezing steps $f \lesssim 5$.
}

\myadd{
Additionally, \autoref{fig:PODacceleration_error} illustrates the CPU time in relation to the different freezing steps $f$. The CPU time decreases similar to $f ^{-1}$, indicating that updating the frozen terms accounts for most of the computational time, whereas the update of the linear problem contributes only a small amount. For example, even at $f = 1$, the POD-MOR consumes approximately $80\%$ of the computational time compared to the full model without POD, as updating the reduced coordinates costs much less.
}

\myadd{
\begin{figure}[!htp]
  \centering
  \begin{subfigure}{\textwidth}
    \centering
    \includegraphics[width = \linewidth]{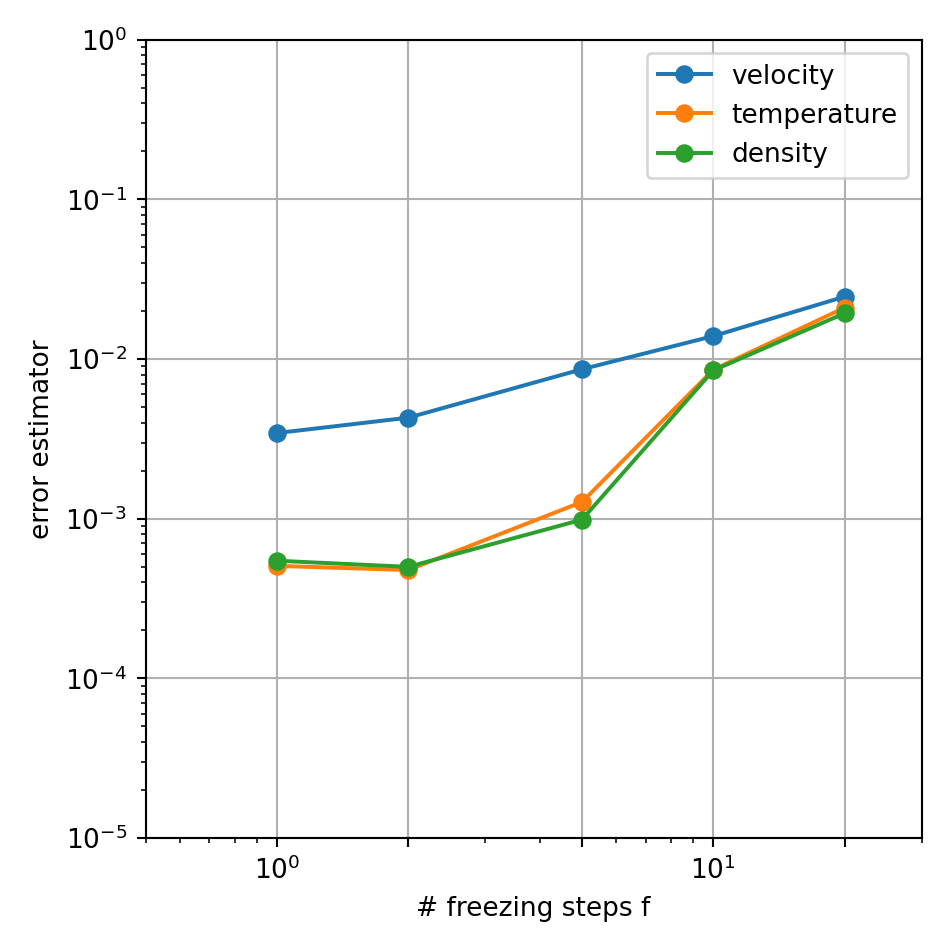}
  \end{subfigure}
  \caption{POD error vs. number of freezing steps with coupled model. Initial particle distribution \gridB{}, $\Theta = \[ 0, 2 \] \[ \textup{s} \]$, $\lambda = 2 \e{-3}$, $\zeta = 4 \e{-1}$. $\Delta t = 1 \e{-5} \[ \textup{s} \]$. Ratio of POD modes $10\%$. These results indicate that we can accelerate the POD-MOR for SPH simulation through the linearization approach. For example, in this setup, choosing $f = 2$ nearly doubles computational efficiency while allowing for approximately $25\%$ additional error.
}
  
  \label{fig:PODacceleration_error}
\end{figure}

\begin{figure}[!htp]
  \centering
  \begin{subfigure}{\textwidth}
    \centering
    \includegraphics[width = \linewidth]{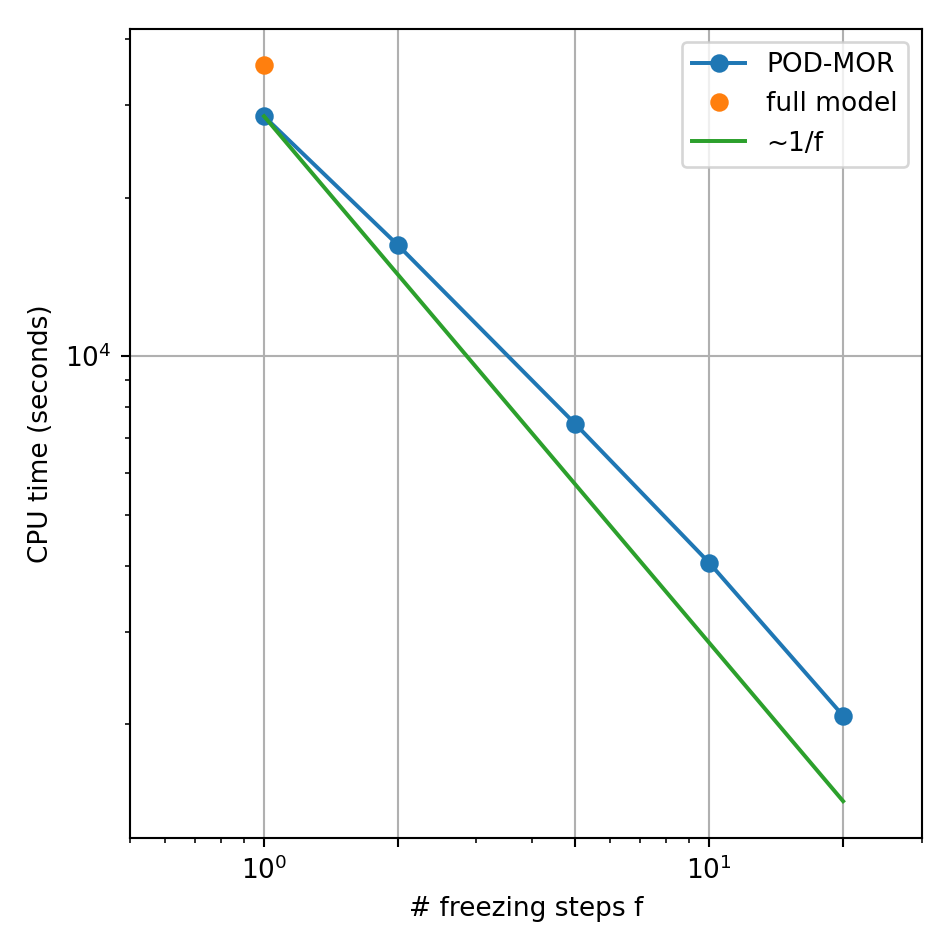}
  \end{subfigure}
  \caption{CPU time vs. number of freezing steps with coupled model. Initial particle distribution \gridB{}, $\Theta = \[ 0, 2 \] \[ \textup{s} \]$, $\lambda = 2 \e{-3}$, $\zeta = 4 \e{-1}$. $\Delta t = 1 \e{-5} \[ \textup{s} \]$. Ratio of POD modes $10\%$.}
  \label{fig:PODacceleration_time}
\end{figure}
}

\myadd{
We note that POD-MOR is commonly applied in scenarios requiring rapid computation while accepting certain level of errors, such as in solving inverse problems or parameter search problems. The full model can be utilized when higher accuracy is necessary, but, this acceleration of POD-MOR for SPH simulation holds significant value in various other scenarios.
}

\myadd{
\subsection{Predictive Setting with Various Tool Rotation Speeds}
}

\myadd{
The numerical results discussed above are based on a reproducible setting, where the physical process is fixed. To demonstrate the capability of our POD-MOR in a predictive setting, we choose the tool rotational speed (angular velocity $\omega$) as a parameter of the system.
}

\myadd{
The tool rotational speed is a crucial parameter in practice that controls the quality of welding in FSSW problems. In the previous setup, we fixed the RPM of the tool at $240$ in \autoref{tab:numerical_sphSetup_parameters}. In this subsection, we vary this value to study its impact on POD-MOR, implementing the linearization approach 
\autoref{eq:model_POD_linearizedMassConservation}, \autoref{eq:model_POD_linearizedLinearMomentumConservation}, and \autoref{eq:model_POD_linearizedLinearHeatEquation} by freezing some of the terms in most of the SPH iterations.
}

\myadd{
For simplicity, we focus on three different rotational speeds: the fast one $\omega = 360 \[ \textup{RPM} \]$, the medium one $\omega = 240 \[ \textup{RPM} \]$, and the slow one $\omega = 120 \[ \textup{RPM} \]$.
In \autoref{fig:DifferentRPM_DoFratio}, we plot the POD error as a function of the percentage of POD modes used relative to the total DoF, where a uniform percentage ($10\%$, $20\%$, ...) of position, velocity, temperature, and density modes is selected for each data point. We take both the rotational speed in the POD data (for training) and the rotational speed in a new simulation (for testing) from the three $\omega$ values.
}

\myadd{
In all cases, the POD error exhibits a sharp decay when the number of POD modes increases from $40\%$ to $50\%$, indicating that $50\%$ DoF is sufficient for reproducing the system. This is attributed to the centrosymmetric nature of the geometry and motion of the system.
The case where we choose training $\omega = 120 \[ \textup{RPM} \]$ and testing $\omega = 360 \[ \textup{RPM} \]$ shows a larger POD error. This occurs because the system evolving slowly with a lower tool rotational speed is insufficient for accurately capture the modes for a faster system, in other words, the POD data is not representative in this scenario. For the other three cases, where the testing $\omega$ is smaller than or equal to the training $\omega$, good predictions with similar trends are observed. Among them, the POD error is only slightly larger when the train $\omega = 360 \[ \textup{RPM} \]$ and the test $\omega = 120 \[ \textup{RPM} \]$ have a bigger gap.
}

\myadd{
\begin{figure}[!htp]
  \centering
  \begin{subfigure}{\textwidth}
    \centering
    \includegraphics[width = \linewidth]{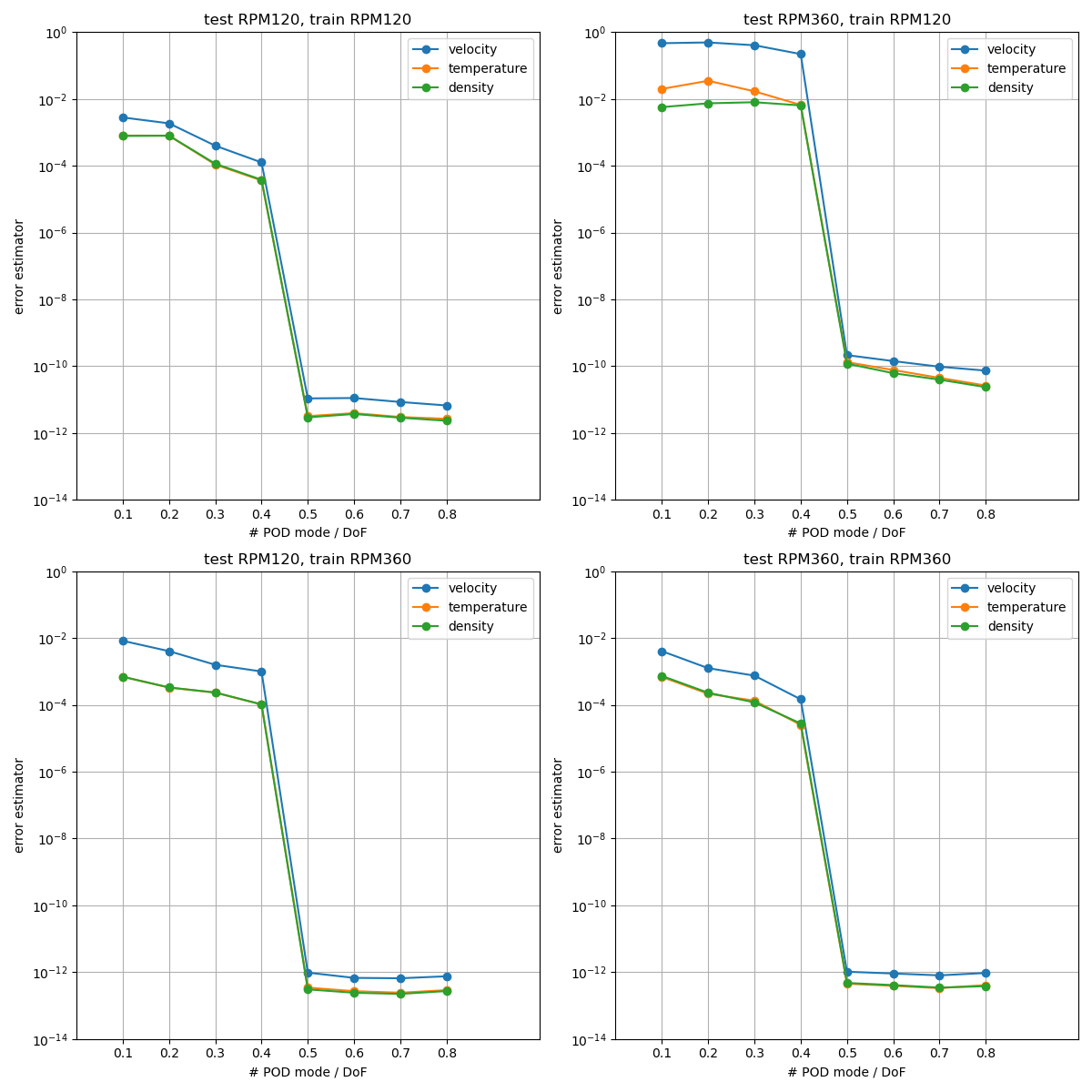}
  \end{subfigure}
  \caption{POD error vs. ratio of POD modes with coupled model. Initial particle distribution \gridB{}, $\Theta = \[ 0, 2 \] \[ \textup{s} \]$, $\lambda = 2 \e{-3}$, $\zeta = 4 \e{-1}$. $\Delta t = 1 \e{-5} \[ \textup{s} \]$. Top left: testing RPM $\omega=120$, training RPM $120$. Top right: testing RPM $360$, training RPM $120$. Bottom left: testing RPM $120$, train RPM $360$. Bottom right: testing RPM $360$, training RPM $360$.}
  \label{fig:DifferentRPM_DoFratio}
\end{figure}
}

\myadd{
In \autoref{fig:DifferentRPM_freezingStep}, we plot the POD error (between linearized models and the full model) against the number of freezing steps $f$, using $5\%$ of POD modes relative to the total DoF. We select the rotational speed in the POD data (train) from the fast, medium, and slow $\omega$ values, while fixing the rotational speed in a new simulation (test) at $\omega = 240 \[ \textup{RPM} \]$.
The case with train $\omega = 120 \[ \textup{RPM} \]$ and test $\omega = 240 \[ \textup{RPM} \]$ exhibits a larger POD error, as explained previously. For the other two cases, the corresponding plots show an increase in POD error with an increasing number of freezing steps $f$, but with decreasing computational time. In both cases, $f = 5$ results in an error that is less than twice the error at $f = 1$, with computaitonal time greatly reduced (similar to \autoref{fig:PODacceleration_time}). Comparing the middle plot with \autoref{fig:PODacceleration_error}, where $10\%$ of POD modes are used, we find that the POD error with $5\%$ of POD modes is larger, but the overall trend is similar, which is expected. When the training and testing $\omega$ values are the same, the POD error is smaller. However, even when they are different, we are still able to apply the acceleration approach of the POD-MOR for SPH simulation in predictive settings. 
}

\myadd{
Both sets of results demonstrate the capability of POD-MOR for SPH simulation in a predictive setting. This indicates that when the POD data does not perfectly align with the simulation we intend to implement, the POD-MOR and its acceleration can still perform effectively.
}

\myadd{
\begin{figure}[!htp]
  \centering
  \begin{subfigure}{\textwidth}
    \centering
    \includegraphics[width = \linewidth]{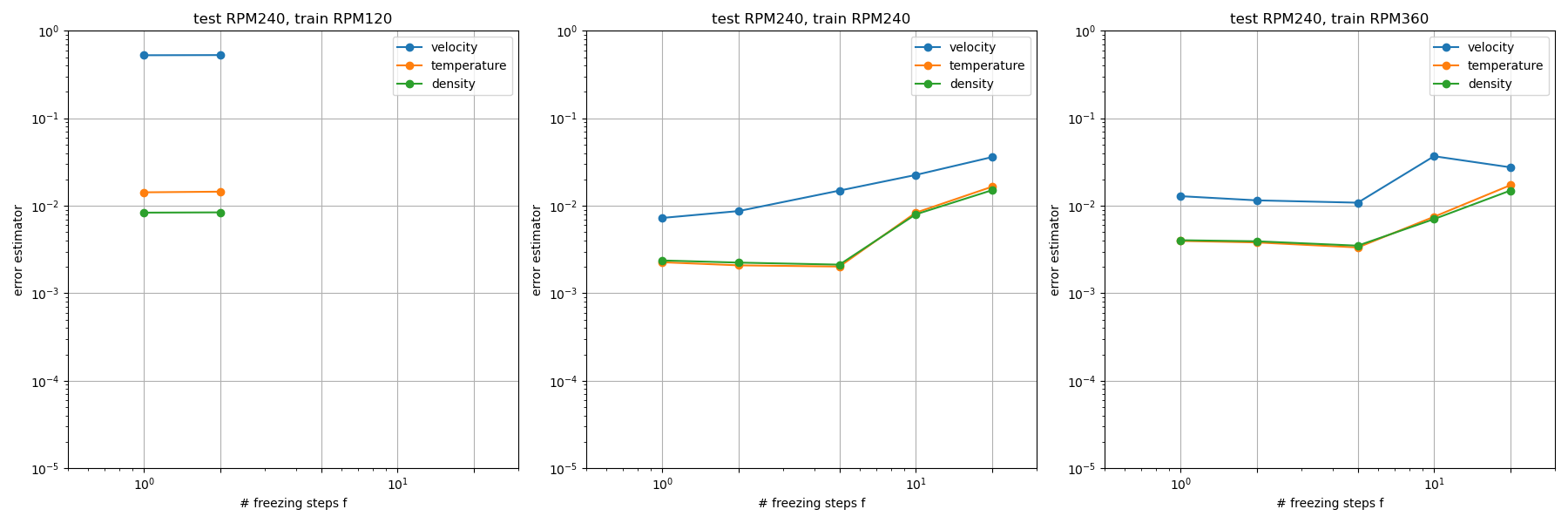}
  \end{subfigure}
  \caption{POD error vs. number of freezing steps with coupled model. Initial particle distribution \gridB{}, $\Theta = \[ 0, 2 \] \[ \textup{s} \]$, $\lambda = 2 \e{-3}$, $\zeta = 4 \e{-1}$. $\Delta t = 1 \e{-5} \[ \textup{s} \]$. Ratio of POD modes is $5\%$. Testing RPM $\omega=240$. Left: training RPM $120$. Middle: training RPM $240$. Right: training RPM $360$.}
  \label{fig:DifferentRPM_freezingStep}
\end{figure}
}


\section{Conclusion and Discussion}
\label{sec:conclusion}

Motivated by the success of POD-MOR techniques in the Eulerian framework, our study aims to explore the applicability of POD-MOR in SPH simulations. We have considered the two-dimensional FSSW problem as a representative example, which can be modeled by the nonlinear system that couples the heat equation and the flow equations.

By carefully selecting appropriate SPH parameters, we are able to quantitatively evaluate the POD error, as well as the error resulting from uniform coarsening, using our proposed error indicator. In the case of prescribed motion, we observe that POD-MOR proves to be highly efficient for heat equation, even when dealing with irregular prescribed motion. This finding aligns with the results obtained in the Eulerian framework.

Furthermore, in the case of prescribed temperature, we demonstrate the advantages of POD-MOR over the uniform reduction of particle numbers, as the POD error is significantly smaller. Since MOR in the mesh-free Lagrangian framework is generally not well understood, POD-MOR offers a promising strategy.

In the scenario involving full coupling, we plot the POD error as a function of the ratio of DoFs utilized, where we achieve a reduction of more than half of the DoFs with an associated relative error of approximately one percent. Therefore, POD provides a effective reduction for SPH simulations, whereas uniformly reducing the number of particles behaves much poorly.

The visualization of POD modes provides valuable physical insights into how POD organizes SPH particles and reduces the model's DoFs. Based on these intriguing observations, our work represents an initial step towards exploring the feasibility of POD-MOR for SPH simulations in other specific scenarios.

\myadd{
We also accelerate the computation in the online stage via  linearizing the nonlinear operators and freezing specific terms. Numerical results show that computational time can be significantly reduced while introducing only a mild additional error. 
It should be noted that other possible approaches may also address this challenge, including machine learning-based methods, non-global bases, and hyper-reduction techniques. 
We note that POD-MOR is frequently utilized in contexts where fast computations are essential and moderate errors are permissible, such as in inverse problem-solving and parameter optimization tasks.  Consequently, the acceleration of POD-MOR for SPH simulations is highly beneficial across a range of applications.
}

\myadd{
Our POD-MOR for SPH simulations has been shown to be effective in both reproducible and predictive settings across different tool rotational speeds. 
This indicates that our method has great potential for parameter study problems.
}

It is important to acknowledge the limitations of this study and possible future research. In this study, we intentionally select SPH parameters that result in regular flow patterns, which is the simplest and reasonable case. Comprehensive tests are required to assess the capability and stability of POD-MOR in various SPH setups.
\myadd{
While we acknowledge the success of Eulerian approaches that enable longer simulation times with POD-MOR, we wish to point out that achieving this in a Lagrangian framework is challenging, as it requires all particles to evole over the entire phase space, which is challenging in the current setup. We will consider this as a direction for future investigation. Another direction for future exploration involves investigating other model problems using SPH. Such an approach would enable a more comprehensive examination of specific challenges and provide deeper insights into the underlying issues.
}


In brief, our study serves as an initial exploration of the potential of POD-MOR for SPH simulations in specific scenarios. While limitations exist, future efforts should address these challenges and expand the application of this approach to more diverse SPH setups.

\section*{CRediT Author Statement}

\textbf{Lidong Fang}: Methodology, Software, Validation, Formal analysis, Investigation, Data Curation, Writing---Original Draft, Writing---Review \& Editing, Visualization.
\textbf{Zilong Song}: Methodology, Software, Formal analysis, Data Curation, Writing---Review \& Editing.
\textbf{Kirk Fraser}: Conceptualization, Methodology, Software, Writing---Review \& Editing, Supervision.
\textbf{Faisal Habib}: Methodology, Software, Resources, Data Curation, Writing---Review \& Editing.
\textbf{Christopher Drummond}: Conceptualization, Writing---Review \& Editing, Funding acquisition.
\textbf{Huaxiong Huang}: Conceptualization, Resources, Writing---Review \& Editing, Supervision, Project administration, Funding acquisition.

\section*{Acknowledgement}

Dr. Hamid Reza Karbasian contributed to this research by developing the initial code for SPH and POD but with a different formulation.

\bibliographystyle{elsarticle-harv} 
\bibliography{notes_FSWSPHPOD}

\end{document}